\documentclass[a4paper, fleqn, usenatbib]{mnras}
\usepackage{newtxtext,newtxmath}
\usepackage[T1]{fontenc}
\usepackage{ae,aecompl,morefloats}
\usepackage{graphicx}	% Including figure files
\usepackage{amsmath}	% Advanced maths commands
\usepackage{amssymb}	% Extra maths symbols
\usepackage{pdflscape} 	%install landscape
\usepackage{times}

\def \aj {AJ}
\def \mnras {MNRAS}
\def \pasp {PASP}
\def \apj {ApJ}
\def \apjs {ApJS}
\def \apjl {ApJL}
\def \aap {A\&A}
\def \nat {Nature}
\def \araa {ARAA}
\def \iaucirc {IAUC}
\def \aaps {A\&A Suppl.}
\def \pasa {PASA}

\def\lesssim{\mathrel{\hbox{\rlap{\hbox{\lower4pt\hbox{$\sim$}}}\hbox{$<$}}}}
\def\gtrsim{\mathrel{\hbox{\rlap{\hbox{\lower4pt\hbox{$\sim$}}}\hbox{$>$}}}}

\long\def\symbolfootnote[#1]#2{\begingroup%
\def\thefootnote{\fnsymbol{footnote}}\footnote[#1]{#2}\endgroup}

\title[The sites of stripped-envelope SNe]{The very young resolved stellar populations around stripped-envelope supernovae}

\author[Maund]{Justyn R. ~Maund\thanks{email: j.maund@sheffield.ac.uk}\thanks{Royal Society Research Fellow}\\
Department of Physics and Astronomy, University of Sheffield, Hicks Building, Hounsfield Road, Sheffield S3 7RH, U.K.\\
}

\date{Accepted XXX. Received YYY; in original form ZZZ}

\pubyear{2016}

\begin{document}
\label{firstpage}
\pagerange{\pageref{firstpage}--\pageref{lastpage}}
\maketitle
%%%%%%%%%%%%%%%%%%%%%%%%%%%%%%%%
%ABSTRACT
%ABSTRACT
%ABSTRACT
%%%%%%%%%%%%%%%%%%%%%%%%%%%%%%%%

\begin{abstract}
The massive star origins for Type IIP supernovae (SNe) have been established through direct detection of their red supergiants progenitors in pre-explosion observations; however, there has been limited success in the detection of the progenitors of H-deficient SNe.  The final fate of more massive stars, capable of undergoing a Wolf-Rayet phase, and the origins of Type Ibc SNe remains debated, including the relative importance of single massive star progenitors or lower mass stars stripped in binaries.
We present an analysis of the ages and spatial distributions of massive stars around the sites of 23 stripped-envelope SNe, as observed with the Hubble Space Telescope, to probe the possible origins of the progenitors of these events.   Using a Bayesian stellar populations analysis scheme, we find characteristic ages for the populations observed within $150\,\mathrm{pc}$ of the target Type IIb, Ib and Ic SNe to be $\log (t) = 7.20$, $7.05$ and $6.57$, respectively.  The Type Ic SNe in the sample are nearly all observed within $100\,\mathrm{pc}$ of young, dense stellar populations.  The environment around SN 2002ap is an important exception both in terms of age and spatial properties.  These findings may support the hypothesis that stars with $M_{init} > 30M_{\odot}$ produce a relatively large proportion of Type Ibc SNe, and that these SN subtypes arise from progressively more massive progenitors.   Significantly higher extinctions are derived towards the populations hosting these SNe than previously used in analysis of constraints from pre-explosion observations. The large initial masses inferred for the progenitors are in stark contrast with the low ejecta masses estimated from SN light curves.
\end{abstract}
%%%%%%%%%%%%%%%%%%%%%%%%%%%%%%%%
%KEYWORDS
%KEYWORDS
%KEYWORDS
%%%%%%%%%%%%%%%%%%%%%%%%%%%%%%%%

%%%%%%%%%%%%%%%%%%%%%%%%%%%%%%%%
%INTRODUCTION
%INTRODUCTION
%INTRODUCTION
%INTRODUCTION
%INTRODUCTION
%INTRODUCTION
%INTRODUCTION
%%%%%%%%%%%%%%%%%%%%%%%%%%%%%%%%
\section{Introduction} 
\label{sec:intro}
Core-collapse supernovae (CCSNe) are expected to be the end products of the evolution of massive stars with $M_{init} > 8M_{\odot}$.  There has been some success in the direct identification of the red supergiant (RSG) stars responsible for the hydrogen-rich Type IIP SNe \citep[see e.g.][]{2008arXiv0809.0403S}, however these represent only the least massive of the stars that might explode as CCSNe ($M_{init} \lesssim 16M_{\odot}$).   The canonical prediction of single-star stellar evolution models is that higher mass stars will experience strong mass loss through stellar winds, removing their outer hydrogen envelopes prior to explosion and, ultimately, yielding hydrogen-poor CCSNe \citep{eld04,nombin96}.  The progenitors of such SNe might, therefore, be expected to be Wolf-Rayet (WR) stars with $M_{init} \gtrsim 30M_{\odot}$.  Alternatively, such mass loss might also occur through mass transfer from the progenitor onto a massive binary companion \citep{izzgrb}, meaning that lower mass stars ($8 \lesssim M_{init} \lesssim 30M_{\odot}$) might also produce H-poor SNe.  Statistical studies of the local SN rate \citep{2011MNRAS.412.1522S} and constraints on the progenitors from non-detections in pre-explosion observations \citep{2013MNRAS.436..774E} suggest that both the single and binary progenitor channels are required to produce the observed population of Type Ibc SNe.

The workhorse facility for the acquisition of pre-explosion images, in which the bulk of progenitors have been detected, has been the {\it Hubble Space Telescope} (HST).  This is due to the ability of HST to resolve individual stars in other galaxies out to large distances ($\lesssim 30\,\mathrm{Mpc}$) and the high-quality of its well maintained public archive\footnote{https://archive.stsci.edu/hst/}.  The search for the progenitors of Type Ibc SNe has, however, yielded only a single detection in pre-explosion observations \citep{2013ApJ...775L...7C, 2016MNRAS.461L.117E, 2016ApJ...825L..22F}.  The general dearth of detections is expected to be primarily due to either the high temperature of the progenitor, requiring deep ultraviolet (UV) observations sensitive to the high temperatures of WR stars ($ \geq 50\,\mathrm{kK}$), or the relative faintness of lower mass progenitors in binary systems \citep{2016MNRAS.461L.117E}.  As shown by \citet{2012A&A...544L..11Y}, the interpretation of detection limits on the progenitors of Type Ibc SNe from pre-explosion observations, or the detection of a hypothetical high temperature progenitor, would not necessarily yield a unique solution for the initial mass.  For example, the study of \citet{2013MNRAS.436..774E} just compared the available detection limits with observations of Galactic and Large Magellanic Cloud (LMC) WR stars, rather than placing a definitive mass limit on the progenitors.  The interpretation of the detection limits for the progenitors is complicated by the high levels of extinction observed towards Type Ibc SNe \citep{2011ApJ...741...97D}, which would preferentially diminish their blue SEDs.  Furthermore, the two progenitor channels are not mutually exclusive, as massive stars that undergo a WR phase might also have a binary companion. Indeed the majority of massive stars are expected to have a binary companion, with which they are expected to strongly interact \citep{2012Sci...337..444S}.  In the case of the Type Ibn SN~2006jc, a possible binary companion may have been recovered in late-time observations by \citet{2016ApJ...833..128M}.  

The paucity of directly identified progenitors for Type Ibc SNe is in stark contrast to the relative success in the detection of the progenitors of the hydrogen-poor Type IIb SNe, for which luminous cool progenitors (both yellow supergiants (YSGs) and RSGs) have been identified in pre-explosion images \citep{alder93j,2008MNRAS.391L...5C,2011ApJ...739L..37M,2011ApJ...741L..28V,2013ApJ...772L..32V,2014AJ....147...37V}.  In the case of SN~1993J, a binary companion has been found in late-time spectroscopic and photometric observations \citep{maund93j,2014ApJ...790...17F}.  In the case of the Type IIb SN~2011dh, however, the presence of a binary companion associated with the YSG progenitor is still hotly debated \citep{2014ApJ...793L..22F,2015MNRAS.454.2580M}.  As discussed by \citet{2012A&A...538L...8G}, specifically in the case of the progenitor of SN~2011dh, and \citet{2014ARA&A..52..487S}, the nature of mass loss from massive stars is still highly uncertain, even without considering binary interactions.

Given the difficulties in assessing the evolutionary pathways giving rise to H-poor SNe and, particularly, given the lack of a large number of detections of the most H-deficient SN progenitors, it has become common for alternative measures of the nature of the progenitor to be sought.  Recent studies of photometric observations of stripped-envelope SNe \citep[see e.g.][and references therein]{2013MNRAS.434.1098C,2016MNRAS.457..328L, 2016MNRAS.458.2973P,2017arXiv170707614T} have used light curve modelling \citep{1982ApJ...253..785A} to determine properties such as the ejected mass.  Mostly recently, \citet{2017arXiv170707614T} claim the the ejecta masses determined for their sample of stripped-envelope SNe to be $M_{ej} \lesssim 5 - 6M_{\odot}$; implying that the progenitors are nearly all lower mass stars ($M_{init}<20M_{\odot}$) in binaries. Studies of late-time nebular spectra can yield abundances which have been used to constrain the oxygen mass of the progenitors' cores; such as \citet{2015A&A...573A..12J} who derive similar progenitor masses to those derived from pre-explosion observations.

Alternatively, analysis of the host environment may provide clues to the nature of progenitors that do not depend on the interpretation of observations of the SNe themselves.  Studies of the metallicities of SN sites, derived from {\sc H ii} regions in the host galaxies, reveal a possible trend for more H-deficient SNe arising in more metal rich environments \citep{2008ApJ...673..999P,2010MNRAS.407.2660A,2011ApJ...731L...4M}, which could in turn be related to the mass loss rates of the progenitors \citep{2014ARA&A..52..487S}. More recently, this type of analysis has progressed to spatially resolved scales using integral field spectroscopy to determine ages through the strength of $\mathrm{H\alpha}$ emission at or in close proximity to the SN site \citep{2013AJ....146...30K,2013AJ....146...31K}.   In addition, the relative association of SN positions using pixel statistic measures and flux proxies of star formation rate (e.g. $\mathrm{H\alpha}$)  has led to the suggestion that more H-deficient SNe are found in closer proximity to recent star formation and, hence, are associated with higher mass progenitors \citep[for a review see][]{2015PASA...32...19A}.  To be applicable to providing direct constraints on the progenitors of SNe, however, these types of studies are missing fundamental information about how environmental properties, that can only be measured in bulk (such as emission line strengths or integrated broad-band fluxes) are coupled to the evolution of the individual star that exploded.

The resolved stellar population around a SN position provides an alternative proxy for the age of the progenitor, from the age of the surrounding stars, and hence its initial mass.  The underlying assumption of the analysis is that the progenitor is coeval with the massive star population observed at the SN position and relies on well-established methods for the analysis of ensembles of stars. Previously \citet{1999AJ....118.2331V}, \citet{2005astro.ph..1323M}, \citet{2009ApJ...703..300G}, \citet{2011ApJ...742L...4M} and \citet{2014ApJ...791..105W} have considered the resolved stellar populations associated with SNe for which a search for the progenitor was also attempted; and, as this approach does not rely on fortuitous pre-explosion observations, progenitor constraints have also been provided from the populations associated with SN remnants in the LMC \citep{2009ApJ...700..727B} and M31 and M33 \citep{2012ApJ...761...26J, 2014ApJ...795..170J}, as well as for SN 1979C \citep{1999PASP..111..313V}.  More recently \citet{2016MNRAS.456.3175M} and \citet{2017arXiv170401957M} have considered a Bayesian scheme for the analysis of stellar populations around a Type Ic and 12 Type IIP SNe, respectively.  For the Type Ic SN 2007gr, \citeauthor{2016MNRAS.456.3175M} found that the SN position was located at the centre of a large massive star complex, with an approximate diameter of $\sim 300\,\mathrm{pc}$ and concluded that a single star, having undergone a WR phase, was a viable candidate for this SN despite no detection of the progenitor in pre-explosion observations \citep{2008ApJ...672L..99C,2014ApJ...790..120C}.  For Type IIP SNe, with identified progenitors, \citet{2017arXiv170401957M} was able to show that, while there were mixed aged populations within $100\,\mathrm{pc}$ of the SN positions, with significant levels of differential extinction, there were clear populations with ages directly correlated with the initial mass inferred from pre-explosion images.  

Here we present an analysis of the resolved stellar populations around the sites of 23 Type IIb, Ib and Ic SNe, as imaged using HST. In Section \ref{sec:obs} we present the sample, the observations and their analysis.  In Sections \ref{sec:res:age} and \ref{sec:res:spatial} we analyse the age and spatial distributions of the surrounding resolved stellar populations and in section \ref{sec:disc} we discuss the overall properties of the ensemble of environments.

%%%%%%%%%%%%%%%%%%%%%%%%%%%%%%%%
%OBSERVATIONS AND ANALYSIS
%OBSERVATIONS AND ANALYSIS
%OBSERVATIONS AND ANALYSIS
%OBSERVATIONS AND ANALYSIS
%OBSERVATIONS AND ANALYSIS
%OBSERVATIONS AND ANALYSIS
%OBSERVATIONS AND ANALYSIS
%OBSERVATIONS AND ANALYSIS
%%%%%%%%%%%%%%%%%%%%%%%%%%%%%%%%
\section{Sample and Observations}
\label{sec:obs}
%
%Sample Selection
%
\subsection{Sample Selection}
\label{sec:obs:sample}
We consider a sample of stripped-envelope SNe that have been observed either before or after explosion with HST in at least two filters.  For this study, we have primarily considered those stripped-envelope SNe observed out to a distance of $\sim 30\,\mathrm{Mpc}$, with contemporaneous observations in at least two filters acquired with the Wide Field Channel 3 Ultraviolet-Visible Channel (WFC3 UVIS), Advanced Camera for Surveys (ACS) High Resolution Channel (HRC) and Wide Field Channel (WFC) and the Wide Field Planetary Camera 2 (WFPC2) instruments.  For those SN locations covered by WFPC2 observations, we only considered those observations that were purposefully acquired for imaging the SN location to ensure that the depth of the observations was comparable to those acquired with the newer, more efficient HST instrumentation.  Furthermore, while it might have been possible to extend this study to consider ground based observations of the sites of nearer SNe ($\lesssim10\,\mathrm{Mpc}$), we have elected to restrict our sample SN sites to those observed with HST to ensure a  degree of uniformity in the data quality across the sample.  Using these criteria we made an initial selection of 23 SNe with suitable HST observations as presented in Table \ref{tab:obs:snlist}; although this is not an exhaustive list of all possible targets.

\begin{table*}
\caption{The sample of 23 stripped-envelope SNe with Hubble Space Telescope observations considered in this study. \label{tab:obs:snlist}}
\begin{tabular}{ccccccc}
\hline
Object		&	Type		&	Host		&	$A_{V}$		&	$\mu$	  		&	$i$		& Progenitor \\
			&			&			&	(mags)$^{1}$ 	&	(mags)	  		& (degrees)$^{2}$&Ref.\\ 
\hline
1993J$^{3}$ 	&	IIb 		& 	M81 		& 0.220 			& 27.75(0.08)$^{4}$ 		& 	62.7		&$5$ \\
1994I$^{6}$	&  	Ic$^{7}$	&	M51 		& 0.096			& 29.26(0.37)$^{8}$		& 	32.6		&$9$\\
1996aq$^{10}$ 	&	Ic$^{10}$	& NGC 5584 	& 0.107			& 31.72 (0.03)$^{11}$ 	& 	42.4		&$\ldots$\\
1996cb$^{12}$ 	&	IIb$^{12}$	&NGC 3510 	& 0.084 			&30.55 (0.40)$^{13}$ 	&	78.1		&$\ldots$ \\
2000ew$^{14}$ & 	Ic$^{15}$	&NGC3810 	& 0.121 			& 30.85(0.40)$^{16}$ 	& 	48.2		&17 \\
2001B$^{18}$ 	& 	Ib$^{19}$	&IC 391 		& 0.349 			& 32.03(0.40)$^{20}$	& 	18.1		&17\\
2001gd$^{21}$ 	& 	IIb$^{22}$	&NGC 5033 	& 0.031 			& 31.34(0.13)$^{13}$	& 	64.6		&$\ldots$\\
2002ap$^{23}$ 	& 	Ic$^{24}$	&M74 		& 0.192 			& 29.84(0.42)$^{25}$	& 	19.8		&26\\
2004gn$^{27}$ 	& 	Ib$^{28}$	&NGC 4527 	& 0.060 			& 30.66(0.10)$^{29}$	& 	81.2		&$\ldots$\\
2004gq$^{30}$ 	& 	Ib$^{31}$	&NGC 1832 	& 0.200 			& 32.00(0.24)$^{13}$	& 	71.8		&$\ldots$\\
2004gt$^{32}$ 	& 	Ic$^{33}$	& NGC 4038 	& 0.127 			& 31.57(0.07)$^{34}$	& 	51.9		& 35\\
2005at$^{36}$	&	Ic$^{37}$	&NGC 6744	& 0.118			& 29.81(0.09)$^{38}$	&	53.5		& 39\\
2007fo$^{40}$		&	Ib$^{41}$	& NGC 7714	&0.144			& 32.25(0.27)$^{42}$ & 	45.1		&$\ldots$\\
2008ax$^{43}$ 		& 	IIb$^{44}$	& NGC 4490 	& 0.060 			& 30.02(0.40)$^{42}$ &  	90.0		&45\\
2009jf$^{46}$		& 	Ib$^{47}$	& NGC 7479	& 0.309			& 32.49(0.30)$^{13}$ & 	43.0		&48 \\
2011dh$^{49}$ 		& 	IIb$^{50}$	& M51 		& 0.099			& 29.26(0.37)$^{8}$ & 	32.6		& 51\\

2012P$^{52}$ 		& 	IIb$^{53}$	&NGC 5806 	& 0.139 			& 32.09(0.12)$^{13}$ & 	60.4		&$\ldots$\\

2012au$^{54}$ 		& 	Ib$^{55}$	&NGC 4790	& 0.132 			& 31.74(0.3)$^{42}$ & 	58.8		& 56\\

2012fh$^{57}$ 		& 	Ic$^{57}$	&NGC 3344 	& 0.090 			& 29.96(0.10)$^{38}$ & 	18.7		& 58\\

iPTF13bvn$^{59}$	 & 	Ib$^{60}$	&NGC 5806 	& 0.139 			& 32.09(0.12)$^{13}$  &	60.4		& 61  \\
2013df$^{62}$ 		& 	IIb$^{62}$	&NGC 4414 	& 0.053 			& 31.20(0.11)$^{63}$ & 	56.6		& 64\\ 

2013dk$^{65}$ 		& 	Ic$^{66}$	&NGC 4038 	& 0.127 			& 31.57(0.07)$^{34}$ & 	51.9		& 67\\
2016bau$^{68}$ 		&	Ib$^{69}$	&NGC 3631	& 0.045			& 29.72(0.27)$^{42}$ & 	34.7	& $\ldots$\\
\hline
\end{tabular}\\
$^{1}$ Foreground extinction towards the SN from \citet{2011ApJ...737..103S} as quoted by the NASA/IPAC Extragalactic Database (https://ned.ipac.caltech.edu); 
$^{2}$ Host galaxy inclination quoted by HyperLEDA (http://leda.univ-lyon1.fr/) \citep{2014A&A...570A..13M}; $^{3}$ \citet{1993IAUC.5731....1R}; $^4$ \citet{2001ApJ...553...47F}; $^5$ \citet{alder93j}, \citet{2002PASP..114.1322V}, \citep{maund93j}, \citet{2014ApJ...790...17F}; $^{6}$ \citet{1994IAUC.5961....1P}; $^{7}$ \citet{1994ApJ...436L.135W}; $^{8}$ \citet{2006MNRAS.372.1735T}; $^{9}$ \citet{1996AJ....111.2047B}, \citet{2016ApJ...818...75V}; $^{10}$ \citet{1996IAUC.6454....1N}; $^{11}$ Weighted average of the Cepheid distance measurements reported by \citet{2011ApJ...730..119R}, \citet{2013MNRAS.434.2866F} and \citet{2013AJ....146...86T}; $^{12}$ \citet{1996IAUC.6524....1N}; $^{13}$ \citet{2009AJ....138..323T}; $^{14}$ \citet{00ewiauc1}; $^{15}$ \citet{00ewiauc3}; $^{16}$ \citet{2014MNRAS.444..527S}; $^{17}$ \citet{2005astro.ph..1323M}; $^{18}$ \citet{01biauc1}; $^{19}$ \citet{01biauc3}; $^{20}$  \citet{1988ang..book.....T}; $^{21}$ \citet{2001IAUC.7761....1N}; $^{22}$ \citet{2001IAUC.7765....2M}; $^{23}$ \citet{2002IAUC.7810....1N}; $^{24}$   \citet{2002IAUC.7811....1K}, \citet{2002IAUC.7811....2M}, \citet{2002IAUC.7811....3G}; $^{25}$ \citet{2005MNRAS.359..906H}; $^{26}$ \citet{2007MNRAS.381..835C}, \citet{2013MNRAS.436..774E}; $^{27}$ \citet{2004CBET..100....1L}; $^{28}$ \citet{2005PASP..117..773V};  $^{29}$ Weighted average of Cepheid distance measurements reported by \citet{2006ApJS..165..108S}, \citet{2001ApJ...551..973S},\citet{2013AJ....146...86T}, \citet{2003A&A...411..361K} and \citet{2001ApJ...547L.103G};  $^{30}$  \citet{2004IAUC.8452....2P}; $^{31}$ \citet{2004IAUC.8452....3F}; $^{32}$ \citet{2004IAUC.8454....1M}; $^{33}$ \citet{2004IAUC.8456....4G}; $^{34}$ \citet{2008AJ....136.1482S}; $^{35}$ \citet{2005ApJ...630L..33M, 2005ApJ...630L..29G}; $^{36}$ \citet{2005CBET..119....1M}; $^{37}$ \citet{2005CBET..122....1S}; $^{38}$ \citet{2009AJ....138..332J}; $^{39}$ \citet{2014A&A...572A..75K}; $^{40}$ \citet{2007CBET..997....1K}; $^{41}$ \citet{2007CBET.1001....2D}; $^{42}$ Weighted average of Tully-Fisher distances reported by \citet{2007A&A...465...71T}; $^{43}$ \citet{2008CBET.1280....1M}; $^{44}$ \citet{2008CBET.1285....1B} and \citet{2008CBET.1298....1C}; $^{45}$ \citet{2008MNRAS.391L...5C}; $^{46}$  \citet{2009CBET.1952....1L}; $^{47}$ \citet{2009CBET.1955....1K} and \citet{2009CBET.1955....2S}; $^{48}$  \citet{2011MNRAS.416.3138V}; $^{49}$ \citet{CBET2736}; $^{50}$ \citet{CBET2736a} and \citet{CBET2736b}; $^{51}$ \citet{2011ApJ...739L..37M} and \citet{2011ApJ...742L...4M}; $^{52}$ \citet{2012ATel.3881....1A}; $^{53}$ \citet{2012CBET.2993....1D}; $^{54}$ \citet{2012ATel.3967....1H}; $^{55}$ \citet{2012ATel.3968....1S}; $^{56}$ \citet{2012ATel.3971....1V}; $^{57}$ \citet{2012CBET.3263....1N}; $^{58}$ \citet{2012ATel.4502....1P}; $^{59}$ \citet{2013ATel.5137....1C}; $^{60}$ \citet{2013ATel.5142....1M}; $^{61}$ \citet{2013ApJ...775L...7C}; $^{62}$ \citet{2013CBET.3557....1C}; $^{63}$ \citet{2003A&A...411..361K}; $^{64}$ \citet{2014AJ....147...37V}; $^{65}$ \citet{2013CBET.3565....1C}; $^{66}$ \citet{2013ATel.5165....1H}; $^{67}$ \citet{2013MNRAS.436L.109E}; $^{68}$ \citet{2016ATel.8875....1R}; $^{69}$  \citet{2016ATel.8818....1G}.
\end{table*} 
%
%Observations
%
\subsection{Observations and Photometric Analysis}
\label{sec:obs:observations}

\begin{figure*}
\includegraphics[width=18.0cm]{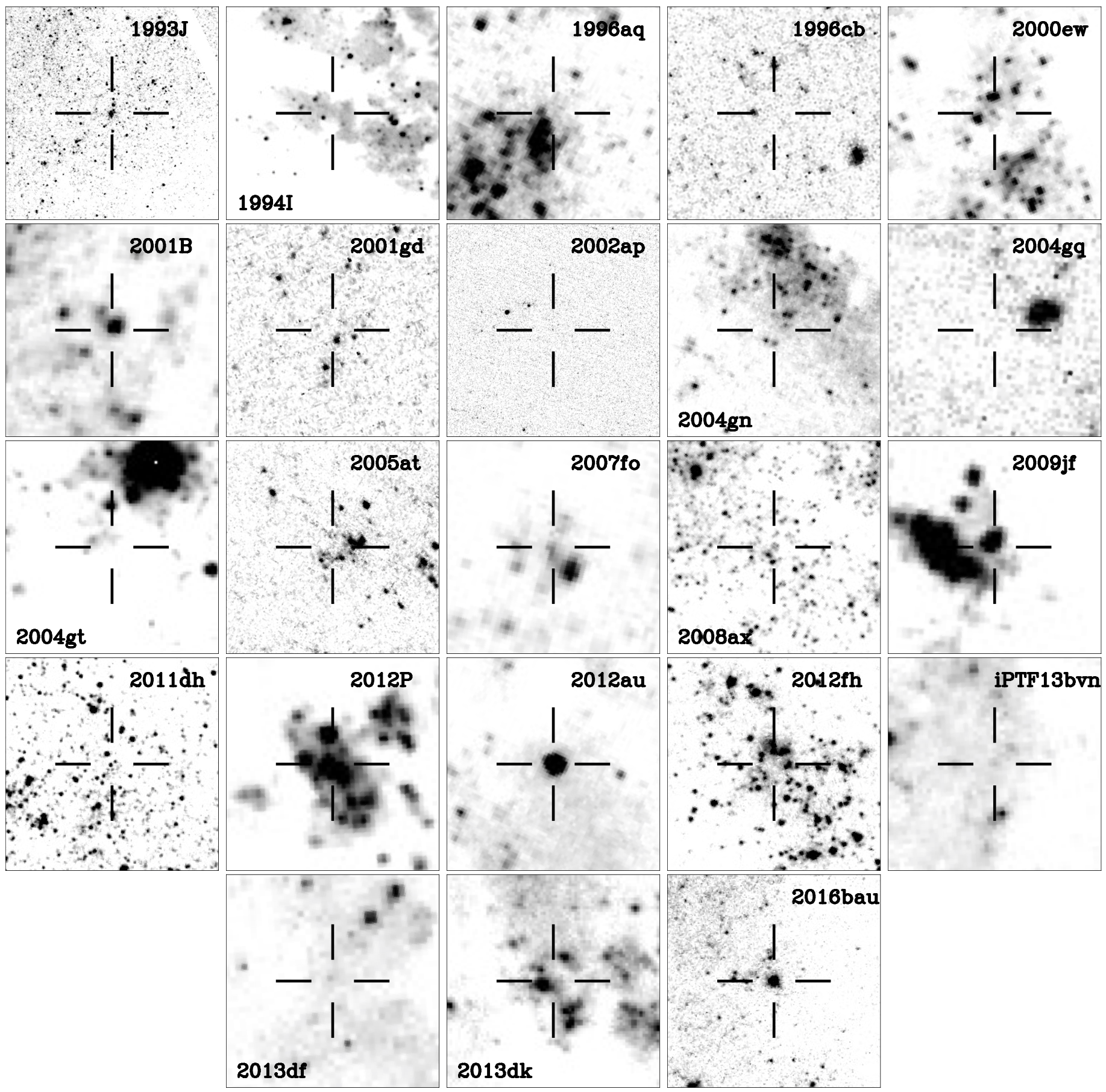}
\caption{$V$-band ($F555W$/$F606W$) Hubble Space Telescope images of the sites of the sample of 23 stripped-envelope SNe.  All images have been scaled to have dimensions $300 \times 300\,\mathrm{pc}$, are centred on the position of the target SN and are oriented such that North is up and East is left.}
\label{fig:res:stamp}
\end{figure*}

Observations of the sites of the target SNe were retrieved from the Mikulski Archive for Space Telescopes, having been processed with the latest calibrations as part of the On-the-fly recalibration pipeline.  A detailed log of the observations for the target SNe is presented in Table \ref{tab:obs:observations}, however specific characteristics of each set of observations are discussed in Section \ref{sec:res:individual}.

In general, for each SN, the available observations were acquired at the same pointing and same epoch, permitting directly correlated photometry on all the frames.  In instances where there were slight spatial offsets between the images, the deepest image was used to define a reference astrometric coordinate system to which the other images were aligned using the {\tt tweakreg} task as part of the {\tt drizzlepac} package running in the PyRAF environment \footnote{http://www.stsci.edu/institute/software\_hardware/pyraf - PyRAF is product of the Space Telescope Science Institute, which is operated by AURA for NASA}. Photometry of these observations was conducted using the DOLPHOT photometric package\footnote{http://americano.dolphinsim.com/dolphot/} \citep{dolphhstphot}, which automatically refines the alignment solution between the input images, conducts point-spread function (PSF) fitting photometry, determines empirical aperture corrections and, where required, corrected for charge transfer inefficiency (CTI).  In some cases, for WFC3 UVIS and ACS/WFC observations, the images were subject to correction for CTI using the pixel-based scheme of \citet{2010PASP..122.1035A} as part of the data retrieval process and so further CTI corrections using DOLPHOT were not applied.  To ensure uniformity in the photometry across all observations for all the SN locations in the sample, we used the DOLPHOT settings for the ACS/WFC and WFC3/UVIS observations proposed by \citet{2014ApJS..215....9W} for the Panchromatic Hubble Andromeda Treasury\footnote{https://archive.stsci.edu/prepds/phat/} data.  For the small amount of ACS/HRC and WFPC2 data in the sample, we used the standard settings listed in the respective DOLPHOT manuals.

The amount of supporting observations for each of the SNe varied widely depending on when the SN occurred, the brightness of the SN and the availability of follow-up observations.   The most immediate impact of this diversity was in the availability of high quality observations from which to identify the SN position in the HST observations.  Depending on the available observations of the SNe in our sample we adopted three different approaches to the determining the SN positions: 1) differential astrometry using a previously acquired HST observations of the SN; 2) differential astrometry using ground or other space-based observations of the SN; or 3) using the reported Right Ascension and Declination of the SN coupled with the  recalibration of the absolute astrometric coordinate solution of the HST observations with respect to the positions of catalogued stars \citep[see e.g.][]{2005astro.ph..1323M}, queried using the ALADIN tool\footnote{aladin.u-strasbg.fr/} \citep{2000A&AS..143...33B,2014ASPC..485..277B}.   Image stamps of the sites of the 23 target SNe are presented in Figure \ref{fig:res:stamp}.  

For the purposes of our analysis, we constructed photometric catalogues for all sources within a projected radius of $150\,\mathrm{pc}$ of the position of each SN.  The size of this selection region was chosen to be comparable to the previous analysis of \citet{2016MNRAS.456.3175M} and the scales of large stellar complexes; as, unlike other studies \citep[e.g.][]{2009ApJ...703..300G}, we can incorporate background stellar populations in our fitting procedure.  For the purposes of comparing stellar sources with theoretical isochrones, the photometric catalogues were restricted to those sources determined to be stellar or spatially ``point-like" by DOLPHOT (assigned class ``1");  although we note that we have not made any corrections for crowding and, hence, our catalogues contain all recovered sources \citep{2009AJ....137..419W}.  If the SN or pre-explosion source was present in the images, all sources within 3 pixels of the SN position (uniformly applied to all data) were excluded from the catalogue. For stars that were not detected in a given band, we followed the approach of \citet{2016MNRAS.456.3175M} and used artificial star tests to determine individual detection probability functions, parameterised by a cumulative normal distribution, for each star.  Artificial stars tests were also used, across the circular region in which stars were selected from each of the images, to determine an additional scaling factor for the photometric uncertainty to account for the effects of crowding.

\begin{table*}
\caption{Hubble Space Telescope observations of the sites of the sample of 23 stripped-envelope SNe. \label{tab:obs:observations}}
\begin{tabular}{lccccl}
            \hline  
Object & Date & Exp & Instrument & Filter & Program \\
            &          & Time(s) &         &        & \\
            \hline            
1993J&2012 Feb 19&3000&WFC3/UVIS&F336W&12531$^{1}$\\ 
	&2012 Feb 19&836&WFC3/UVIS&F814W&12531\\ 
	&2011 Dec 24&856&WFC3/UVIS&F438W&12531 \\ 
	&2011 Dec 24&856&WFC3/UVIS&F555W&12531\\ 
	&2011 Dec 24&856&WFC3/UVIS&F625W&12531 \\ 
1994I&2005 Jan 20&2720&ACS/WFC&F435W&10452$^{2}$ \\ 
	&2005 Jan 20&1360&ACS/WFC&F555W&10452\\ 
	&2005 Jan 20&1360&ACS/WFC&F814W&10452 \\ 
	&2005 Jan 20&2720&ACS/WFC&F658N&10452 \\ 
1996aq&2010 Mar 19&3600&WFC3/UVIS&F555W&11570$^{3}$ \\ 
	&2010 Mar 19&2400&WFC3/UVIS&F814W&11570\\ 
1996cb&2001 Feb 13&700&WFPC2 &F555W&8602$^{4}$ \\ 
	&2001 Feb 13&700&WFPC2 &F555W&8602\\ 
	&2001 Apr 30&700&WFPC2 &F814W&8602 \\ 
	&2001 Apr 30&700&WFPC2 &F814W&8602 \\ 
	&2001 May 13&700&WFPC2 &F814W&8602 \\ 
	&2001 May 13&700&WFPC2 &F814W&8602 \\ 
2000ew&2013 Dec 19&956&WFC3/UVIS2&F475W&13350$^{5}$ \\ 
	&2013 Dec 19&956&WFC3/UVIS2&F606W&13350\\ 
	&2002 Jun 26&430&ACS/WFC1&F555W&9353 $^{6}$\\ 
	&2002 Jun 26&430&ACS/WFC1&F814W&9353 \\ 
	&2002 Jun 26&400&ACS/WFC1&F435W&9353 \\ 
2001B&2002 Jun 09&700&ACS/WFC1&F555W&9353 \\ 
	&2002 Jun 09&800&ACS/WFC1&F435W&9353 \\ 
	&2002 Jun 09&700&ACS/WFC1&F814W&9353 \\ 
2001gd&2005 Jun 28&480&ACS/HRC&F555W&10272 $^{7}$\\ 
	&2005 Jun 28&720&ACS/HRC&F814W&10272 \\ 
2002ap&2004 Aug 31&840&ACS/HRC&F435W&10272 \\ 
	&2004 Aug 31&360&ACS/HRC&F625W&10272\\ 
	&2004 Jul 06&480&ACS/HRC&F555W&10272 \\ 
	&2004 Jul 06&720&ACS/HRC&F814W&10272 \\ 
2004gn&2013 Dec 10&956&WFC3/UVIS2&F606W&13350$^{5}$\\ 
	&2013 Dec 10&956&WFC3/UVIS2&F475W&13350\\ 
2004gq&2008 Jan 11&460&WFPC2 &F555W&10877 $^{8}$\\ 
	&2008 Jan 11&700&WFPC2 &F814W&10877 \\ 
2004gt&2005 May 16&1530&ACS/HRC&F555W&10187$^{9}$ \\ 
	&2005 May 16&1860&ACS/HRC&F814W&10187\\ 
	&2005 May 16&1672&ACS/HRC&F435W&10187 \\ 
	&2010 Jan 22&1016&WFC3/UVIS&F625W&11577$^{10}$\\ 
	&2010 Jan 22&1032&WFC3/UVIS&F555W&11577 \\ 
	&2010 Jan 22&1032&WFC3/UVIS&F814W&11577 \\ 
	&2010 Jan 27&5534&WFC3/UVIS&F336W&11577 \\ 
2005at & 2014 Jul 26 & 1131 & WFC3/UVIS & F336W & 13364$^{11}$ \\
		& 2014 Jul 26 & 977 & WFC3/UVIS & F438W & 13364 \\
		& 2014 Jul 26 & 1155 & WFC3/UVIS & F555W & 13364 \\		
		& 2014 Jul 26 & 1001 & WFC3/UVIS & F814W & 13364 \\
2007fo & 2011 Oct 16 &1472 & WFC3/UVIS & F390W & 12170$^{12}$ \\
		& 2011 Nov 24 & 437 & ACS/WFC & F814W & 12170 \\
		& 2011 Nov 24 & 1360 & ACS/WFC & F606W & 12170 \\
2008ax&2011 Jul 12&1704&WFC3/UVIS2&F336W&12262$^{13}$ \\ 
	&2011 Jul 12&1029&WFC3/UVIS2&F625W&12262 \\ 
	&2011 Jul 12&1684&WFC3/UVIS2&F438W&12262 \\ 
	&2011 Jul 12&1310&WFC3/UVIS2&F606W&12262 \\ 
	&2011 Jul 12&1936&WFC3/UVIS2&F814W&12262 \\ 
2009jf & 2010 Oct 25 & 520  & ACS/WFC & F555W & 11575$^{14}$ \\
		& 2010 Oct 25 & 520 & ACS/WFC & F814W & 11575 \\
2011dh&2014 Aug 07&3772&WFC3/UVIS2&F225W&13433$^{15}$ \\ 
	&2014 Aug 07&1784&WFC3/UVIS2&F336W&13433 \\ 
	&2014 Aug 10&1072&ACS/WFC1&F435W&13433 \\ 
	&2014 Aug 10&1232&ACS/WFC1&F555W&13433 \\ 
	&2014 Aug 10&2176&ACS/WFC1&F814W&13433\\ 
\hline
\end{tabular}	
\end{table*}

\begin{table*}
\contcaption{}
\label{tab:obs:observations:contd}
\begin{tabular}{lccccl}
            \hline  
Object & Date & Exp & Instrument & Filter & Program \\
            &          & Time(s) &         &        & \\
            \hline   
2012P&2005 Mar 10&1600&ACS/WFC1&F435W&10187$^{9}$ \\ 
	&2005 Mar 10&1400&ACS/WFC1&F555W&10187\\ 
	&2005 Mar 10&1700&ACS/WFC1&F814W&10187 \\ 
2012au&2013 Dec 03&1100&WFC3/UVIS2&F606W&12888$^{16}$ \\ 
	&2013 Dec 03&1100&WFC3/UVIS2&F438W&12888 \\ 
2012fh&2014 Mar 02&980&WFC3/UVIS&F814W&13364 $^{11}$\\ 
	&2014 Mar 02&1110&WFC3/UVIS&F336W&13364 \\ 
	&2014 Mar 02&956&WFC3/UVIS&F438W&13364 \\ 
	&2014 Mar 02&1134&WFC3/UVIS&F555W&13364 \\ 
iPTF13bvn&2005 Mar 10&1600&ACS/WFC1&F435W&10187$^{9}$\\ 
	&2005 Mar 10&1400&ACS/WFC1&F555W&10187 \\ 
	&2005 Mar 10&1700&ACS/WFC1&F814W&10187 \\ 
2013df&1999 Apr 29&2000&WFPC2 &F439W&8400$^{17}$\\ 
	&1999 Apr 29&1600&WFPC2 &F555W&8400 \\ 
	&1999 Apr 29&1600&WFPC2 &F814W&8400 \\
2013dk & 2010 Jan 22 &1016&	WFC3/UVIS&	F625W&	11577$^{10}$\\	
	&2010 Jan 22&	1032&	WFC3/UVIS	&F555W&11577\\	
	&2010 Jan 22&	1032	&WFC3/UVIS	&F814W&11577\\	
	&2010 Jan 27&	5534	&WFC3/UVIS	&F336W&11577\\	
2016bau & 2016 Oct 21 & 780 & WFC3/UVIS & F814W& 14668$^{18}$\\
		& 2016 Oct 21 & 710 & WFC3/UVIS & F555W & 14668\\	
	\hline
\end{tabular}\\
Principal Investigator: $^{1}$ A. Filippenko; $^{2}$ S. Beckwith; $^{3}$ A. Riess; $^{4}$ A. Filippenko; $^{5}$ A. Fruchter; $^{6}$ S. Smartt; $^{7}$ A. Filippenko; $^{8}$ W. Li; $^{9}$ S. Smartt; $^{10}$ B. Whitmore; $^{11}$ D. Calzetti; $^{12}$ A. Gal-Yam; $^{13}$ J. Maund; $^{14}$ S. Van Dyk; $^{15}$ J. Maund; $^{16}$ S. Van Dyk; $^{17}$ K. Noll; $^{18}$ A. Filippenko. 

\end{table*}

%%%%%%%%%%%%%%%%%%%%%%%%%%%%%%%%
%RESULTS
%RESULTS
%RESULTS
%RESULTS
%RESULTS
%RESULTS
%RESULTS
%RESULTS
%%%%%%%%%%%%%%%%%%%%%%%%%%%%%%%%

\section{The ages of the surrounding stellar populations}
\label{sec:res:age}

%
%Bayesian stellar populations analysis
%
\subsection{Stellar Populations Analysis}
\label{sec:res:stellpops}
In order to analyse the properties of the ensemble of stars at each SN location in comparison with the predictions of stellar evolution models, given by isochrones, we followed the Bayesian technique presented by \citet{2016MNRAS.456.3175M} in conjunction with the Padova group isochrones based on the PARSEC stellar evolution models \footnote{http://stev.oapd.inaf.it/cgi-bin/cmd  - version 2.5} \citep{2002A&A...391..195G,2012MNRAS.427..127B}.  We have utilised the isochrones presented in the sets of filters available for each of the HST instruments to avoid unnecessary photometric transformations to the more usual Johnson-Cousins filter system.

The hierarchical Bayesian approach of \citeauthor{2016MNRAS.456.3175M} considers the properties of the ensemble of stars, in all photometric bands simultaneously, as being drawn from a mixture of underlying stellar population components with age $\tau = \log (t /\mathrm{years})$ and age width $\sigma_{\tau}$ (parameterised by normal distributions).  In addition to calculating the posterior probabilities for the ages of the stellar population,  \citeauthor{2016MNRAS.456.3175M} also calculated the Bayesian Evidence, or marginal likelihood, using Nested Sampling techniques \citep{2004AIPC..735..395S,2013arXiv1306.2144F},  which permits the identification of the number of components ($N_{m}$) that appropriately describes the population without overfitting.  We consider the relative merits of two models using the Bayes factor $K$, the ratio of the evidences of the models, and for this study we adopt the model selection criteria presented by \citet[][although see also \citealt{2008ConPh..49...71T}]{Jeffreys61}.  To streamline the calculation of the Bayesian evidence, we also fix the age width of each stellar population component $i$ (with age $\tau_{i}$) to  $\sigma_{\tau}  = 0.05$.  This adjustment dramatically speeds up the calculation, but has no impact on the final result as if a broader stellar population age component is required, the Bayesian scheme merely considers it to be a product of multiple adjacent populations.  For each population component, we determine the ``weight" $w_{i}$ which indicates the relative probability of stars belonging to that population.  As with the analysis of \citeauthor{2016MNRAS.456.3175M}, our analysis also considers the observed stars to be a mixture of single and ``composite" binaries (that are the photometric sum of the brightnesses of two stars of the same age that have otherwise evolved separately with no interaction).

We assume that reddening and extinction follow a standard \citet{ccm89} Galactic reddening law with $R_{V} = 3.1$.  For the full range of stellar population fits, we consider the extinction $A_{V}$, rather than the reddening, in the range $0.0 \leq A_{V} \leq 4.0\,\mathrm{mags}$.  Further to \citeauthor{2016MNRAS.456.3175M}, we also include the consideration of differential extinction $\mathrm{d}A_{V}$, allowing stars in the population to have individual extinctions $A^{\prime}_{V}$ with a probability  $p(A^{\prime}_{V}) \sim N(A_{V}, \left(\mathrm{d}A_{V}\right)^{2})$; where $A_{V}$ is the mean extinction towards the stellar population, and $\mathrm{d}A_{V}$ parameterises the width of the distribution of extinctions.  As noted in Section \ref{sec:res:1994I}, a major exception to this form of differential extinction is required for the analysis of the population around the position of SN~1994I.

We also select isochrones for metallicity $Z$ that follow reported oxygen abundances for SN locations, where available.  While the metallicity could be included as an additional parameter, due to the relatively small number of stars at each SN location we treat the metallicity as a fixed quantity such that $p(Z^{\prime}) = \delta(Z^{\prime} - Z)$.  Assuming a solar oxygen abundance of 8.69 \citep{2009ARA&A..47..481A}, we select isochrones for metallicity from the coarse grid available following the oxygen-metallicity scheme presented by \citet{2007A&A...466..277H} and used by \citet{2008arXiv0809.0403S}.

Sources that could be considered extended, e.g. possible compact clusters, were identified by considering their spatial extent using the sharpness and $\chi^2$ parameters reported by DOLPHOT.  Problematically, for complex backgrounds the $\chi^2$ parameter is a particularly poor discriminant for extended sources, especially for bright sources where minor variations in the background can induce large values of $\chi^2$.   We considered sources with absolute sharpness values $> 0.3$ and $\chi^{2} > 2.5$ to be extended and excised them from our photometric catalogues.  In addition, we combined the spatial criteria with the brightness limit proposed by \citet{2005A&A...443...79B}, where sources with $M_{V} < -8.6\,\mathrm{mag}$ are equally likely to be clusters.  For objects that meet both of the spatial and brightness criteria, we considered likely cluster properties through comparison of the photometry with theoretical STARBURST99 \citep{1999ApJS..123....3L} spectral energy distributions (SEDs).  We note that the limitations of these selection criteria were previously discussed by \citet{2017arXiv170401957M}.

\begin{figure}
\includegraphics[width=12cm,angle=270]{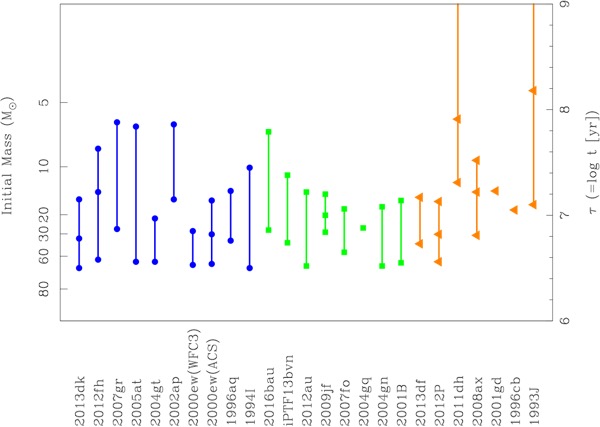}
\caption{Ages of the stellar populations observed around the sample of 23 SNe of types IIb (orange triangle; $\blacktriangle$), Ib (green square; $\blacksquare$) and Ic (blue circle; $\bullet$), as listed in Table \ref{tab:res}.}
\label{fig:res:panel}
\end{figure}

The results of our analysis of the ages of the stellar populations around our sample of 23 stripped-envelope SNe are presented in Table \ref{tab:res} and Figure \ref{fig:res:panel}, and results for individual SNe, including colour-magnitude diagrams (CMDs), are presented in Section \ref{sec:res:individual} (see also Appendix \ref{sec:app:cmd}).

\begin{landscape}
\begin{table}
\caption{The results of the analysis of the ages of the stellar populations within $150\,\mathrm{pc}$ of the sample of 23 stripped-envelope SNe. \label{tab:res}}
\def\arraystretch{1.5}% 
\begin{tabular}{lcccccccccccc}
\hline
SN 		& Type & $N_{m}$  & $A_{V}$ & $\mathrm{d}A_{V}$ &  \multicolumn{2}{c}{Bayes Factors} & $\tau_{1}$ & $w_{1}$ & $\tau_{2}$ & $w_{2}$& $\tau_{3}$ & $w_{3}$ \\ 
\cline{6  - 7}
      &          &                  &                &     & $\ln K(2,1)$ & $\ln K(3,2)$ \\
\hline
1993J 		& IIb &3 & $ 1.04 \pm 0.02 $ & $ 0.06 \pm 0.01 $ & $ 3256.7 \pm  0.3 $  & $ 652.2 \pm  0.3 $ &  $ 7.10_{- 0.02}^{+0.02}$ & 0.17 & $ 8.18_{- 0.02}^{+0.01}$ & 0.46 & $ 9.23_{- 0.02}^{+0.02}$ & 0.37 \\
1994I$^{\ast}$ 	& Ic 	&2 & $ 1.48 \pm 0.02 $ & $ 0.20 \pm 0.00 $ & $ 500.4 \pm  0.3 $  & $ -1.7 \pm  0.4 $ &  $ 6.50_{- 0.01}^{+0.01}$ & 0.88 & $ 7.45_{- 0.01}^{+0.02}$ & 0.12 & \ldots & \ldots \\
1996aq 		& Ic 	&2 & $ 0.64 \pm 0.05 $ & $ 0.03 \pm 0.01 $ & $ 132.1 \pm  0.2 $  & $ -1.4 \pm  0.2 $ &  $ 6.76_{- 0.03}^{+0.03}$ & 0.73 & $ 7.23_{- 0.02}^{+0.03}$ & 0.25 & \ldots & \ldots \\
1996cb 		& IIb 	& 1 & $ 0.13 \pm 0.20 $ & $ 0.02 \pm 0.02 $ & $ 0.2 \pm  0.1 $  & $ -1.1 \pm  0.1 $ &  $ 7.05_{- 0.10}^{+0.11}$ & 0.00 & \ldots & \ldots & \ldots & \ldots \\
2000ew(ACS)	& Ic 	&3 & $ 1.25 \pm 0.05 $ & $ 0.10 \pm 0.04 $ & $ 55.8 \pm  0.2 $  & $ 11.5 \pm  0.2 $ &  $ 6.54_{- 0.02}^{+0.01}$ & 0.73 & $ 6.82_{- 0.01}^{+0.03}$ & 0.17 & $ 7.14_{- 0.02}^{+0.04}$ & 0.11 \\
2000ew(WFC3)	& Ic 	&2 & $ 1.26 \pm 0.06 $ & $ 0.02 \pm 0.06 $ & $ 13.3 \pm  0.2 $  & $ -2.0 \pm  0.2 $ &  $ 6.53_{- 0.02}^{+0.01}$ & 0.91 & $ 6.85_{- 0.02}^{+0.04}$ & 0.08 & \ldots & \ldots \\
2001B 		& Ib 	& 2 & $ 1.02 \pm 0.06 $ & $ 0.03 \pm 0.02 $ & $ 49.6 \pm  0.2 $  & $ -1.4 \pm  0.2 $ &  $ 6.55_{- 0.02}^{+0.02}$ & 0.62 & $ 7.14_{- 0.03}^{+0.02}$ & 0.38 & \ldots & \ldots \\
2001gd 		& IIb & 1 & $ 0.14 \pm 0.10 $ & $ 0.02 \pm 0.02 $ & $ -0.7 \pm  0.1 $  & $ -1.6 \pm  0.1 $ &  $ 7.23_{- 0.11}^{+0.04}$ & 0.00 & \ldots & \ldots & \ldots & \ldots \\
2002ap 		& Ic 	& 2 & $ 0.44 \pm 0.24 $ & $ 0.04 \pm 0.01 $ & $ 9.6 \pm  0.1 $  & $ -0.5 \pm  0.1 $ &  $ 7.15_{- 0.10}^{+0.11}$ & 0.41 & $ 7.86_{- 0.12}^{+0.09}$ & 0.72 & \ldots & \ldots \\
2004gn 		& Ib 	& 2 & $ 1.62 \pm 0.07 $ & $ 0.02 \pm 0.02 $ & $ 58.2 \pm  0.2 $  & $ -1.9 \pm  0.3 $ &  $ 6.52_{- 0.01}^{+0.02}$ & 0.76 & $ 7.08_{- 0.02}^{+0.02}$ & 0.24 & \ldots & \ldots \\
2004gq 		& Ib	& 1 & $ 0.42 \pm 0.20 $ & $ 0.04 \pm 0.01 $ & $ 0.5 \pm  0.1 $  & $ -1.9 \pm  0.1 $ &  $ 6.88_{- 0.03}^{+0.04}$ & 1.00 & \ldots & \ldots & \ldots & \ldots \\
2004gt 		& Ic 	&2 & $ 0.76 \pm 0.03 $ & $ 0.20 \pm 0.03 $ & $ 86.1 \pm  0.3 $  & $ -0.4 \pm  0.3 $ &  $ 6.56_{- 0.04}^{+0.02}$ & 0.91 & $ 6.97_{- 0.02}^{+0.02}$ & 0.09 & \ldots & \ldots \\
2005at 		& Ic 	& 2 & $ 0.85 \pm 0.04 $ & $ 0.20 \pm 0.02 $ & $ 305.9 \pm  0.3 $  & $ 0.3 \pm  0.3 $ &  $ 6.56_{- 0.04}^{+0.03}$ & 0.60 & $ 7.84_{- 0.02}^{+0.01}$ & 0.40 & \ldots & \ldots \\
2007fo 		& Ib 	& 2 & $ 0.71 \pm 0.04 $ & $ 0.06 \pm 0.01 $ & $ 35.8 \pm  0.2 $  & $ -0.4 \pm  0.3 $ &  $ 6.65_{- 0.03}^{+0.03}$ & 0.80 & $ 7.06_{- 0.03}^{+0.03}$ & 0.20 & \ldots & \ldots \\
2008ax 		& IIb 	& 3 & $ 1.20 \pm 0.04 $ & $ 0.10 \pm 0.01 $ & $ 534.5 \pm  0.2 $  & $ 79.6 \pm  0.3 $ &  $ 6.81_{- 0.02}^{+0.03}$ & 0.19 & $ 7.22_{- 0.02}^{+0.01}$ & 0.20 & $ 7.52_{- 0.01}^{+0.01}$ & 0.59 \\
2009jf 		& Ib 	& 3 & $ 0.06 \pm 0.04 $ & $ 0.02 \pm 0.01 $ & $ 11.4 \pm  0.2 $  & $ 2.3 \pm  0.3 $ &  $ 6.84_{- 0.05}^{+0.03}$ & 0.27 & $ 7.00_{- 0.03}^{+0.03}$ & 0.50 & $ 7.20_{- 0.11}^{+0,05}$ & 0.17 \\
2011dh 		& IIb 	& 3 & $ 0.23 \pm 0.02 $ & $ 0.10 \pm 0.01 $ & $ 240.3 \pm  0.2 $  & $ 30.9 \pm  0.2 $ &  $ 7.31_{- 0.01}^{+0.04}$ & 0.28 & $ 7.91_{- 0.02}^{+0.02}$ & 0.60 & $ 9.16_{- 0.08}^{+0.04}$ & 0.08 \\
2012P 		& IIb 	& 3 & $ 1.01 \pm 0.02 $ & $ 0.06 \pm 0.01 $ & $ 98.2 \pm  0.2 $  & $ 22.0 \pm  0.2 $ &  $ 6.56_{- 0.01}^{+0.02}$ & 0.48 & $ 6.82_{- 0.01}^{+0.02}$ & 0.33 & $ 7.13_{- 0.01}^{+0.02}$ & 0.22 \\
2012au 		& Ib 	& 2 & $ 0.69 \pm 0.28 $ & $ 0.03 \pm 0.03 $ & $ 16.2 \pm  0.2 $  & $ -1.5 \pm  0.2 $ &  $ 6.52_{- 0.04}^{+0.08}$ & 0.48 & $ 7.22_{- 0.13}^{+0.03}$ & 0.53 & \ldots & \ldots \\
2012fh 		& Ic 	& 3 & $ 0.47 \pm 0.02 $ & $ 0.06 \pm 0.02 $ & $ 409.7 \pm  0.2 $  & $ 56.2 \pm  0.2 $ &  $ 6.58_{- 0.09}^{+0.03}$ & 0.78 & $ 7.22_{- 0.03}^{+0.03}$ & 0.12 & $ 7.63_{- 0.03}^{+0.03}$ & 0.08 \\
iPTF13bvn	& Ib 	& 2 & $ 0.98 \pm 0.02 $ & $ 0.06 \pm 0.01 $ & $ 12.4 \pm  0.2 $  & $ -2.9 \pm  0.2 $ &  $ 6.74_{- 0.02}^{+0.02}$ & 0.71 & $ 7.38_{- 0.06}^{+0.01}$ & 0.34 & \ldots & \ldots \\
2013df 		& IIb 	& 2 & $ 0.30 \pm 0.03 $ & $ 0.05 \pm 0.03 $ & $ 85.4 \pm  0.2 $  & $ -1.6 \pm  0.2 $ &  $ 6.73_{- 0.02}^{+0.03}$ & 0.54 & $ 7.17_{- 0.02}^{+0.01}$ & 0.45 & \ldots & \ldots \\
2013dk 		& Ic 	& 2 & $ 1.07 \pm 0.03 $ & $ 0.06 \pm 0.01 $ & $ 155.6 \pm  0.3 $  & $ -1.3 \pm  0.3 $ &  $ 6.50_{- 0.02}^{+0.02}$ & 0.38 & $ 6.78_{- 0.01}^{+0.03}$ & 0.35 & $ 7.15_{- 0.02}^{+0.02}$ & 0.26 \\
2016bau 		& Ib 	& 2 & $ 0.30 \pm 0.60 $ & $ 0.06 \pm 0.03 $ & $ 109.2 \pm  0.2 $  & $ -1.0 \pm  0.2 $ &  $ 6.86_{- 0.06}^{+0.06}$ & 0.55 & $ 7.79_{- 0.02}^{+0.02}$ & 0.45 & \ldots & \ldots \\
\hline
\end{tabular}\\
$^{\ast}$ See section \ref{sec:res:1994I} issues concerning the quality of the fit and alternative generative models
\end{table}
\end{landscape}

\subsection{Individual Supernovae}
\label{sec:res:individual}
%%%%%%%%%%%%%%%%%%%%%%%%
%SN 1993J
%%%%%%%%%%%%%%%%%%%%%%%%
\subsubsection{SN 1993J}
\label{sec:res:1993J}
Given the offset of SN~1993J from the centre of M81 ($0.32R_{25}$; \citealt{2013MNRAS.428.1927C}), using the metallicity gradient determined by \citet{metapil} we estimate the metallicity at the location of SN~1993J to be $\sim 0.75Z_{\odot}$ which, for the purposes for comparison with isochrones, we consider to be consistent with an approximately solar metallicity. 

For our analysis of the surrounding stellar population, we have utilised the 2011-12 WFC3/UVIS observations that cover UV to optical wavelengths \citep{2014ApJ...790...17F}.  The position of SN~1993J measured on 2002 ACS HRC observations \citep{maund93j} was used to identify the SN on the WFC3 observations.  The proximity of M81 and the quality of the available observations means that the stellar population in the vicinity of SN 1993J is probed to very deep levels.   For reasons of computational efficiency, for SN 1993J we have decided to limit our analysis to those stars within a radius of only $70\,\mathrm{pc}$, whilst excluding the late-time SN from the photometric catalogue.

From the UV and optical CMDs (see Fig. \ref{fig:obs:93Juv:cmd}) we can see evidence for at least 3 age components to the surrounding stellar population, with the youngest corresponding to the lifetime of a $M_{init} = 15.7^{+1.6}_{-1.3}M_{\odot}$ star, which is consistent with the masses previously derived by  \citet{alder93j}, \citet{maund93j}, \citet{1993Natur.364..509P} and \citet{1993Natur.364..507N}.  The other stellar population components are consistent with much older stars with $M_{init}  < 4.5M_{\odot}$.  For the entire population  we find a relatively high level of extinction of $A_{V} = 1.4\,\mathrm{mags}$, which is $\sim 0.4\,\mathrm{mags}$ higher than \citet{maund93j} previously estimated using the surrounding stars and $0.5\,\mathrm{mags}$ higher than  \citet{2014ApJ...790...17F} estimated towards the SN itself.
As a check for our analysis, we conducted a qualitative comparison with the stellar ages determined from the UV and optical data with infrared WFC3/IR observations, acquired at the same time with the $F105W$, $F125W$ and $F160W$ filters (as part of program GO-12531; PI Fox) as shown on Fig. \ref{fig:obs:93Jir:cmd}.  We find excellent agreement for the older populations and the reddest evolved stars arising from the younger age component.

\begin{figure*}
\begin{center}
\includegraphics[width=6.0cm, angle=270]{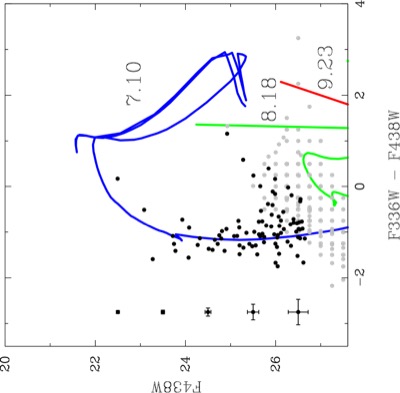}
\includegraphics[width=6.0cm, angle=270]{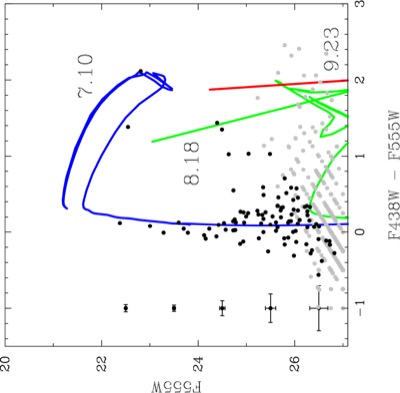}\\
\includegraphics[width=6.0cm, angle=270]{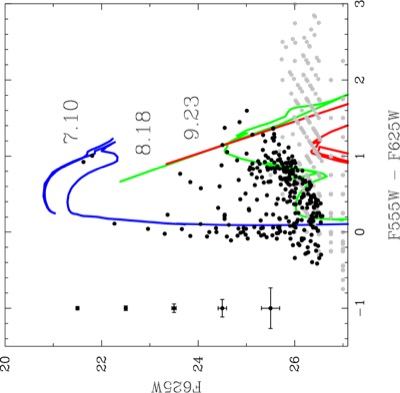}
\includegraphics[width=6.0cm, angle=270]{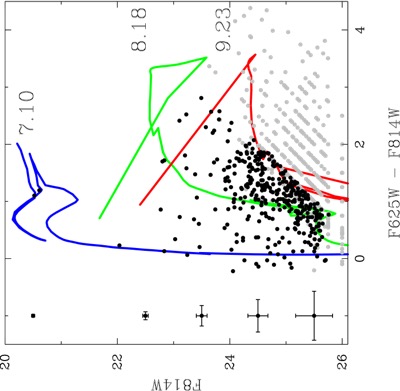}\\
\end{center}
\caption{CMDs showing the stellar population around the site of SN 1993J in M81.  Black points indicate stars detected in both filters, while grey points indicate those stars detected in only one of the filters.  The column of points with error bars on the left-hand side of the plots shows the average uncertainties for stars at that brightness.  Due to the proximity of the host galaxy, we only show the population within $70\,\mathrm{pc}$ of the SN position. \label{fig:obs:93Juv:cmd}}
\label{fig:obs:93Juv:cmd}
\end{figure*}

\begin{figure*}
\includegraphics[width=6.0cm, angle=270]{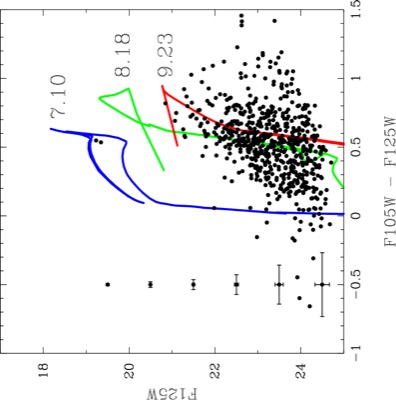}
\includegraphics[width=6.0cm, angle=270]{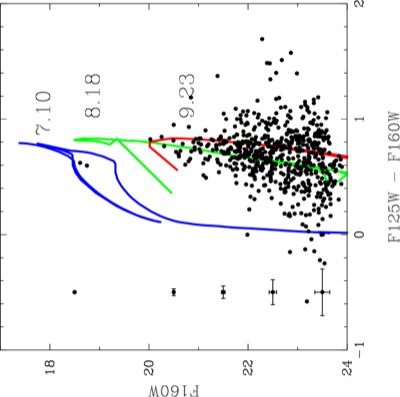}
\caption{The stellar population around the site of SN 1993J in M81 observed at IR wavelengths.  Overlaid are the isochrones representing the stellar population components derived from the analysis of the optical observations (see Fig. \ref{fig:obs:93Juv:cmd}).}
\label{fig:obs:93Jir:cmd}
\end{figure*}

%%%%%%%%%%%%%%%%%%%%%%%%
%SN 1994I
%%%%%%%%%%%%%%%%%%%%%%%%
\subsubsection{SN 1994I}
\label{sec:res:1994I}

\begin{figure}
\begin{center}
\includegraphics[width=6.0cm, angle=270]{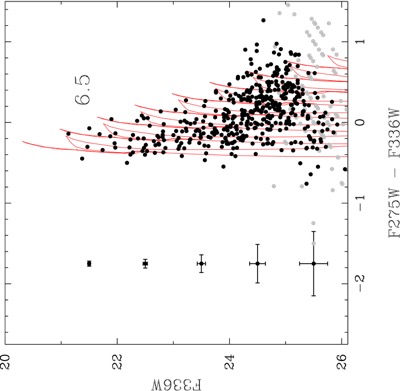}
\vspace{0.5cm}
\includegraphics[width=6.0cm, angle=270]{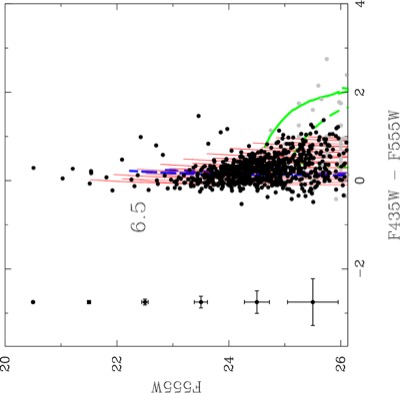}
\vspace{0.5cm}
\includegraphics[width=6.0cm, angle=270]{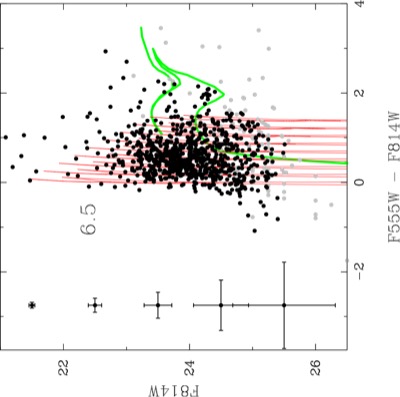}
\vspace{0.2cm}

\end{center}
\caption{The same as Figure \ref{fig:obs:93Juv:cmd} but for stars within $150\,\mathrm{pc}$ of SN 1994I in M51.}
\label{fig:obs:94I:cmd}
\end{figure}

For our analysis we considered both UV (WFC3/UVIS) and optical datasets (ACS/WFC); however due to the large period between their respective acquisitions (almost 10 years) we analysed them separately.  We utilised the results of \citet{2016ApJ...818...75V} to locate the position of SN~1994I in both datasets.  CMDs for both sets of observations, covering the stellar populations observed within $150\,\mathrm{pc}$ of the SN position are presented in Fig. \ref{fig:obs:94I:cmd}.

In comparing the distribution of stars on the CMD derived from both the WFC3 and ACS observations with the shapes of suitable isochrones, it is clear to the eye that the underlying generative model we have assumed (see Section \ref{sec:res:stellpops}) is not sufficient to describe the large range in colours.  The central concentration of points on the CMDs is not well described by the superposition of multiple age components, for which the extinction (even with normally distributed differential extinction) tends to zero.   As an alternative, we also consider a model with two extinction components: differential extinction following a uniform distribution (i.e. a range) defined by a minimum extinction $A_{V, min}$ and maximum range of differential extinction $\mathrm{d}A_{V}$, for which  we adopt a prior of $p(\mathrm{d}A_{V}) \propto 1/\mathrm{d} A_{V}$ to penalise large range values; and $A_{V}(2)$ which follows the original normal distribution form for the differential extinction.  The results of these two approaches are summarised in Table \ref{tab:res:1994I:evid}, from which it is clear that a uniform form for $\mathrm{d}A_{V}$ is preferred for both the UV and optical datasets (see Fig. \ref{fig:obs:94I:cmd}).  At UV wavelengths, a single age component is sufficient to describe the observed CMD, however at optical wavelengths the presence of a small number of outliers ($<1\%$) requires the incorporation of an additional age component.   We note that the minimum amount of extinction derived from the WFC3 observations is lower than that found from the ACS observations, which may be the result of the UV colours being more sensitive to extinction than the combination of the ACS filters or that the ACS observations are skewed by more heavily reddened objects. 
 
Overall the bulk of the stars are well described by a single age population of $3.2\,\mathrm{Myr}$, with a large range of differential extinction.  This corresponds to the lifetime of a star with initial mass $M_{init} \approx 100M_{\odot}$.

\begin{table*}
\caption{Results for the different differential extinction models applied to the stellar population around SN~1994I. \label{tab:res:1994I:evid}}
\def\arraystretch{1.5}% 
\begin{tabular}{lccccccccc}
\hline\hline
		&$N_{m}$		& Evidence	&$A_{V,min}$ &  $\mathrm{d}A_{V}$ & $A_{V}(2)$ & $\tau_{1}$  &  $w_{1}$  & $\tau_{2}$  & $w_{2}$ \\  
\hline
\multicolumn{10}{c}{{\bf WFC3/UVIS F275W/F336W}}\\
Gaussian	& 2 & $-4681.02\pm0.24$ & $0.08\pm0.02$ & $0.02\pm0.01$ & \ldots & $6.50^{+0.01}_{-0.02}$ & 0.97 & $6.85^{+0.07}_{-0.33}$ & 0.03\\
Uniform	& 1 & $-2088.52\pm0.21$ & $0.46^{+0.06}_{-0.02}$ & $2.00\pm0.01$ & \ldots & $6.49^{+0.02}_{-0.01}$ & 1.00 & \ldots & \ldots \\
Uniform	& 2 & $-2093.26\pm0.25$ & $0.48^{+0.04}_{-0.03}$ & $2.00\pm0.01$ & $2.58^{+0.91}_{-2.42}$&$6.49\pm0.01$ & 1.00 & $7.57^{+2.11}_{-0.81}$ & 0.00\\
\hline
\multicolumn{10}{c}{{\bf ACS/WFC F435W/F555W/F814W}}\\
Gaussian	& 2 & $-5592.00\pm0.25$ & $1.48^{+0.01}_{-0.02}$ & $0.10$ & \ldots & $6.50 \pm 0.01$ & 0.88 & $7.45^{+0.02}_{-0.01}$ & 0.12\\
Uniform	& 1 & $-2533.47\pm0.21$ & $0.78^{+0.04}_{-0.02}$ & $2.00\pm0.01$ & \ldots & $6.50 \pm {0.01}$ & 1.00 & \ldots & \ldots \\
Uniform	& 2 & $-2446.45\pm0.29$ & $0.72^{+0.04}_{-0.02}$ & $2.00\pm0.01$ & $0.07 \pm 0.01$ &$6.49\pm0.01$ & 0.98 & $7.57^{+2.11}_{-0.81}$ & 0.00\\
\hline\hline
\end{tabular}
\end{table*}

%%%%%%%%%%%%%%%%%%%%%%%%
%SN 1996AQ
%%%%%%%%%%%%%%%%%%%%%%%%
\subsubsection{SN~1996aq}
\label{sec:res:1996aq}
We used five bright foreground stars on the WFC3/UVIS images with coordinates present in the Sloan Digital Sky Survey \citep{2011AJ....142...72E} to recalibrate the absolute astrometry and determine the SN position on the HST WFC3/UVIS observations to within $\pm 3$ pixels. For the site of SN 1996aq, \citet{2011ApJ...731L...4M} measured an oxygen abundance using $O3N2$ that implied a metallicity $\sim 0.75Z_{\odot}$.  From the two colour photometry of the stars in the vicinity of SN 1996aq, we find evidence for two population components with ages corresponding to initial masses of $33^{+8}_{-5}$ and $13\pm1M_{\odot}$.

%%%%%%%%%%%%%%%%%%%%%%%%
%SN 1996CB
%%%%%%%%%%%%%%%%%%%%%%%%
\subsubsection{SN~1996cb}
\label{sec:res:1996cb}
The position of SN 1996cb was found using the reported position of the SN \citep{1996IAUC.6524....1N}.  The astrometric solution of the WFPC2 PC chip F555W observation was recalibrated with respect to 21 USNO-B \citep{monet03} and Sloan Digital Sky Survey sources, with a final precision of $0.02\,\mathrm{arcsec}$.  We expect the uncertainty on the position to be dominated by the precision of the reported position for the SN.
In the absence of a reported metallicity measurement appropriate for SN 1996cb, we assume a solar metallicity.   \citet{1999AJ....117..736Q} estimated an upper limit to the extinction of $A_{V} < 3.1 \times 0.12 = 0.37$ mags, from comparison of SN~1996cb with observations of SN 1993J, which is consistent with our lower estimate of $A_{V}=0.13\,\mathrm{mags}$, derived from the surrounding stellar population.  The stellar population is well fit by a single age component, although we note that the uncertainty on the population age is correlated with the uncertainties for the extinction, arising due to the limited depth and colour information provided by the two-colour WFPC2 observations.  The brightest source on the CMD is likely to be a cluster, given its brightness, although at the resolution and depth of these WFPC2 observations it is difficult to discern if this object is truly extended.  The age of the population corresponds to the lifetime of a star with initial mass $17^{+4}_{-3}M_{\odot}$.

%%%%%%%%%%%%%%%%%%%%%%%%
%SN 2000ew
%%%%%%%%%%%%%%%%%%%%%%%%
\subsubsection{SN~2000ew}
\label{sec:res:2000ew}
We use the position of the SN previously reported by  \citet{2005astro.ph..1323M}, which was determined using post-explosion HST WFPC2 observations containing the SN.  In addition, more recent, deeper WFC3 observations covering the site of SN 2000ew were acquired in 2013.  We consider the stellar population around SN 2000ew as imaged in both datasets, but due to the time difference between their respective acquisitions we analyse them separately.  

In the ACS observations we find evidence for three stellar population components.  We also find the two youngest components in the analysis of the WFC3 observations, however the third and oldest component is not detected due to the lack of a redder $F814W$ WFC3 observation.  The ages of the two youngest populations in the ACS and WFC3 observations agree to within the uncertainties, and correspond to the lifetimes of stars with initial masses of $93^{+7}_{-23}$ and $28\pm 4M_{\odot}$.  The oldest stellar population  discernible in the ACS $F555W/F814W$ CMD corresponds to the lifetime of stars with $M_{init} \sim 15M_{\odot}$.

DOLPHOT does not report any of the sources in the vicinity of SN~2000ew to be noticeably extended.  Given the brightness distribution of the stars on the CMDs, the likely maximum mass for any of these sources if they are clusters is $\sim 10^{4}M_{\odot}$.   As there is no clear, offset population of bright sources, we conclude that any sources that might be clusters are mixed, on the CMDs, with a larger, general population of individual stars.

%%%%%%%%%%%%%%%%%%%%%%%%
%SN 2001B
%%%%%%%%%%%%%%%%%%%%%%%%
\subsubsection{SN~2001B}
\label{sec:res:2001B}
For the position of the SN, we follow the identification of a source present in post-explosion ACS observation which \citet{2005astro.ph..1323M} concluded was the fading SN.  \citet{2010MNRAS.407.2660A} found an approximately solar metallicity for the site of SN~2001B.  Similarly to SN~2000ew, we find evidence for significant extinction, well in excess of the value of $E(B-V) = 0.14\,\mathrm{mags}$ adopted by \citet{2005astro.ph..1323M} and \citet{2013MNRAS.436..774E}.  Our analysis of the population suggests two age components consistent with the lifetimes of stars with initial mass $90^{+9}_{-23}$ and $14.7\pm1.5M_{\odot}$.  As with SN~2000ew, there are no obviously very bright or extended sources and, similarly to SN~2000ew (see Section \ref{sec:res:2000ew}), any clusters must share the same locus on the CMD with individual stars.

%%%%%%%%%%%%%%%%%%%%%%%%
%SN 2001gd
%%%%%%%%%%%%%%%%%%%%%%%%
\subsubsection{SN~2001gd}
\label{sec:res:2001gd}
The site of SN~2001gd was observed with the ACS HRC.  The small size of the field limited the number of the potential catalogue stars available for recalibrating the astrometric solution, and hence we rely on the astrometric accuracy of HST to approximately locate the SN.  \citet{2010MNRAS.407.2660A} report an approximately solar metallicity, estimated from the $N2$ indicator.  Due to the paucity of stars in the vicinity of SN~2001gd, mostly detected in the $F814W$ image, we have limited evidence for a single stellar population component with an age of $\sim 17\,\mathrm{Myr}$ corresponding to the lifetime of a relatively low mass progenitor of $M_{init} = 12.9^{+1.6}_{-1.1}M_{\odot}$.  There is a large degeneracy between the inferred age and the extinction, as the fit was limited by the availability of only two-colour photometry.
%%%%%%%%%%%%%%%%%%%%%%%%
%SN 2002ap
%%%%%%%%%%%%%%%%%%%%%%%%
\subsubsection{SN~2002ap}
\label{sec:res:2002ap}

From the analysis of M74 presented by \citet{metapil}, we assume an LMC metallicity for the site of SN~2002ap.  We identified the SN position on 2004 ACS HRC  observations through comparison with an earlier set of observations acquired in 2003 (as part of programme GO-9114; PI R. Kirshner) when the SN was still relatively bright.  SN~2002ap is notable for having the fewest number of stars in its locality (5) in the sample of SNe considered here, which previously prohibited a corresponding analysis of the stellar population by \citet{2014ApJ...791..105W}.  Our Bayesian methodology is still applicable, but is more sensitive to the properties of individual stars and uncertainties in the adopted distance.  The isochrone fitting procedure identified two age components for the stars, corresponding to the ages of stars with initial mass $14\pm3$ and $6.2^{+0.9}_{-0.6}M_{\odot}$.  The upper mass constraint is consistent with the initial mass limits for the progenitor of $\sim 15 - 20M_{\odot}$ estimated by \citet{2007MNRAS.381..835C} and $\lesssim 23M_{\odot}$ by \citet{2017ApJ...842..125Z}.  
%%%%%%%%%%%%%%%%%%%%%%%%
%SN 2004gn
%%%%%%%%%%%%%%%%%%%%%%%%
\subsubsection{SN~2004gn}
\label{sec:res:2004gn}

\citet{2010MNRAS.407.2660A} estimated a metallicity at the SN position of approximately solar metallicity.  We identified the SN position on the 1996 WFPC2 observations, with recalibrated astrometric solution, from the reported RA and Dec of the SN \citep[see][]{2013MNRAS.436..774E}, and then calculated a full geometric transformation to the 2014 WFC3 observations.   The position of the SN is just offset from a large association (see Fig. \ref{fig:res:stamp}).  On the CMD the limited colour information available from the 2014 WFC3 observations probes the main sequence turn off of a young population, but cannot detect redder, older populations.  We tentatively conclude that there are sufficient blue stars to support a young stellar population with an age of $\sim3.3\,\mathrm{Myr}$, but also possibly detect stars on the blue loop from a slightly older population with age $12\,\mathrm{Myr}$.  The ages correspond to the lifetimes of stars with initial masses of $97^{+18}_{-3}$ and $16.2 \pm 1.5 M_{\odot}$.  We determine a large estimate for the extinction of $A_{V} = 1.62 \pm 0.07\,\mathrm{mags}$; which we note is significantly higher than the extinction ($A_{V} \sim 0.21\,\mathrm{mags}$) assumed by \citet{2013MNRAS.436..774E}.
%%%%%%%%%%%%%%%%%%%%%%%%
%SN 2004gq
%%%%%%%%%%%%%%%%%%%%%%%%
\subsubsection{SN~2004gq}
\label{sec:res:2004gq}

To identify the SN position, as imaged on the WFPC2 PC chip, we recalibrated the astrometric solution of the WFPC2 mosaic, using 6 USNO B1.0 stars \citep{monet03} and the reported RA and Dec, with a final precision of 0.35 arcsec.   We assume a solar metallicity given the oxygen abundance measured by  \citet{2010MNRAS.407.2660A}.  The CMD of the stellar population around SN~2004gq is relatively sparse, due to the limited depth and colour information of the WFPC2 observations, and in our analysis we find a single stellar population, with age $7.6\,\mathrm{Myr}$ corresponding to the lifetime of a star with initial mass of $24 \pm 3M_{\odot}$.
\newpage
%%%%%%%%%%%%%%%%%%%%%%%%
%SN 2004gt
%%%%%%%%%%%%%%%%%%%%%%%%
\subsubsection{SN~2004gt}
\label{sec:res:2004gt}

\begin{figure*}
\includegraphics[width=10.5cm, angle=270]{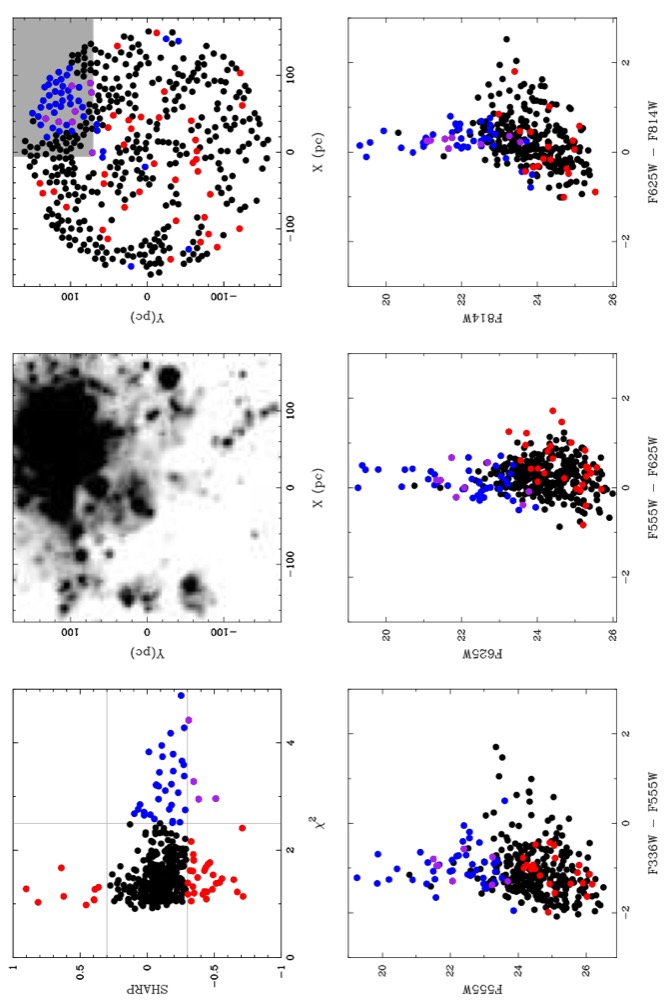}
\caption{The shape and size properties of sources identified within $150\,\mathrm{pc}$ of the site of SN~2004gt.  {\it Top Left)} the shape parameters (sharpness and $\chi^{2}$) from DOLPHOT photometry.  Sources in black are considered to have stellar point-like PSFs (with $| \mathrm{sharp} | < 0.3$ and $\chi^{2} < 2.5$).  {\it Top Middle)}  The location of SN~2004gt in NGC 4038 as imaged with WFC3 UVIS ($F555W$). {\it Top Right)}  The corresponding shape classifications, following the colour scheme in the top left panel, for sources around the SN position.  The grey shaded region corresponds to the centre of knot S and sources in this area are excluded from the analysis of the stellar population. {\it Bottom Row)}  Corresponding CMDs for the population around SN 2004gt, colour-coded following the scheme in the top left panel according to their shape properties.}
\label{fig:obs:04gt:shape}
\end{figure*}

SN 2004gt was located just South East of Knot S in NGC 4038 \citep{1970ApJ...160..801R}, and a precise position for the SN on  2010 WFC3 observations was determined with respect to post-explosion ACS HRC observations acquired in 2005 and the position determined by \citet{2005ApJ...630L..33M}.  \citet{2013AJ....146...30K} measure an oxygen abundance of 8.71 for the site of SN~2004gt, such that we assume a solar metallicity in our analysis.  In considering the dense stellar population around SN~2004gt, we note that a number of bright sources are associated with the centre of Knot S and that the site of SN~2004gt falls inside the tidal radius \citep[$> 450\,\mathrm{pc}$;][]{1999AJ....118.1551W}.  The brightest sources are clearly non-stellar, with large values for both the $\chi^{2}$ and sharpness parameters (see Fig. \ref{fig:obs:04gt:shape}).  As the centre of Knot S does not constitute a resolved population we exclude sources in this region from our analysis.  We also find that other fainter sources identified by DOLPHOT as having non-stellar shapes are found along the edges of dust lanes, where the brightness gradient of the background is large.

We identify two major age components in the stellar population, although with a large degree of differential extinction which is not unexpected due to the density of stars in this field.  The $F336W-F555W$ CMD is crucial for parameterising the youngest population, and breaking the degeneracy between age and extinction.  The two age components of $3.6$ and  $9.3\,\mathrm{Myr}$ are consistent with the life-times of stars with initial masses of $\sim 85$ and $20M_{\odot}$, and straddle the single age estimate derived by \citet{1999AJ....118.1551W} of $7 \pm 1\,\mathrm{Myr}$ from the integrated colours of Knot S.  The degree of extinction we derive towards the resolved stellar population is, however, substantially larger than measured by \citet{1999AJ....118.1551W} to Knot S itself ($A_{V} = 0.03\pm 0.12 \,\mathrm{mag}$).  \citet{2005ApJ...630L..33M} concluded the SN itself was subject to minimal extinction, with the shape of the continuum in early spectra being similar to SN~2002ap with $A_{V} \sim 0.3 \,\mathrm{mags}$; and this conclusion was also shared by \citet{2005ApJ...630L..29G}.  Using a crude  $F555W -F814W$ CMD based on earlier 1996 WFPC2 observations,  \citet{2005ApJ...630L..33M} and \citet{mythesis} estimated an extinction of $A_{V} = 0.22 \pm 0.03\,\mathrm{mag}$ and were able to identify an older age component of $8.5\,\mathrm{Myr}$. 
%%%%%%%%%%%%%%%%%%%%%%%%
%SN 2005at
%%%%%%%%%%%%%%%%%%%%%%%%
\subsubsection{SN~2005at}
\label{sec:res:2005at}

For the position of SN~2005at we utilised the location of the SN presented by \citet{2014A&A...572A..75K} and bootstraped it, using 12 stars from 2007 WFPC2 observations (GO-10877l; PI Li) to the 2014 WFC3 observations with a precision of 0.24 pixels or 10 milliarcsec.  We adopt the Tip of Red Giant Branch distance for NGC 6744 reported by \citet{2009AJ....138..332J} and, after \citet{2014A&A...572A..75K}, assume a solar metallicity for the site of the SN based on the metallicity gradient presented by \citet{metapil}.  The $F336W$ observation is crucial for probing the youngest, most massive stars around the site of SN~2005at.  None of sources around SN~2005at appear extended in the DOLPHOT photometry nor are  brighter than $M_{V} < -8.6\,\mathrm{mags}$, from which we conclude there is no strong evidence for any of the sources being compact clusters.  The ages of these two populations correspond to the lifetimes of stars with initial masses of $85^{+14}_{-23}$ and $6.3\pm0.3M_{\odot}$.

We note that the determination of the extinction towards SN~2005at by \citet[$A_{V}=1.9$ mags;][]{2014A&A...572A..75K}, based on the assumption that SN~2005at is a clone of SN~2007gr, is significantly higher than we determine here from our analysis of the surrounding stellar population.  This may reflect a real amount of extinction that was local to the SN, that is not found in the wider surrounding stellar population.  If, as \citet{2014A&A...572A..75K} hypothesise, SN~2005at is a copy of SN~2007gr, then the high mass derived for that progenitor \citep{2016MNRAS.456.3175M} would still imply a high-mass for the progenitor of SN~2005at.
%%%%%%%%%%%%%%%%%%%%%%%%
%SN 2007fo
%%%%%%%%%%%%%%%%%%%%%%%%
\subsubsection{SN~2007fo}
\label{sec:res:2007fo}
 Late-time post-explosion HST observations of the site of SN~2007fo were acquired in 2011 using the $F390W$, $F606W$ and $F814W$ filters (in addition there was a shorter $F300X$ observation which we have not considered here).  In order to identify the SN position on the HST frames, we utilised a 2000s Gemini NIRI+ALTAIR observation of SN 2007fo acquired, using adaptive optics, on 2007 Aug 21 (Program GN-2006B-Q-8; PI Gal-Yam).  Using 22 stars common to both the Gemini NIRI+ALTAIR $K^{\prime}$ and  HST ACS $F814W$ observations, we determined the SN position on the late-time images to within 0.54$\,\mathrm{pix}$ or 27 milli-arcsecs.  The SN position is coincident with a source in the HST observations with brightness $m_{F390W} = 24.30 \pm 0.04$, $m_{F606W} = 23.86 \pm 0.02$ and $m_{F814W} = 23.73 \pm 0.05$ mags.  SN 2007fo lies in a crowded region of NGC 7714. With correction for extinction the source coincident with SN 2007fo is at the star/cluster brightness threshold.  
 
In the absence of a confident metallicity measurement appropriate for SN 2007fo, we assume a default solar metallicity.   For determining the properties of the population at the SN position we have made the assumption that all the sources are stellar, however there is at least one object with brightness $M_{F606W} + A \approx -9.9$ mags, which is incompatible with single and binary stellar isochrones and is most likely a cluster.  Due to the distance and degree of crowding we cannot make a definite classification based on spatial extent criteria alone.  This bright object is excluded from our analysis.  Under the assumption that the remaining population are stellar sources, we find evidence for two young age components, subject to a significant degree of extinction $A_{V} = 0.71$ mags.  We note, however, that this overall fit to the population is dominated by the $F390W$ and $F606W$ observations, which are substantially deeper than the corresponding $F814W$ observation.  These ages correspond to initial masses of $52.2^{+16.7}_{-11.1}M_{\odot}$ and $16.8^{+1.7}_{-1.5}M_{\odot}$.  We compared the SED of the source at the SN position with \citet{2004astro.ph..5087C} ATLAS9 stellar SEDs and STARBURST99 cluster SEDs, with the marginal likelihoods of the two fits favouring the stellar SED by $\Delta \ln Z  = 11.8 $ consistent with a luminous B-supergiant.
 
%%%%%%%%%%%%%%%%%%%%%%%%
%SN 2008ax
%%%%%%%%%%%%%%%%%%%%%%%%
\subsubsection{SN~2008ax}
\label{sec:res:2008ax}

The position of SN~2008ax was recovered on  WFC3/UVIS observations from 2011, using post-explosion WFPC2 observations from 2008 (GO-11119; PI Van Dyk), to within a final precision of 13 milliarcsec.  We assume a solar metallicity following \citet{2008MNRAS.391L...5C}.   We find evidence for three age components, with a level of extinction comparable to that derived by \citet{2008MNRAS.389..955P} towards the SN itself, but with a large degree of differential extinction.  The case of SN~2008ax highlights the importance of simultaneously considering multiwavelength observations, spanning the ultraviolet to near-infrared, in order to disentangle the effects of extinction in instances where there is a large spread of ages.  The ages of the three components correspond to stars with initial mass of $19.4^{+2.6}_{-1.9}$, $12\pm 1$ and $8.5\pm0.5M_{\odot}$; however, the relatively narrow range of ages of the three components may be suggestive of continuous star formation, rather than discrete episodes such that the initial mass of the progenitor may lie in the range $8 - 20M_{\odot}$.

%%%%%%%%%%%%%%%%%%%%%%%%
%SN 2009jf
%%%%%%%%%%%%%%%%%%%%%%%%
\subsubsection{SN~2009jf}
\label{sec:res:2009jf}
In analysing the host environment, we have used post-explosion ACS/WFC observations (with the $F555W$ and $F814W$ filters)  acquired in 2010.  Using the position for the SN presented by \citet{2011MNRAS.416.3138V}, we recover a faint source at the SN position which, through comparison with the pre-explosion WFPC2 observations, is likely to be the SN itself. We consider all stars within a radius of 20 pixels of the SN position but exclude a circular region with radius of 3 pixels centred on the SN position. We identify 58 separate sources in the vicinity of the SN.  From integral field spectroscopy presented by \citet{2013AJ....146...30K}, the average metallicity at the location of SN~2009jf is $0.68\pm0.05Z_{\odot}$ and we therefore compare the observed stellar population against LMC metallicity isochrones. A single source, located at a projected distance of $\mathrm{70pc}$ from the SN position has $(M_{F555W} + A) =  -11.3\, \mathrm{mags}$ and is likely to be a cluster based on its brightness and is excluded from our analysis.  We note, however, that given the distance to NGC~7479 and the significant degree of crowding around the SN position, no sources appear to deviate from the expected values for point-like sources and, as such, spatial criteria are not able to clarify the possible cluster-like nature of 9 other sources closer to the $M_{F555W}  =  -8.6\,\mathrm{mags}$ threshold.    The derived extinction toward to the population  ($A_{V} = 0.04\,\mathrm{mag}$) is less then the estimated foreground extinction, but consistent with the low amounts of extinction found by \citet{2011MNRAS.413.2583S} and \citet{2011MNRAS.416.3138V} to arise in NGC~7479 itself.  From the ages of the stellar population components, we estimate corresponding initial masses of $26^{+6}_{-3}$, $19\pm2$ and $13^{+3}_{-1}M_{\odot}$.

%%%%%%%%%%%%%%%%%%%%%%%%
%SN 2011dh
%%%%%%%%%%%%%%%%%%%%%%%%
\subsubsection{SN~2011dh}
\label{sec:res:2011dh}
For the observations, position of SN~2011dh and metallicity of the host environment we follow the previous analyses presented by \citet{2011ApJ...739L..37M} and \citet[][and references therein]{2015MNRAS.454.2580M}.  We note that, using the subset of $F555W$ and $F814W$ images from the available pre-explosion observations, \citet{2011ApJ...742L...4M} previously estimated the mass of the progenitor from the properties of the surrounding stellar population.  We find that the maximum number of 3 components are required to describe the observed stellar population in the multiwavelength data, ranging from the ultraviolet to the near-infrared; although the assumption of the maximum number of age components is discussed in more detail in Section \ref{sec:disc}.   The low level of extinction is consistent with the reddening towards the progenitor and the SN previously estimated by \citet{2011ApJ...739L..37M} and \citet{2011ApJ...741L..28V}.  The ages of the three components correspond to the lifetimes of stars with initial masses $11.5 \pm 0.9$, $6\pm0.3$ and $2.0 \pm0.2 M_{\odot}$, with the youngest component yielding a similar mass to the that inferred for the progenitor directly from pre-explosion observations \citep{2011ApJ...739L..37M}.

%%%%%%%%%%%%%%%%%%%%%%%%
%SN 2012P
%%%%%%%%%%%%%%%%%%%%%%%%
\subsubsection{SN~2012P}
\label{sec:res:2012P}
\begin{figure}
\begin{center}
\includegraphics[width=6.0cm, angle=270]{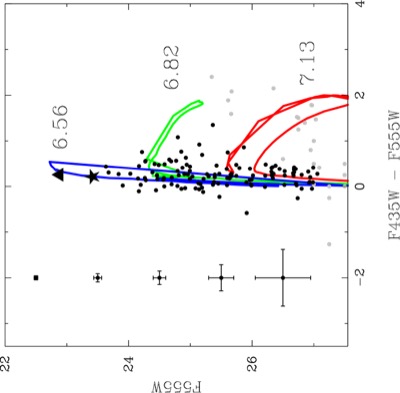}
\vspace{0.5cm}
\includegraphics[width=6.0cm, angle=270]{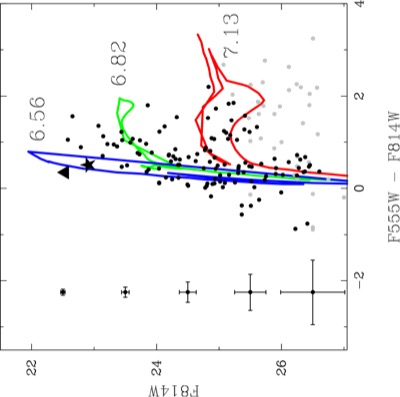}
\vspace{0.2cm}
\end{center}
\caption{Colour magnitude diagrams for the stellar population around SN 2012P. The source coincident with the SN position is indicated by the black star ($\bigstar$) and the nearby brighter source is indicated by the black triangle ($\blacktriangle$) (see text)} 
\label{fig:cmd:12PACS}
\end{figure}

The position of SN 2012P was identified on the pre-explosion ACS observations with respect to a 72s Gemini GMOS $i$-band image of the SN acquired on 2012 Jan 17 (as part of program GN-2011B-Q-215; PI D. Howell), with a final precision of 0.05 arcsec (or 1 ACS/WFC pixel).  The SN was found to be coincident with a bright source, which was found to still be present at the same brightness in late-time WFC3 UVIS observations of the SN location acquired on 2015 Jun 26 (as part of program GO-13684; PI Van Dyk).  The continued presence of the source at the SN location suggests it was a likely host cluster, that is a component of a larger association (see Fig. \ref{fig:res:stamp}).   \citet{2016A&A...593A..68F} estimated an oxygen abundance at the site of SN 2012P of $12 + \log (O/H) = 8.61$, such that for our analysis we use solar metallicity isochrones. 
We determine a large degree of extinction towards the surrounding stellar population (see Fig. \ref{fig:cmd:12PACS}), that is consistent with the reddening estimates determined by \citet{2016A&A...593A..68F}  through comparison of SN 2012P with 2011dh.  The stellar population covers a range of ages, with the three components used in our analysis corresponding to the life-times of stars with initial masses of $85^{+12}_{-22}$,  $28^{+5}_{-3}$ and $14.9^{+1.5}_{-1.1}M_{\odot}$.

The two brightest sources in the field, located just North of the SN position and at the SN position, are spatially extended and above the cluster brightness limit.   In contrast to other SNe in the sample and, in particular iPTF13bvn (in the same galaxy), it is surprising that there is such as confluence of evidence that supports the cluster-like nature of these objects given the distance to NGC~5806.  From  DOLPHOT we find that the $\chi^{2}$ and sharpness parameters are significantly different from the parameters derived for the rest of the sources in the area around the SN (with $\chi^{2} > 2.7$).  While \citet{2016A&A...593A..68F} report photometry for the source coincident with the SN,  we note that the photometric magnitudes are actually consistent with those of the other bright source offset from the SN location by 8.1 pixels (0.41 arcsec or 51pc) North of the SN position (see Figure \ref{fig:res:stamp}).  This source is the brightest source in the field, while for the source coincident with the SN position we measure the brightness to be $m_{F435W} = 23.63 \pm 0.05$, $m_{F555W} = 23.42 \pm 0.05 $ and $m_{F814W} = 22.91 \pm 0.04 \, {\mathrm{mags}}$.

%%%%%%%%%%%%%%%%%%%%%%%%
%SN 2012au
%%%%%%%%%%%%%%%%%%%%%%%%
\subsubsection{SN~2012au}
\label{sec:res:2012au}

We considered the stellar population observed in post-explosion WFC3 observations of SN 2012au from 2013, in which the SN was still bright.  Using the position of the SN, we confirmed that the progenitor was not detected in the relatively shallow pre-explosion WFPC2 observations (with a positional uncertainty of 33 milliarcsec on WF3 chip) acquired on 2001 Aug 17 (for program GO-9042; PI  S. Smartt; see also \citealt{2012ATel.3971....1V}).  \citet{2013ApJ...770L..38M} determined a metallicity for the location of SN~2012au of $\sim 1 - 2Z_{\odot}$, and for our analysis we adopt a solar metallicity.  Due to the limited colour information and the small number of stars around the SN site, there is significant uncertainty on the extinction, but the overall level of extinction we estimate for the stellar population is a factor of $\sim 2$ higher than estimated by \citet{2013ApJ...770L..38M} and \citet{2013ApJ...772L..17T} towards the SN itself.  The complex background structure at the site of the SN, as seen in Fig. \ref{fig:res:stamp}, may suggest that some members of the observed stellar population are embedded in a dust sheet.  This may also induce a large uncertainty on the extinction, and could explain the discrepancy between the different extinctions estimates if the SN occurred above the dust sheet.  The large uncertainties on the extinction suggest caution with the interpretation of the properties of the stellar population.
The ages we derive for the two stellar population components are consistent with the lifetimes of stars with initial masses $97^{+10}_{-36}$ and $13^{+3}_{-1}M_{\odot}$.
%%%%%%%%%%%%%%%%%%%%%%%%
%SN 2012fh
%%%%%%%%%%%%%%%%%%%%%%%%
\subsubsection{SN~2012fh}
\label{sec:res:2012fh}
To determine the position of SN~2012fh on the post-explosion WFC3/UVIS observations, we utilised a Swift Ultraviolet and Optical Telescope (UVOT) observation of the SN acquired on 2012 Oct 26.   An image acquired with the UVM2 filter was aligned with an $F275W$ WFC3 observation (which was not used in the final analysis of the stellar population) allowing us to identify the SN position to within 5.33 WFC3 pixels or 0.21 arcsecs.  The SN is located within a dense stellar association, associated with young, blue stars (see Fig. \ref{fig:res:stamp}), with similar spatial extent to the association observed at the position of the Type Ic SN 2007gr \citep{2016MNRAS.456.3175M}.  Using the reported position of SN~2012fh, we used the metallicity gradient of the host galaxy NGC 3344 \citep{metapil} to estimate the oxygen abundance at the SN location of $12 + \log (O/H) = 8.36$, corresponding to an approximately half-solar or LMC metallicity.  The multi-colour coverage of SN~2012fh yields relatively well constrained ages and extinction.  We find evidence that the surrounding stellar population is composed of three age components, corresponding to the lifetimes of stars with initial masses of  $75^{+25}_{-20}$, $13\pm1$ and $7.9\pm0.5M_{\odot}$.

%%%%%%%%%%%%%%%%%%%%%%%%
%iPTF13bvn
%%%%%%%%%%%%%%%%%%%%%%%%
\subsubsection{iPTF13bvn}
\label{sec:res:13bvn}

For our analysis of the stellar population around iPTF13bvn, we utilise the pre-explosion observations that were used to the identify the progenitor candidate \citep{2013ApJ...775L...7C} and the properties of the SN and its location in NGC~5806 that were previously used by \citet[][see also \citealt{2016ApJ...825L..22F}]{2016MNRAS.461L.117E}.  We infer a significant degree of extinction towards the surrounding stellar population, with $A_{V} = 0.98\,\mathrm{mags}$ almost a factor of $\sim 2$ than the value proposed by \citet{2014AJ....148...68B}  towards the SN itself and significantly higher than the value assumed by \citet{2013ApJ...775L...7C} and \citet{2016A&A...593A..68F}.  There are two age components that can be used to describe the surrounding stellar population, with the younger solution driven by a number of bright blue sources evident in the $F435W - F555W$ CMD.  The ages of the two components correspond to the lifetimes of stars with initial masses of $36^{+9}_{-6}$ and $10.5^{+1.0}_{-0.7}M_{\odot}$, with the age of the latter yielding a mass consistent with estimates for the proposed binary progenitor system derived from the combination of pre-explosion and late-time observations of the SN location \citep{2016MNRAS.461L.117E}.
%%%%%%%%%%%%%%%%%%%%%%%%
%SN 2013df
%%%%%%%%%%%%%%%%%%%%%%%%
\subsubsection{SN~2013df}
\label{sec:res:2013df}

 To determine the position of SN~2013df on the pre-explosion WFPC2 observations, we utilised a post-explosion observation of the SN acquired with the HST WFC3 on 2013 Jul 17 (as part of program GO-12888; PI S. Van Dyk).  The precision of the geometric transformation between the pre- and post-explosion observations was determined to be 0.7 WFPC2 WF pixels, and the position of the SN was found to be offset from a source in the pre-explosion observations by only 0.2 WF pixels.  In their analysis, \citet{2014AJ....147...37V} claim a non-detection of a source in the pre-explosion $F439W$ at the SN position, which they confirmed by visual inspection of their data.  We find there is a flux excess visible at the SN position in the $F439W$ image that is detected by DOLPHOT at $2.7\sigma$; while this does not meet the criteria for a formal detection, it is discernible in the image and, hence, we incorporate it into our analysis with the commensurate large photometric uncertainties.  Using DOLPHOT, we measured the brightness of the source in the pre-explosion images to be: $m_{F439W}=25.61\pm0.40$, $m_{F555W}=24.45\pm0.07$, $m_{F606W}=24.49\pm0.23$ and $m_{F814W}=23.19\pm0.05$ mags.  We incorporate the progenitor into our analysis of the surrounding stellar population.  We find two stellar population components with ages corresponding to $5.4$ and $14.8\,\mathrm{Myr}$, or the lifetimes of stars with initial masses of $37^{+9}_{-7}$ and $14\pm 1M_{\odot}$; with the progenitor belonging to the older component.

%%%%%%%%%%%%%%%%%%%%%%%%
%SN 2013dk
%%%%%%%%%%%%%%%%%%%%%%%%
\subsubsection{SN~2013dk}
\label{sec:res:2013dk}
We use the position for SN 2013dk on the WFC3/UVIS frames determined and reported by \citet{2013MNRAS.436L.109E}. \citeauthor{2013MNRAS.436L.109E} analysed pre-explosion observations of the site of SN~2013dk and, while not recovering a progenitor, noted that the SN was offset from a likely cluster; for which they determined an age in the range $15 - 22\,\mathrm{Myr}$ and extinction $A_{V} \sim 0.6\,\mathrm{mags}$, although we note the pre-explosion $F336W$ photometry was discrepant with STARBURST99 models.
In considering the surrounding stellar population, we use the properties of NGC~4038 previously adopted in our analysis of SN~2004gt (see Section \ref{sec:res:2004gt}).  The cluster close to the SN is the brightest source within $150\,\mathrm{pc}$ of the SN position and we exclude it from our analysis of the population.  We find evidence for three stellar population components correspond to the lifetimes of stars with masses of $98\pm10$ and $31^{+6}_{-4}$ and $14\pm1\,\mathrm{M_{\odot}}$.

%%%%%%%%%%%%%%%%%%%%%%%%
%SN 2016bau
%%%%%%%%%%%%%%%%%%%%%%%%
\subsubsection{SN~2016bau}
\label{sec:res:2016bau}
Given the offset of SN~2016bau from the centre of the host galaxy, we determined the oxygen abundance for the SN location of $12 + \log(O/H) = 8.60\pm0.04$, following \citet{metapil}; corresponding to just below solar metallicity.   We used WFC3/UVIS observations of the host galaxy acquired in 2016 with the SN still present in the images.  These observations were substantially deeper than the available three-colour pre-explosion observations of NGC  3631 acquired with HST WFPC2 in 2001 (as part of program GO-9042).   We determined the location of SN~2016bau on the pre-explosion observations, with respect to the SN position measured on the 2016 post-explosion images, to within 0.59 pixels on the PC chip or 0.03 arcsec.  Using DOLPHOT we rule out the presence of a pre-explosion source within 3 pixels of the SN position with a $\mathrm{S/N > 3}$ with detection limits of $m_{F450W} > 23.68\pm0.86$, $m_{F606W} > 25.23\pm0.62$ and $m_{F814W} > 23.75\pm0.66\,\mathrm{mags}$. 

The interpretation of the CMD to derive an age for the two apparent stellar populations is complicated by the limited colour information and depth of the WFC3 observations, as well as by the paucity of stars around the SN position.  The case for a young stellar population is dominated by a number of faint, blue sources; however, the degrees of extinction and differential extinction are also uncertain, which limits the robustness of our analysis, and one additional source could, such as for SN~2012aw  \citep{2017arXiv170401957M}, skew the interpretation of the blue sources.   A key test of the presence of the youngest stellar population will be observations of the SN site at UV wavelengths, which may also be able to resolve the degeneracies between age and extinction.
The ages we find for the two, sparse population components corresponding to the lifetimes of stars with initial masses $25\pm5$ and $6.7\pm 0.3 M_{\odot}$.

%%%%%%%%%%%%%%%%%%%%%%%%%%%%%%%%%%%%%%%%%%%%%%%%
%%%%%%%%%%%%%%%%%%%%%%%%%%%%%%%%%%%%%%%%%%%%%%%%
%SPATIAL DISTRIBUTIONS
%%%%%%%%%%%%%%%%%%%%%%%%%%%%%%%%%%%%%%%%%%%%%%%%
%%%%%%%%%%%%%%%%%%%%%%%%%%%%%%%%%%%%%%%%%%%%%%%%
\section{Spatial distribution of surrounding stars}
\label{sec:res:spatial}
From Figure \ref{fig:res:stamp}, it is clear that there is large variation in the number and spatial distribution of stars around the sites of the SNe in the sample. \citet{2016MNRAS.456.3175M}, for example, noted that the position of SN~2007gr lay almost at the direct centre of a massive star complex with spatial extent  $\sim 300\,\mathrm{pc}$.    The measure of the number of massive stars, their spatial extent and the relative offset of the SN position from these massive stars provide important clues into the amount of star formation that occurred, the spatial scales over which it took place, the possible offset of the site of the SN location from the position of the birth of the progenitor and the relative correlation of different SN types with massive stars.  In order to quantify the difference in spatial distributions of massive stars around the site of the SN locations, we consider the numbers counts within circular regions centred on the SN position.   We use three apertures of fixed physical size: 100, 150 and 300pc, to make the results for individual SNe directly comparable to each other; however, we do not correct for effects due to inclination.  Given the heterogeneity of the available observations for the SNe considered here (see Table \ref{tab:obs:observations}), we are limited in this study to conducting this analysis using $\sim V$-band observations (such as the $F555W$ and $F606W$ filters), which are available for all of the target SN sites in our sample.  We use the extinction $A_{V}$ derived from our analysis of the stellar population to calculate absolute magnitudes for all objects that fall within the selection apertures.  We consider the number count of all sources above an absolute brightness threshold of $M_{V} < -6\,\mathrm{mags}$, which we have found to be complete for the entirety of our sample out to the most distant SNe.  A benefit of this approach is that we are not concerned with which sources might be clusters or unresolved binaries, but merely assessing the relative numbers of objects brighter than the cutoff magnitude.  The number of stars, within the set distances with respect to each SN (including SN 2007gr from \citealt{2016MNRAS.456.3175M}), is presented in Table \ref{tab:res:dens}.  

\begin{figure}
\begin{center}
\includegraphics[angle=270, width=7.0cm]{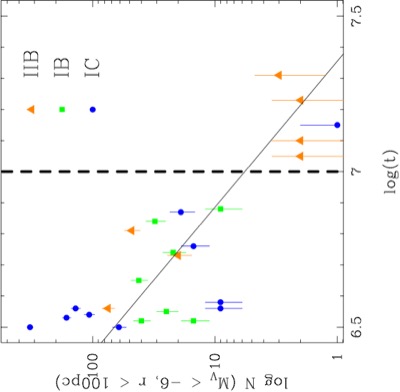}
\end{center}
\caption{The number of objects with $M_{V} < -6\,\mathrm{mags}$ within $100\,\mathrm{pc}$ of each SN in the sample with respect to the age of the youngest identified stellar population component ($\tau_{1}$).}
\label{fig:res:npop}
\end{figure}

In Figure \ref{fig:res:npop}, we show the number counts within the $100\,\mathrm{pc}$ region around each SN with respect to the age determined for the youngest stellar population component ($\tau_{1}$).  We find that there is a trend for increasing numbers of bright, massive stars in close proximity to the SN position with decreasing population age.  We identify a rough relation between the age of the number of bright stars of $\log_{10} (N; M_{V} < -6\,\mathrm{mag}) \approx 15.7 - 2.13 \times \tau$ and find an approximate division between two groups of objects, with ages less than and larger than $10\,\mathrm{Myr}$.  For ages $\tau_{1}>7$ or $10\,\mathrm{Myr}$, the sparsest environments are associated with some of the Type IIb SNe in our sample and the Type Ic SN 2002ap; however the vast majority of the Type Ibc SNe, in high density environments are associated with at least one age component with age $ < 10\,\mathrm{Myr}$.  We note that, by selecting an absolute magnitude cutoff for which we have established complete detections, this dichotomy is not due to distance; for instance, SN 1993J is one of the closest SNe in our sample and has the largest numbers of stars detected within $\mathrm{150\,pc}$, but only 2 bright sources are found within $\mathrm{100\,pc}$ of the SN position.  

To assess the relative concentration of stars around the SN location, we also considered the ratios in number densities between the circular regions of different sizes, as shown in Figure \ref{fig:res:clus_sep}.  A detailed interpretation of Figure \ref{fig:res:clus_sep} is presented in Appendix \ref{sec:app:spatial}; however, more generally, SNe located towards the centre of an association will appear in the top, right-hand quadrant of the figure, while those on the outskirts of an association will appear in the bottom, left-hand quadrant.  The youngest populations, associated with the progenitors of Type Ic SNe, lie in close proximity ($< 100\,\mathrm{pc}$; which, given the size of the smallest aperture used for this analysis, is the smallest length scale we can consider) of a close association.  Although SN 2002ap clearly deviates from this trend, its spatial properties are  consistent with the older age inferred for the surrounding stars.  The Type Ib and IIb environments exhibit a wider spread in properties despite, in some cases, having young populations components with similar ages to those associated with Type Ic SNe, suggesting that the underlying age component responsible for the progenitor may not be spatially correlated with the stars arising from the youngest age component.

\begin{figure}
\begin{center}
\includegraphics[angle=270, width=7.0cm]{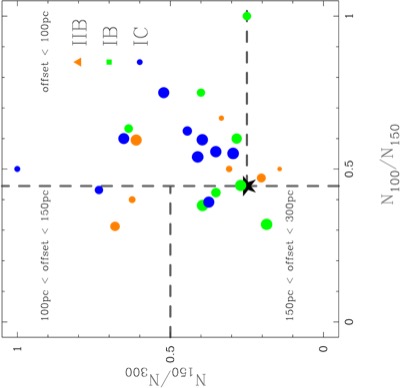}
\includegraphics[angle=270, width=1.3cm]{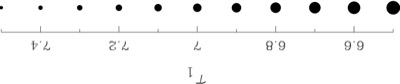}
\end{center}
\caption{Ratios of the number counts of bright ($M_{V} < -6\,\mathrm{mags}$) sources within $100$, $150$ and $300\,\mathrm{pc}$ of the position of each of the sample SNe.  The size of point is related to the age of the youngest component ($\tau_{1}$) following the scale bar on the right.  The position of the uniform field density is indicated by the star ($\star$).  For a description of the underlying model and the origins of the divisions marked on the figure see Appendix \ref{sec:app:spatial}.}
\label{fig:res:clus_sep}
\end{figure}

%\begin{landscape}
\begin{table*}
\caption{ The number of sources with  $M_{V} < -6\,\mathrm{mags}$ within a distance $r$ of the positions of the 23 stripped-envelope SNe in the sample. \label{tab:res:dens}}
\def\arraystretch{1.5}% 
\begin{tabular}{lcccccc}
\hline\hline
SN   & $N_{\star}(\mathrm{150\, pc})$ & \multicolumn{1}{c}{$r= 100\,\mathrm{pc}$} &\multicolumn{1}{c}{$r = 150\,\mathrm{pc}$} & \multicolumn{1}{c}{$r = 300\,\mathrm{pc}$} & $N_{100}/N_{150}$ & $N_{150}/N_{300}$\\
%\cline{4-8}\cline{10-14}\cline{16-20}
%		&		&&-5&-6&-7&-8&-8.5		&&-5&-6&-7&-8&-8.5		&&-5&-6&-7&-8&-8.5	\\
\hline
1993J	&3919$^{\dagger}$		&2		&4		&13		&	$0.5\pm0.4$	&$0.31\pm0.18$ \\
1994I 	&920					&326		&591		&2003	&	$0.55\pm0.04$&$0.30\pm0.01$ \\
1996aq	&156					&15		&24		&54		&	$0.63\pm0.21$&$0.44\pm0.11$ \\
1996cb	&36					&2		&28		&8		&	$0.40\pm0.33$&$0.63\pm0.36$ \\
2000ew(ACS)&264				&107 	&192		&546		&$0.56\pm0.07$&$0.35\pm0.03$ \\
2000ew(WFC3)&394				&164		&275		&695		&$0.60\pm0.06$&$0.40\pm0.03$ \\
2001B	&77					&25		&56		&207		&$0.45\pm0.11$&$0.27\pm0.04$ \\
2001gd	&27					&2		&3		&9		&$0.67\pm0.61$&$0.33\pm0.22$ \\
2002ap	&5					&1		&2		&2		&$0.50\pm0.61$&$1.00\pm1.00$ \\
2004gn	&173					&40		&105		&266		&$0.38\pm0.07$&$0.39\pm0.05$ \\
2004gq	&15					&9		&12		&30		&$0.75\pm0.33$&$0.40\pm0.14$ \\
2004gt	&440					&138		&352		&940		&$0.39\pm0.04$&$0.37\pm0.02$ \\
2005at	&148					&9		&12		&23		&$0.75\pm0.33$&$0.52\pm0.19$ \\
2007fo	&71					&42		&70		&247		&$0.60\pm0.12$&$0.28\pm0.04$ \\
2007gr$^{\ddagger}$	&284			&19		&44		&60		&$0.43\pm0.12$&$0.73\pm0.15$ \\
2008ax	&861					&48		&102		&503		&$0.47\pm0.08$&$0.20\pm0.02$ \\
2009jf	&58					&31		&49		&77		&$0.63\pm0.15$&$0.64\pm0.12$ \\
2011dh	&544					&3		&6		&42		&$0.50\pm0.35$&$0.14\pm0.06$ \\
2012P	&162					&75		&126		&206		&$0.60\pm0.09$&$0.61\pm0.07$ \\
2012au	&59					&15		&47		&253		&$0.32\pm0.09$&$0.19\pm0.03$ \\
2012fh	&243					&9		&15		&23		&$0.60\pm0.25$&$0.65\pm0.22$ \\
iPTF13bvn&85					&22		&52		&148		&$0.42\pm0.11$&$0.35\pm0.06$ \\
2013df	&71					&20		&64		&94		&$0.31\pm0.08$&$0.68\pm0.11$ \\
2013dk	&170					&61		&113		&275		&$0.54\pm0.09$&$0.41\pm0.05$ \\
2016bau	&84					&1		&1		&4		&$1.00\pm1.41$&$0.25\pm0.28$ \\
\hline
\end{tabular}\\
$^{\dagger}$ Number of stars detected within $150\,\mathrm{pc}$ of the SN position (note: the stellar population analysis was conducted for 808 stars within $\mathrm{70\,pc}$ of the SN; see section \ref{sec:res:1993J})\\
$^{\ddagger}$ See \citet{2016MNRAS.456.3175M}\\
\end{table*}
%\end{landscape}

%%%%%%%%%%%%%%%%%%%%%%%%%%%%%%%%%%%%%%%%%%%%%%%%
%%%%%%%%%%%%%%%%%%%%%%%%%%%%%%%%%%%%%%%%%%%%%%%%
%DISCUSSION
%%%%%%%%%%%%%%%%%%%%%%%%%%%%%%%%%%%%%%%%%%%%%%%%
%%%%%%%%%%%%%%%%%%%%%%%%%%%%%%%%%%%%%%%%%%%%%%%%
\section{Discussion}
\label{sec:disc}

\subsection{Ages and spatial properties of the resolved stellar populations}
\label{sec:disc:age}

\begin{table*}
\caption{Average stellar population properties and their dependence on SN type.\label{tab:disc:aver}}
\begin{tabular}{cccccccccccccccc}
\hline
       & \multicolumn{3}{c}{IIP$^{\dagger}$}&& \multicolumn{3}{c}{IIb} & & \multicolumn{3}{c}{Ib} & & \multicolumn{3}{c}{Ic} \\
    \cline{2 - 4}  \cline{6 - 8}  \cline{10 - 12} \cline{14 - 16} \\
    & Mean & St. Dev. & N & & Mean & St. Dev. & N & & Mean & St. Dev. & N & & Mean & St. Dev. & N \\
\hline
Age \\

$\tau_{1}$ & 6.80& 0.40 & 12 && 6.97 & 0.26 & 7 & & 6.70 & 0.15 & 8 & & 6.67 & 0.21 & 9 \\
$\tau_{2}$ &7.68&0.53& 12 && 7.46 & 0.50 & 5 & & 7.24 & 0.25 & 7 & & 7.34& 0.42 & 9\\
$\tau_{3}$ &8.05 & 0.99& 6&& 8.26 & 0.85 & 4 & & 7.20 & $\cdots$& 1 & & 7.31 & 0.23 & 3\\
\hline
$A_{V}$ & 0.65 & 0.36 & 12&& 0.58 & 0.47 & 7 & & 0.73 & 0.49 & 8 & & 0.83 & 0.37 & 9 \\
$N(M_{V} \leq -6)_{100}$ &$\cdots$ &$\cdots$ &$\cdots$ & & 44.3 & 52.8 &7 & & 49.0 & 32.3 & 8 &  & 105.7 & 188.3 & 9  \\
$N_{100}/N_{150}$ &$\cdots$ &$\cdots$ &$\cdots$ & & 0.49 & 0.11 & 7 & & 0.60 & 0.22 & 8 & & 0.55 & 0.11 & 9\\
$N_{150}/N_{300}$ &$\cdots$ &$\cdots$ &$\cdots$ & & 0.41 & 0.22 & 7 & & 0.35 & 0.14 & 8 & & 0.56 & 0.22& 9\\

\hline
\end{tabular}
\\
$^{\dagger}$  from \citet{2017arXiv170401957M}.
\end{table*}

\begin{figure}
\includegraphics[angle=270,width=8cm]{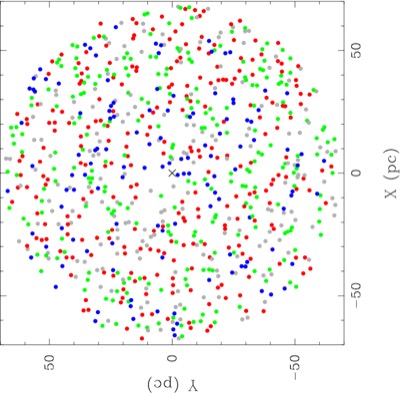}
\caption{Classifications for the stars with $\mathrm{70\,\mathrm{pc}}$ of the position of the SN~1993J (indicated by $\times$).  Stars are colour coded according to their posterior probability of belonging to the age components of $\tau = 7.10$ (blue), 8.18 (green) and 9.23 (red) (or grey, if $p(class) < 0.75$).}
\label{fig:disc:93Jclass}
\end{figure}

The average properties of the stellar populations, as a function of SN type, are presented in Table \ref{tab:disc:aver}.  Similar to the inference about the progenitors of Type Ia and CC SNe from their host galaxies, (for an overview see \citealt{2005AJ....129.1369P} and references therein), we note that the progenitors arising from older stellar populations, may arise in mixtures of young and old stellar populations, but SNe arising from young stellar populations must always arise in environments hosting young stellar populations.   This is most easily apparent when we consider the average value of the youngest stellar population component ($\tau_{1}$) summarised in Table \ref{tab:disc:aver}.  We find that the average age of the youngest components increases corresponding to the scheme $\mathrm{Ic} \rightarrow \mathrm{Ib} \rightarrow \mathrm{IIb}$, which might naively suggest lower mass progenitors associated with progressively older populations.  We note, however, that the average ages for the second age components ($\tau_{2}$) do not show such a sequence.  A key feature of this analysis is the standard deviation of $\tau_{1}$, which also shows an increasing scattering with stellar population age.  Incorporating the results from \citet{2017arXiv170401957M}, we find strong evidence that Type IIP SNe might arise in mixtures of populations (with young and old components), while at the other extreme the ages for the populations around Type Ic SNe are much more tightly grouped.  An outlier from this inference for the Type Ic SNe is SN 2002ap ($2.3 \sigma$) which, as discussed in Section \ref{sec:disc:preexp}, is consistent with results from pre-explosion observations.   Without appealing to complex generative models (which might attempt to consider SNe arising from multiple mass ranges) we can determine a characteristic age for each SN type, under the assumption that each SN arises from a population component common to all populations seen around a given subtype (and penalising population components that are not observed in other examples of that subtype).   The characteristic age can be defined as the age at which the combined probability for the age distributions of all SNe of a given subtype is maximised: $\tau_{C} = \tau \left( \mathrm{max} \prod_{i}^{N} \left(p\left(\tau_{i,1},\tau_{i,2},\tau_{i,3}\right)\right)\right)$.    We find $\tau_{C}(\mathrm{IIb}) = 7.20$, $\tau_{C}(\mathrm{Ib}) = 7.05 $ and $\tau_{C}(\mathrm{Ic}) = 6.57$.  Using the results of \citet{2017arXiv170401957M}, we find that the characteristic age for Type IIP SNe is $7.10$.  Both the average and characteristic ages suggest an association for Type Ic SNe with younger populations and, hence, these SNe having more massive progenitors that would have undergone a WR phase.  We also find that that the Type IIb and IIP SNe have similar characteristic ages, indicating they arise from similar mass progenitors which is consistent with pre-explosion observations.  Alternatively, the stellar populations associated with Type Ib SNe show much more spread and this may indicate that this subtype of SNe has progenitors with properties intermediate between the extremes of the Type IIb and Ic SNe.

We caution that our analysis of the properties of the ensemble of stellar populations, for a given SN type, are made on a statistical basis over the entire sample, and may not be applicable to individual objects.  For individual SNe, the probability distribution for ages (especially in the presence of multiple age components) are complex and may not indicate directly from which age component a given SN progenitor arose.  \citet{2009ApJ...703..300G} suggested that the contribution of stars from unrelated age components to the derived star formation history could be reduced by restricting the analysis to only those stars within $50\,\mathrm{pc}$ of the SN position.  This would minimise what they consider to be contamination from a ``background" population. Our approach means that, should there be a background stellar population, with a significantly different age to the massive star population associated with the SN, then it should appear as an additional separate age component.  In our analysis of the closest SN in our sample, SN 1993J (for which we selected stars within $\mathrm{70\,\mathrm{pc}}$ of the SN position), we find that the different population age components overlap spatially (see Fig. \ref{fig:disc:93Jclass}).  The nearest young objects around SN 1993J are within a few {\it projected} parsecs of the SN position, significantly closer than the brightest, blue sources previously identified by \citet{2002PASP..114.1322V} and \citet{maund93j}; but other sources of similar young age are found at larger distances.  These objects would all fall within the area of one WFC3 pixel at distances $>5\,\mathrm{Mpc}$, and such detail would be beyond resolution capabilities and the detection limits for the remainder of the more distant SNe in the sample.  Restrictive selection of the stars around a SN location may not, therefore, efficiently remove contamination from a background population nor necessarily preferentially select stars with similar age properties to the progenitor.

As demonstrated in Section \ref{sec:res:spatial}, the apparent, projected spatial distribution of stars with respect to the SN position may provide an alternative clue as to which stellar population component a progenitor belonged to.  The majority of the Type Ic SNe are located within 100pc of a massive star complex, implying a strong association between the progenitor and very young, nearby stars.  An obvious exception is SN 2002ap, which appears dislocated from the other Type Ibc SN in terms of both age, the number of bright massive stars in its vicinity and their spatial distribution (see Fig. \ref{fig:res:clus_sep}).  The apparent scatter in the properties of the spatial locations increases as one progresses from Type Ib to Type IIb SNe.   For those Type IIb SNe with seemingly young host stellar populations (such as 2008ax and 2013df), the spatial distribution of surrounding stars are very different.  Around the position of SN 2008ax there does not appear to be any nearby association, rather the distribution of stars is consistent with a uniform spatial distribution, which may be correlated with the overlapping ages of the observed stellar populations components. The location of SN 2013df in Figure \ref{fig:res:clus_sep},  would suggest there is an association of massive stars but that the SN position is offset  (which is apparent from Fig. \ref{fig:res:stamp}) and, hence, the progenitor was not associated with the youngest stellar population.  The offset of the position of iPTF13bvn from the majority of bright young stars might would also support the hypothesis that the progenitor arose from an older component, which would be consistent with pre-explosion constraints on the progenitor (see Section \ref{sec:disc:preexp}).  The difficulty of drawing specific conclusions for individual objects, however, is reflected by SN~2012P, where the age and spatial properties of the region (and, indeed, the coincidence of the SN with a probable cluster) would support the interpretation that it came from a high mass progenitor, which would seem to be inconsistent with the analysis presented by \citet{2016A&A...593A..68F}.

%
%EXTINCTION
%

\subsection{Extinction}
\label{sec:disc:ext}

\begin{figure}
\begin{center}
\includegraphics[width = 7.0cm, angle = 270]{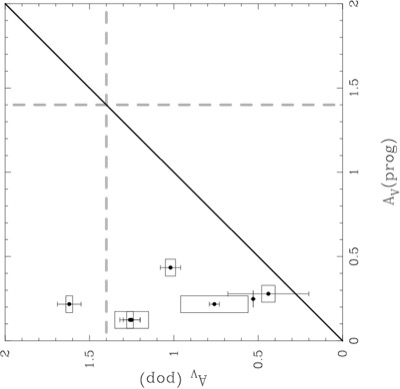}
\end{center}
\caption{A comparison between the extinctions derived here and those previously used for the analysis of detection limits for the progenitors in pre-explosion images for SNe shared between our sample and those of \citet{2013MNRAS.436..774E}.  The boxes indicate the degree of differential extinction $\mathrm{d}A_{V}$.  The grey dashed lines indicate the mean extinction of $A_{V} =  1.4\,\mathrm{mags}$ inferred towards the sample of Type Ibc SNe presented by \citet{2011ApJ...741...97D}.}
\label{fig:disc:eldext}
\end{figure}

The high levels of extinction associated with these SNe, as inferred from the stellar populations, are in direct contrast with the low levels of extinction used in the interpretation of detection limits for the progenitors in pre-explosion observations by \citet{2013MNRAS.436..774E}, which are lower by a factor of $\sim 2-3$ (see Fig. \ref{fig:disc:eldext}).  While the extinction inferred toward the surrounding stellar population need not necessarily be applicable to the progenitor (which could be located in front or behind the observed population), higher levels of extinction towards the SNe, and by extension their progenitors, may have important consequences for the luminosity constraints placed on the progenitors by \citet{2013MNRAS.436..774E} using pre-explosion images.  The reddening/extinction estimates assumed by \citet{2013MNRAS.436..774E} were derived from a range a different sources.  For example, for SN~2000ew \citeauthor{2013MNRAS.436..774E} adopted the foreground extinction previously used in the analysis of \citet[][$A_{V} = 0.14\,\mathrm{mags}$]{2005astro.ph..1323M}; however, in their original analysis of the CMD for the region around SN~2000ew there is evidence for a significant red offset in the locus of the stellar population with respect to the isochrones indicating significant reddening \citep[see][and their Figure 19]{2005astro.ph..1323M}.  

The average reddening towards the large sample of Type Ibc SNe presented by \citet{2011ApJ...741...97D} is $\overline{E(B-V)} = 0.45 \pm 0.22\,\mathrm{mags}$ which, for a Galactic $R_{V} = 3.1$ reddening law \citep{ccm89} corresponds to $\overline{A_{V}} \sim  1.4\,\mathrm{mags}$.    This level of extinction is higher than the average degree of extinction inferred for our sample of SNe of $\overline{A_{V}} = 0.7 \pm 0.4\,\mathrm{mags}$, although this figure neglects the role of differential extinction, but supports the association of high levels of extinction with the sites of stripped-envelope SNe.  In their sample of stripped-envelope \citet{2017arXiv170707614T} found a large average extinction $\overline{A_{V}} = 0.6 \pm 0.5\,\mathrm{mags}$, although with a significant scatter and, we note, they were able to assess the impact of different values of host $R_{V}$ (which we do not consider).  The average extinction assumed in the study of \citet{2013MNRAS.436..774E} was $\overline{A_{V}} = 0.7 \pm 0.9\,\mathrm{mags}$, however this relatively high-value is dominated by three highly extinguished SNe (2005V, 2011hm and 2011hp), which are not included in the sample presented here.  If the high extinction SNe are excluded, the average extinction towards the \citeauthor{2013MNRAS.436..774E} sample significantly decreases to $\overline{A_{V}} = 0.2 \pm 0.1\,\mathrm{mags}$.

Alternative sources for reddening/extinction estimates towards SNe in our sample do support higher values for $A_{V}$ consistent with estimates for the stellar population. Using a spectrum of SN~2000ew presented by \citet{2009MNRAS.397..677T}, from 17 Mar 2001, we measure an equivalent width for Na I D of $1.6\pm0.3$\AA\, corresponding to $E(B-V) = 1\pm0.3$ mags \citep[][; although we note that the measured equivalent width is at the extreme of the abscissa of the defined relationship]{2012MNRAS.426.1465P}\footnote{We also note that \citet{2013ApJ...779...38P} advise caution in the use of Na I D line strengths to derive reddenings.}. In addition, the extinction found from the stellar population analysis would correspond to an equivalent width of $1.2$\AA\, which, given the noise of the spectrum, could be consistent with our measurement of the line strength. While the equivalent width of sodium does not provide a precise estimate of the extinction towards SN~2000ew, it does support the high level of extinction inferred from the CMDs constructed from both the ACS and WFC3 observations.  Similarly, \citet{2017arXiv170703828} only assumed Galactic reddening of $E(B-V) = 0.03\,\mathrm{mags}$, in their analysis of pre-explosion observations of SN~2012fh.  From a spectrum of SN~2012fh, from the Asiago Supernova classification program \citep{2014AN....335..841T}, we find a strong Na I D absorption feature, at the recessional velocity of the host galaxy, for which we measure an equivalent width of $3.8$\,\AA.   As with SN 2000ew, this is beyond the range established for the \citeauthor{2012MNRAS.426.1465P} relation, but supports high levels of extinction arising in the host galaxy.   It is important to note, however, that the determination of a different extinction to the host stellar population than was derived for the SN itself, however, does not mean that the progenitor was necessarily subject to the same extinction as the population.  We have found significantly lower and higher extinctions for the populations around SNe 2005at and 2004gt, respectively, than was determined for the SNe  \citep{2014A&A...572A..75K, 2005ApJ...630L..33M}.

\begin{figure}
\includegraphics[angle=270,width=8cm]{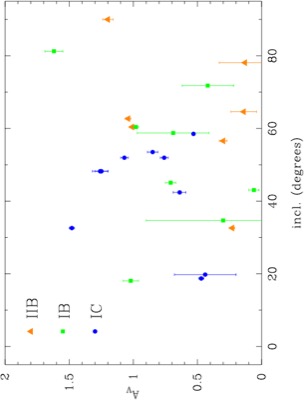}
\caption{The relinferred level of extinction ($A_{V}$) for the stellar populations in our sample against host galaxy inclination.}
\label{fig:disc:inc}
\end{figure}

For our analysis we have considered stars within a fixed projected radius of the target SNe, but this may sample very different column depths through the host galaxies depending on their inclinations.  It might be expected that larger degrees of extinction will be inferred for galaxies observed at large inclination.  From our analysis of this sample of stripped-envelope SNe we find there is no systematic increase in $A_{V}$ with galaxy inclination (see Table \ref{tab:obs:snlist} and Figure \ref{fig:disc:inc}).  The stellar populations observed in galaxies with higher inclinations might also be expected to be more complicated, with a larger number of components, at varying depths, appearing to overlap on the sky.  This might, for example, explain the complex interpretation of the stellar population observed around SN~2008ax, for which the host galaxy NGC~4490 is observed edge-on.  For our sample of SNe (and also including those of \citealt{2017arXiv170401957M}), we find that the average inclinations for stellar population fits requiring $N_{m} = 1$, $2$ and $3$ components are $71.5$ (although for only 3 SNe), 47.3 and 47.4 degrees, respectively; indicating no preference for more complex stellar population configurations with larger inclinations.

\begin{figure}
\includegraphics[angle=270,width=8cm]{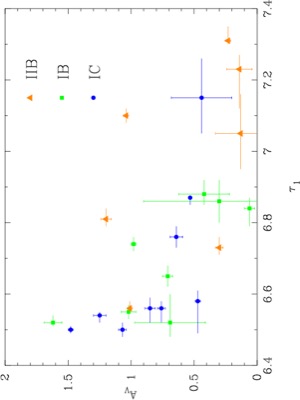}
\caption{The relationship between the extinction ($A_{V}$) and age of the youngest stellar population component ($\tau_{1}$) for each SN in the sample.}
\label{fig:disc:ageext}
\end{figure}

In Fig. \ref{fig:disc:ageext}, we show the relationship between the extinction and the age of the youngest stellar population.  While the older populations of two outlying Type IIb SNe might also be subject to high levels of extinction, Fig. \ref{fig:disc:ageext} would suggest that the younger stellar populations associated with the Type Ibc SNe, that as established in Sections \ref{sec:res:spatial} and \ref{sec:disc:age} are more spatially concentrated, are found in dustier, high extinction environments.  While the progenitors of these SNe may not have arisen from the youngest stellar population component, this relation even extends to those Type IIP SNe that are also found mixed with much younger stellar populations \citep[e.g. 2004et and 2005cs,][]{2017arXiv170401957M}.

%
%Other age estimates
%
\subsection{Age estimates from {\sc H ii} regions}
\label{sec:disc:hiiage}
\begin{figure}
\begin{center}
\includegraphics[width = 7.0cm, angle = 270]{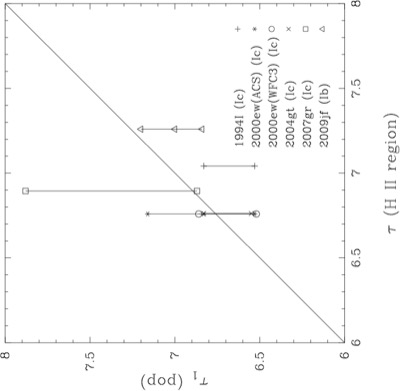}
\end{center}
\caption{A comparison between age estimates derived from the analysis of the resolved stellar population around five Type Ibc SNe, presented here, and those derived from integral field spectroscopy of {\sc H ii} regions in the vicinity of the same SNe presented by \citet{2013AJ....146...30K}.}
\label{fig:disc:othage}
\end{figure}

\citet{2013AJ....146...30K} used integral field spectroscopy to measure the properties of {\sc H ii} regions in proximity to the sites of five of the SNe considered in our sample.  A comparison of the age estimates derived from the resolved stellar populations and those derived from spectroscopic measures  is shown on Figure \ref{fig:disc:othage}.  In general, there is some level of agreement between at least one of the resolved populations identified in the HST observations with an age derived for an {\sc H ii} region in proximity to the SN.  In the case of SN~2007gr, the ages of the two methodologies agree for the youngest age component with $\tau \sim 6.7$, however this does not apply to the other SNe with even younger age components.  This may reflect the limited lifetimes of natal {\sc H ii} regions associated with the formation of massive stars \citep{2013MNRAS.428.1927C}, indicating that the strength of $\mathrm{H\,\alpha}$ emission may overestimate the ages of the progenitors and, hence, underestimate their initial masses in some cases.  \citeauthor{2013AJ....146...30K} compared their measured values of the equivalent width of $\mathrm{H\,\alpha}$ against predictions from STARBURST99 models \citep{1999ApJS..123....3L}, which are based on single star evolution models.  Recently, \citet{2017arXiv171002154E} presented comparable predictions based on binary star evolution models (Binary Population and Spectral Synthesis - BPASS), and \citet{2017arXiv170503606} found the ages for the majority of {\sc H ii} regions in proximity to Type Ibc SNe to have ages $> 10\,\mathrm{Myr}$ .  With BPASS, line strengths measured by \citeauthor{2013AJ....146...30K} would imply ages for the environments of SNe 2000ew, 2004gt, 2007gr and 2009jf of $\tau = 7.1 - 7.3$, $7.1$, $7.5 - 7.6$ and $7.2$, respectively.  These are significantly lower than those derived in comparison to single star population models and those found here for the youngest population components.

The presence of multiple age components in the resolved stellar populations around stripped-envelope SNe suggests caution should be exercised in the interpretation of spectroscopic observations and derived proxies for the age of the progenitor.   The discrepancy between the ages derived from resolved stellar populations and those predicted from binary stellar population synthesis models could potentially be resolved if the $\mathrm{H\,\alpha}$ age is associated with the oldest stellar population component, rather than the youngest (see Table \ref{tab:res}).  As noted by \citet{2017arXiv170401957M}, a key problem with spectroscopic estimates of the age of the progenitor is the potentially large offset between the actual SN position and the nearest {\sc H ii} region or emission line source.  In the cases of SNe 1994I, 2000ew and 2004gt, the nearest emission line sources are $\gtrsim 110\,\mathrm{pc}$ from the SN position, while for SN 2007gr it is debatable if the source in close proximity to the SN position is a bonafide cluster as claimed by \citeauthor{2013AJ....146...30K} given the lack of significant $\mathrm{H\alpha}$  emission \citep{2016MNRAS.456.3175M}.   We also note that such population models generally (and this is the case for both STARBURST99 and BPASS) consider only the two extremes of either instantaneous or continuous star formation, and for the young associations observed around some of the SNe considered here, for which we observe an age spread, this may not be appropriate.  Given the complexity of the spatial and age distributions of massive stars around the sites of SNe, as seen in the HST observations presented here, such measurements are likely to be highly dependent on the distribution of ages and the relative number of massive stars around the SNe, and this complexity is unlikely to be captured in an emission line strength measured for a single line.

%
%Previous stellar pops studies
%
\subsection{Previous estimates of the ages of resolved stellar populations around SNe}
\label{sec:disc:othpopage}
\citet{2014ApJ...791..105W} attempted to derive age and progenitor masses from HST observations of three of the SNe contained in our sample: SNe 1993J, 1994I and
2002ap.  Due to paucity of stars in the locality of SN~2002ap, and the corresponding limitations to the possible analysis they were able to conduct, we can only consider a comparison
with their findings for SNe 1993J and 1994I.  For SN 1993J, we find qualitatively similar masses for the progenitor, although \citeauthor{2014ApJ...791..105W} 
estimated a median mass for the progenitor of $\sim 11M_{\odot}$ with significantly lower extinction (by a factor of 2) than we find from our analysis.  We note that \citeauthor{2014ApJ...791..105W} 
utilised an $F438W - F606W$ CMD, whereas in our analysis we have considered photometry extending from  the UV to the near-infrared.  For SN 1994I, \citeauthor{2014ApJ...791..105W} do
not present a CMD for easy comparison with our observations and, while we note their extinction estimate ($A_{V} = 0.6\,\mathrm{mags}$) is similar to the
minimum extinction we derive, they find no evidence for differential extinction. Given the photometry of SN~1994I in both UV and optical bands (see Section \ref{sec:res:1994I}) looks very unlike
a single isochrone or the superposition of multiple age components, we consider it unlikely that there is no differential extinction.  This may mean that the mass that \citeauthor{2014ApJ...791..105W}
derive for the progenitor of SN~1994I is an underestimate; the upper range of the progenitor mass they derive does, however, extend as high as $56M_{\odot}$.  

SN 2011dh provides a useful opportunity to assess the effect of the choice of colours used, the depth of the imaging and the Bayesian methodology employed to calculate the properties of the stellar population.  Previously, \citet{2011ApJ...742L...4M} used ACS/WFC observations of M51 acquired prior to the explosion of SN 2011dh, that included the progenitor, to provide an age estimate for the exploding star.  Their analysis, however, used only the $F555W$ and $F814W$ filters and was restricted to only those stars within $\mathrm{50\,pc}$ of the SN position.  We repeated their analysis using the same observations, the same selection radius for stars (although with new DOLPHOT photometry), but with our own Bayesian tools rather than MATCH \citep{2002MNRAS.332...91D}.  Overall, we find good agreement between our analysis of the older dataset and the previously reported results of \citeauthor{2011ApJ...742L...4M}.  We find that the restricted choice of filters yields a slightly larger extinction ($A_{V} = 0.58\,\mathrm{mags}$) than found by \citeauthor{2011ApJ...742L...4M}, although we note that the maximum extinction they considered was $A_{V}  = 0.5\,\mathrm{mags}$.  Both of these values are significantly larger than we have found with the full multi-wavelength dataset used here.  Our analysis also finds an age component of $18.6\,\mathrm{Myr}$ which is almost identical to the age component they identified as being associated with the progenitor.   This suggests, for limited colours, that the degeneracy between age and reddening (extinction) is difficult to resolve with just $V  - I$ CMDs.  If we restrict the extinction to $A_{V} = 0.12\,\mathrm{mags}$, which constitutes the lower limit of foreground extinction considered by \citeauthor{2011ApJ...742L...4M}, then the overall ages of each population component found in our analysis become older, with the youngest age component having an age of $\mathrm{10\,\mathrm{Myr}}$ and the progenitor component having an age of $\mathrm{25\,\mathrm{Myr}}$ (corresponding to the lifetime of a star with $M_{ZAMS} \sim 11 M_{\odot}$).  

From the UV observations, associated with our multi-wavelength dataset, we have also considered an $N_{m}=4$ component fit to the 2011dh data. We find possible evidence for a fourth stellar population component associated with a young age of $\mathrm{8\,Myr}$.  We note that this component constitutes only $4\pm2\%$ of the stars observed around SN 2011dh and that these stars do not constitute a separate component on the  $F555W - F814W$ CMD.  This may have important implications for understanding the role of differential extinction, especially in the case of small numbers of stars or at redder wavelengths, where the apparent spread of colours of stars might be erroneously interpreted as multiple age populations.

%
%NCR pixel statistics
%
\subsection{Normalised Pixel Statistics}
\label{sec:disc:ncr}

Pixel statistic measures provide an alternative statistical method to characterise the host environment through association of the underlying, unresolved stellar population with a proxy of, for example, star formation rate \citep[for a review see][]{2015PASA...32...19A}.  This technique measures the relative brightness of the pixel containing the SN location, usually in narrow band $\mathrm{H\alpha}$ imaging, to the total brightness of the host galaxy.  Studies, such as \citet{2012MNRAS.424.1372A}, have used this technique with low resolution images of the host galaxies of SNe, acquired from ground-based telescopes, to suggest mass constraints for the progenitors of all types of SNe.   A key benefit of this analysis is that it provides a basic comparative measure of the location properties of different types of sources, such as SNe and the types of stars that are their likely progenitors \citep[see e.g.][]{2010A&A...518A..29L,2017A&A...597A..92K}; however, its main disadvantage is that it only probes the bulk properties of the environment, but not the specific physical processes associated with the evolution of populations of massive stars or their influence on the environments \citep{2013MNRAS.428.1927C}.  

\begin{figure}
\begin{center}
\includegraphics[angle=270, width=7cm]{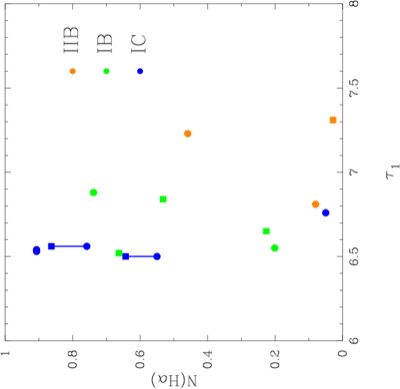}
\end{center}
\caption{The relationship between the age of the youngest stellar population component derived here  and NCR pixel value of $\mathrm{H\alpha}$ presented by \citet[][$\bullet$]{2008MNRAS.390.1527A} and \citet[][$\blacksquare$]{2013MNRAS.436.3464K}. Points connected by lines correspond to the same object with measurements present in both studies.}
\label{fig:disc:ncr:age}
\end{figure}

\begin{figure}
\begin{center}
\includegraphics[angle=270, width=7cm]{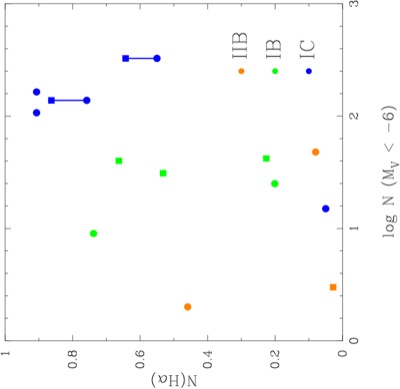}
\end{center}
\caption{The same as for Figure \ref{fig:disc:ncr:age} but comparing the number of stars within 100pc of the SN position with $M_{V} \leq -6\,\mathrm{mags}$ with the NCR pixel value.}
\label{fig:disc:ncr:num}
\end{figure}

In Figures \ref{fig:disc:ncr:age} and \ref{fig:disc:ncr:num} we show the relationships between the properties of the host environment from our analysis of the resolved population and previously published pixel statistic values for $\mathrm{H\alpha}$ emission (derived from narrow-band imaging).  While a number of Type Ic SNe  associated with young stellar populations also have high values, we see there is significant scatter with SNe of other types, with older populations, having similar values (with a Pearson correlation coefficient of -0.35).  There is a more obvious trend and separation of the different SN types, although still with significant scatter, in the relationship between the pixel statistic value and the number of bright stars in the SN locality (yielding a Pearson correlation coefficient of 0.52).  This may reflect that, rather than age, the key characteristic for dictating the pixel statistic value is the number of massive stars, which are responsible for producing the ionising photons required for the production of an {\sc Hii} region.   Some of the scatter present may reflect, however, that we have counted the number of bright sources present in $V$-band images, which includes redder, cooler evolved stars that might not be expected to contribute significantly to the ionizing flux.  As such, the lack of a tighter correlation might not be unexpected, but a similar census at UV and blue wavelengths might be expected to yield a tighter correlation.    As shown in Section \ref{sec:res:spatial}, there is an approximate trend between the age of the host environment and the number of massive stars contained therein;  so it is not unreasonable for there to be a vague trend between $\mathrm{H\alpha}$ and age, but rather as a byproduct of the relationship with the number of massive stars. 

Such analysis, and its cousin ``fractional light" analysis \citep{2010ApJ...717..342H}, has that problem in that it does not explicitly consider extinction, age (and age spread) and stellar density and different studies can measure different values for the same objects \citep[][ as shown on Figs \ref{fig:disc:ncr:age} and \ref{fig:disc:ncr:num}]{2012MNRAS.424.1372A,2017A&A...597A..92K}.   In addition, the reliance on low spatial resolution imaging also means there is ambiguity about the true association between the position of a SN and nearby $\mathrm{H\alpha}$ emission or other star formation proxy \citep{2016MNRAS.456.3175M}.  The lack of a strong trend between the $\mathrm{H\alpha}$ pixel statistic values and the properties of the resolved stellar populations suggest that, while there may be some underlying ``signal", there is also a significant noise associated with the technique.

%
%Pre-explosion observations
%

\subsection{Comparison with pre-explosion observations}
\label{sec:disc:preexp}
The only successful detection of the progenitor of a Type Ib SN to date is for the Type Ib SN iPTF13bvn \citep{2013ApJ...775L...7C}.  For this SN we find two age components, with the older yielding a lifetime consistent with initial masses estimated from the observed pre-explosion SED and binary stellar evolution models.  We note, however, that a key uncertainty in the interpretation of the pre-explosion observations is the extinction, with \citet{2016MNRAS.461L.117E} and \citet{2016ApJ...825L..22F} adopting extinctions of $A_{V} \sim 0.6 \,\mathrm{mags}$, which is $0.4\,\mathrm{mags}$ lower than we find for the surrounding stellar population.  It is unclear what the implications of additional extinction might be for the interpretation of the pre-explosion observations with respect to binary stellar evolution calculations, however the WR stars models presented by \citet{2013A&A...558L...1G} could be applicable (originally thought excluded due to updated photometry of the pre-explosion source; \citealt{2015MNRAS.446.2689E}).

SN~2002ap stands out as having a substantially older age and sparser surrounding stellar population than the other Type Ic SNe in the sample.  This is consistent with the constraints from pre-explosion observations that rule out all single star progenitors with masses $M_{init} \gtrsim 20M_{\odot}$ \citep{2007MNRAS.381..835C,2017ApJ...842..125Z}; and SN 2002ap represents, therefore, a crucial and successful test of the methodology presented here.

Of the 7 Type IIb SNe in our sample, 5 have constraints on the progenitor from pre-explosion observations.  We find excellent agreement between the ages of the stellar population components and the initial masses from direct detection of the progenitors of SNe 1993J, 2008ax, 2011dh and 2013df \citep{alder93j,maund93j,2014ApJ...790...17F,2008MNRAS.391L...5C,2015ApJ...811..147F,2011ApJ...739L..37M,2011ApJ...741L..28V,2014AJ....147...37V}; although we note that for SNe 2008ax, 2012P and 2013df there is evidence for a younger stellar population also in the vicinity of the SN.  Estimates of the initial mass of the progenitors of Type IIb SNe from the measurement of the strength of the oxygen lines in late-time nebular spectra have also yielded consistent results with those derived from pre-explosion observations \citep{2015A&A...573A..12J}.  For SNe 1996cb and 2001gd, which lack pre-explosion observations, we find comparable ages to the other Type IIb SNe, implying initial masses for the progenitors of this type of SN in the range $\sim 8 - 20M_{\odot}$.

For their sample of $N=11$ Type Ibc SNe with pre-explosion observations, \citet{2013MNRAS.436..774E} concluded that there was only small probability ($0.15$) of these objects being drawn from the same sample of WR stars as are found in the LMC \citep{2002ApJS..141...81M}; hence requiring some portion of these SNe to arise from the lower mass binary progenitor channel.  This conclusion was predominantly driven by SN~2002ap which had, at the time, the deepest detection limits available on a progenitor in pre-explosion observations \citep{2007MNRAS.381..835C}.  The later discussion by \citet{2015PASA...32...16S} argued for the complete absence of high-mass progenitors.  These conclusions would, however, seem to be incompatible with both the young ages associated with the sites of Type Ibc SNe found here and, in particular, with the relatively homogeneous spatial properties of the massive star regions hosting Type Ic SNe.  As noted above (see Section \ref{sec:disc:ext}), there are significant differences in the extinctions adopted in \citeauthor{2013MNRAS.436..774E} study and those found here for the SNe shared between the two samples.  If the extinction is systematically underestimated, there is a commensurate overestimation of the sensitivity of the constraints on the progenitor in pre-explosion observations.  An immediate consequence of our derived extinctions is that the fraction of LMC-like WR stars that could have been detected in the available pre-explosion observations, presented by \citeauthor{2013MNRAS.436..774E}, decreases.

Of the original \citeauthor{2013MNRAS.436..774E} sample, we share 7 SNe with our current sample.  The other 4 SNe (2005V, 2010br,  2011am, 2011hp) either suffered from extreme levels of extinction (and, hence, the pre-explosion observations did not provide significant constraints) or had insufficient data to facilitate an analysis of the surround stellar population.  After correcting the pre-explosion limits presented by \citeauthor{2013MNRAS.436..774E} for our revised extinctions and assumed distances (although ignoring differential extinction and colour differences between filter systems, which \citeauthor{2013MNRAS.436..774E} show are relatively small $\sim 0.1\,\mathrm{mag}$), we find that the probability for the non-detection of a WR progenitor increases to $0.44$ (compared to $0.2$ for the \citeauthor{2013MNRAS.436..774E} analysis for the same 7 SNe).  Recently, \citet{2017arXiv170703828} reported new detection limits for the progenitor of SN~2012fh, based on multiple pre-explosion observations.  As noted above, we find that our analysis of the stellar population supports a higher level of extinction than assumed in their study (although, even with additional extinction, the pre-explosion constraints are still  deeper than those for SN~2002ap).  Now including the revised limits for SN~2012fh, we find the probability of not having observed a WR progenitor in the sample of 8 Type Ibc SNe is $0.14$.

\begin{figure}
\includegraphics[width=8cm,angle=270]{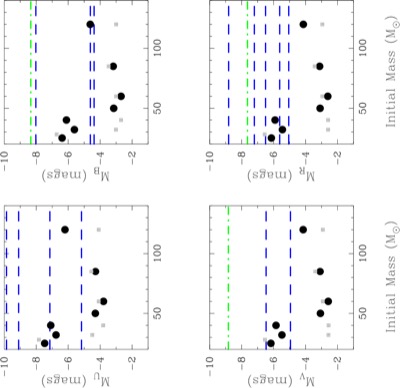}
\caption{$UBVR$ magnitude limits for 8 Type Ib (green; dot-dashed line) and (blue; dashed lines) Ic SNe in the $UBVR$ bands for WR progenitors derived from non-rotating (grey square; $\blacksquare$) and rotating (black circle; $\bullet$) Geneva stellar evolution models \citep{2013A&A...558A.131G}. }
\label{fig:disc:groh}
\end{figure}

Following \citeauthor{2017arXiv170703828}, we compared these revised pre-explosion detection constraints with the absolute magnitudes for progenitors from rotating and non-rotating stellar evolution models presented by \citet{2013A&A...558A.131G}, as shown Figure \ref{fig:disc:groh}.  With revised extinctions, we find that the 8 pre-explosion progenitor constraints would only have yielded detections for a small number of WR stars with initial masses in the range $20 - 40M_{\odot}$, but would not be sensitive to the WR stars arising from the most massive stars.  This comparison also highlights that these detection limits are least sensitive to rotating models which appear, except in one low mass case, to be systematically fainter than the non-rotating models.

In addition, from our measurement of the extinction (see Section \ref{sec:disc:ext}), we also find that younger populations are generally associated with higher levels of extinction, meaning that the effect of enhanced extinction penalises the detection of WR progenitors that would be preferentially associated with younger stellar populations.
%
%Mej
%
\subsection{Implications for ejected masses}
\label{sec:disc:mej}
For their analysis of SN 2007gr, \citet{2016MNRAS.456.3175M} noted that the initial mass inferred from the surrounding stellar population was inconsistent with the apparent ejected mass determined from modelling of the SN \citep{2010MNRAS.408...87M}.  Based on our observations and analysis of the surrounding stellar populations, we have found that most Type Ic SNe (with SN 2002ap being an important exception) are likely to have arisen from massive stars $\gtrsim 20M_{\odot}$, of which some would have been capable of having undergone a WR phase.  This is in contrast to recent studies of the light curves of stripped-envelope SNe \citep[e.g.][]{2013MNRAS.434.1098C,2015MNRAS.450.1295W,2016MNRAS.457..328L,2017arXiv170707614T} that, using semi-analytical \citep{1982ApJ...253..785A} and hydrodynamical models, find low ejecta masses.  

\citet{2017arXiv170707614T} consider the ejecta mass boundary between lower mass progenitors in binaries ($<20M_{\odot}$) or higher mass single ($>28M_{\odot}$) or binary ($>20M_{\odot}$) progenitors as being $M_{ej} \approx 5M_{\odot}$.   Based on early time light curves, covering the rise to and decline from the light curve maximum, \citet{2013MNRAS.434.1098C}, \citet{2016MNRAS.457..328L} and \citet{2017arXiv170707614T}  provide estimates for $M_{ej}$ for 11 SNe in our sample, finding ejecta masses $< 5M_{\odot}$  in all cases except for the Type Ib SN 2009jf \citep[$7.3M_{\odot}$;][]{2013MNRAS.434.1098C}.    The consensus from these studies is that the low ejecta masses associated with normal stripped-envelope SNe are overwhelmingly consistent with lower mass progenitors in binaries; although \citeauthor{2013MNRAS.434.1098C} note the light curves of those Type Ic SNe associated with Gamma Ray Bursts and X-ray Flashes are consistent with higher $M_{ej}$. 

\citet{2015MNRAS.450.1295W} also considered constraints on the ejecta mass from the behaviour of the late-time light curves and found, although in some cases only with upper limits, that significantly higher ejecta masses cannot be excluded.  A significant issue with the application of these models, and in particular the semi-analytical models, is the assumption of a particular value of opacity and its value over the evolution of the SN.  In their study, \citeauthor{2015MNRAS.450.1295W} also noted that the earlier work by \citet{1988ApJ...333..754E}, to model the Type Ib SN 1983N, required two different values of the opacity to successfully reproduce the peak and the late-time light curves with consistent values of $M_{ej}$ and kinetic energy.  \citet{2016MNRAS.457..328L} justified their choice for the value of the opacity by requiring the derived values of $M_{ej}$ to be consistent with the analysis of the pre-explosion observations presented by \citet{2013MNRAS.436..774E}.  As shown above, the latter study may have underestimated the extinction and therefore underestimated the progenitor initial mass constraints.  From their analysis, \citeauthor{2015MNRAS.450.1295W} derived an average value of the opacity of $\kappa = 0.01\, \mathrm{cm^{2} g^{-1}}$, which is lower than assumed in analyses by \citet[][0.07]{2013MNRAS.434.1098C}, \citet[][0.06]{2016MNRAS.457..328L} and \citet[][0.07]{2017arXiv170707614T}, which were based on previous detailed simulations, e.g. \citet{2000AstL...26..797C} and \citet{2003ApJ...593..931M}.  As noted by \citeauthor{2017arXiv170707614T}, \citet{2016MNRAS.458.1618D} show that the assumption of constant $\kappa$ is poor.  For their models, \citeauthor{1988ApJ...333..754E} required an average value of $\kappa = 0.003\,\mathrm{cm^{2} g^{-1}}$, and recovered a larger ejecta mass.  \citeauthor{2016MNRAS.457..328L} estimated that opacity values of $\sim 0.02\,\mathrm{cm^{2} g^{-1}}$, similar to those required by \citeauthor{2015MNRAS.450.1295W}, yielded higher ejecta masses that would be more consistent with those expected for higher mass WR progenitors.  \citeauthor{1988ApJ...333..754E} found that helium and oxygen in the ejecta were recombined prior to maximum light and, therefore, the significant ejecta mass in these elements contributed little to the overall opacity.   Spectropolarimetric observations of Type Ib SNe, for example, reveal the importance of non-thermal excitation for yielding visible He line features, revealing material that would otherwise be invisible \citep{2009ApJ...705.1139M,2016MNRAS.457..288R}.  Resolving uncertainties about the degree of opacity present in stripped-envelope SNe may, therefore, be a route to reconciling the apparent young ages determined for some of progenitors found here and low ejecta masses derived from observations of SN lightcurves.

%
% Caveats and caution
%

\subsection{Future Work}

%
% Reddening/Extinction/Age degeneracy
%

As previously discussed by \citet{2017arXiv170401957M}, there is a potential degeneracy between age and extinction in CMDs (especially if observations are only available over a limited wavelength range).  SN 2016bau provides an obvious example, where the $F555W$ and $F814W$  observations provide limited information on the extinction (yielding large uncertainties) coupled with relatively poor age constraints, in particular for the young age component. Red CMDs (such as $V - I$) do not provide significant diagnostic capability for probing young stellar populations, mainly because young, massive stars, on or just off the main sequence, have high temperatures which may not be detected with these filter combinations or whose temperatures cannot be directly measured with this colour combination; on these CMDs, young isochrones are vertical lines, with almost constant colour. As only the brightest blue stars are detectable at $V$-band wavelengths, a large fraction of young massive stars in the vicinity of CCSNe may go undetected and corresponding ages might be overestimated, and progenitor masses underestimated, by not observing at sufficiently blue wavelengths.   For redder CMDs, the analysis of young stellar populations may be dependent on stars undergoing post-main sequence evolution to constrain the age which, due to the short duration of these phases, constitute only a small fraction of the total number of stars.  Very large numbers of stars are required to appropriately sample the isochrones.  The situation is exacerbated by the role of differential extinction, for which we have had to assume an underlying Gaussian model, but instances such as SN~1994I demonstrate that this form of extinction is not universal.  Previously,  \citet{2012ApJ...761...26J} assumed a flat law with $\mathrm{d}A_{V} \leq 0.5$ for populations with $t \leq 40\,\mathrm{Myr}$, while \citet{2014ApJ...791..105W} were required to increase this limit.  

As demonstrated by such observations as those for SNe 2004gt and 2011dh, UV observations help to provide a better handle on the degree of extinction and simultaneously probe young stellar populations.  Conversely, those SN environments with limited observations at redder wavelengths (e.g. SNe 1996cb, 2004gq and 2016bau) have more poorly constrained ages and extinctions.  In order to probe the young massive star populations associated with core-collapse SNe, especially for stripped-envelope SNe,  observations with the $U$ and $B$ band filters would provide both the depth and dynamical colour range to assess both ages and extinctions.  The short-lived, extreme phases of evolution after the main sequence are difficult to model \citep{2012ARA&A..50..107L}, whereas the main sequence is the longest-lived and best understood phase of the evolution of a massive star that is also best observed at bluer wavelengths.
 The benefit of observing massive star populations at these wavelengths is most easily seen in the analysis for the population around SN~2007gr presented by \citet{2016MNRAS.456.3175M}, for which WFC3 UVIS $F336W$ and $F555W$ observations provided a dynamical colour range of $\Delta (F336W - F555W) \sim 4 - 5\,\mathrm{mags}$ for populations with $t < 10\,\mathrm{Myr}$.   Further studies of the massive star populations associated with CCSNe should, therefore, use bluer wavelength CMDs and this is the {\it raison d'\^{e}tre} behind the HST Cycle 24 program ``A UV census of the sites of core-collapse supernovae" (SNAP-14762; PI Maund). Utilising a single instrument will also overcome any bias that has also been introduced by the use of different instruments and, correspondingly, different depths of the observations for different SN locations (which is particularly obvious for those SNe in our sample for which only WFPC2 observations were available).

%
% Choice of isochrones and the underlying generative model
%

For our analysis we have utilised the Padova isochrones derived from the PARSEC models \citep{2002A&A...391..195G, 2012MNRAS.427..127B}, which cover initial masses up to $\sim 100M_{\odot}$.    As noted by \citet{2017arXiv170401957M}, the Padova isochrones utilise blackbody functions to model the SEDs of high temperature ($>50\,\mathrm{kK}$) stars such as WR stars, which also exhibit significant line emission features that are not included.  This could, potentially, cause large shifts in the locus of these stars in the CMD, which are not accounted for in our analysis.   The most massive star so far discovered \citep[R136a1 with $M_{init} \sim 325M_{\odot}$;][]{2016MNRAS.458..624C} suggests that in especially young stellar populations, such as 30 Dor, any analysis will have to extend to even higher masses, and future studies of this sort may to have adopt more recent isochrones including very massive stars such as \citet{2014MNRAS.445.4287T}. \citet{2016ApJ...823..102C} present a comparison between the PARSEC isochrones and those derived from the Modules for Experiments in Stellar Astrophysics (MESA) stellar evolution package \citep{2011ApJS..192....3P}, finding generally good agreement at the young ages ($\tau = 7.5$) relevant to our study; with PARSEC predicting slightly higher luminosities than the MESA isochrones.  At young ages, the analyses of massive star populations associated with SNe against isochrones generated from different stellar evolution tracks could, potentially, be used to test uncertain physical processes and their different implementations in stellar evolution calculations \citep{2016ApJ...823..102C}.

Given some suggestions that massive stars are preferentially found in binary systems \citep{2012Sci...337..444S}, for which interaction between the binary components is expected to dramatically influence their evolution (in particular through rejuvenation),  it is questionable if our assumed binary fraction (0.5) and use of  ``composite" binaries created from single star isochrones is sufficient to adequately describe very young, massive star populations \citep{2016ApJ...818...75V,2017ApJ...842..125Z,2017arXiv171002154E}.     For those SNe in our sample for which there are also pre-explosion constraints on the progenitors, we find that the ages we derive agree with the inferred initial masses.  In the case of SN 1993J, we find there is a clear stellar population component with an age that is in agreement with the expected progenitor lifetime and that all other components are significantly older.  Despite the progenitor system itself being responsible for the production of a rejuvenated main sequence star \citep{maund93j}, there is no evidence for a significant population of rejuvenated stars in the vicinity of the SN.  Where there are populations that are significantly younger than expected  for a given progenitor (e.g. SN 2012P), we see that the spatial characteristics are consistent with other SNe that are also associated with young stellar populations.  If rejuvenation were a significant source of ``apparently" young stars, we might expect a significant skew in all stellar populations having an apparent ``young" age component, but having a wider range of spatial characterics.  This is not observed here, with the majority of Type IIb SNe apparently arising from older populations and that are also sparser than observed for the Type Ic SNe.  These characteristics also extend to the Type IIP SNe presented by \citet{2017arXiv170401957M}.  

As noted in Section \ref{sec:disc:hiiage}, recent binary population synthesis models suggest ages derived from $\mathrm{H\alpha}$ emission line strength should also be increased \citep{2017arXiv170804618C,2017arXiv170503606}, implying that all SNe (Types IIP, IIb and Ibc) should have progenitors with $M_{init} < 20M_{\odot}$.   If the ages derived from resolved stellar populations (as presented here) are underestimated, due to a contribution from a rejuvenated population, and the ages derived from $\mathrm{H\alpha}$ emission strengths are also underestimated, then it is difficult to understand the different host environments for the different SN types, in particular for the Type Ic SNe.  If ages derived in these ways are under-estimated, then SN subtype is not necessarily just an innate property of the progenitor (such as initial mass, metallicity, binarity etc) but would also have to be a product of environment (in particular proximity of other stars) through some mechanism.  Although the interpretation of  pixel statistic measures is not directly comparable to a direct study of the resolved population (for reasons we address in Section \ref{sec:disc:ncr}), our findings superficially agree in terms of the different types of environments that we infer for the progenitors of the different stripped-envelope SN subtypes.   This may imply that $\mathrm{H\alpha}$ line strength may be insufficient to properly characterise the complex environments around massive stars \citep{2013MNRAS.428.1927C}. A further advancement, in the future, would be to consider full binary stellar evolution models introducing new parameters such as the binary period. Importantly, analysis such as this (using just photometric techniques) could potentially, over a large number of fields, be used to determine the binary fraction probability function.

%%%%%%%%%%%%%%%%%%%%%%%%%%%%%%%%
%CONCLUSIONS
%%%%%%%%%%%%%%%%%%%%%%%%%%%%%%%%

\section{Conclusions}
We have presented an analysis of the resolved stellar populations around the locations of 23 stripped-envelope SNe, as imaged using the Hubble Space Telescope.  Our principal findings are that:
\begin{itemize}
\item{SNe following the scheme $\mathrm{IIb} \rightarrow \mathrm{Ib} \rightarrow \mathrm{Ic}$ follow a general trend of being associated with progressively younger stellar populations and, hence, more massive progenitors.}
\item{There is strong evidence that some of the Type Ib SNe and the majority of Type Ic SNe have arisen from stars massive enough to undergo a WR phase.}
\item{The spatial properties of Type Ic SNe are consistent with the majority of these events occurring within $100\,\mathrm{pc}$ of a dense stellar association which, coupled with young ages, suggests a causal connection between the progenitor and these stellar structures.}
\item{SN~2002ap, which has deep pre-explosion constraints on the progenitor, is a clear deviant from the general age and spatial trends observed for other Type Ic SNe.}
\item{For older populations observed around those Type IIb SNe with pre-explosion constraints on the progenitor, we find no clear evidence for rejuvenated massive stars in their locality.}
\item{The $\mathrm{H\alpha}$  pixel statistic is loosely coupled to age, but is also sensitive to other properties of the environment that means it does vaguely follow some of the trends separating the subtypes of stripped-envelope SNe that we have identified.}
\item{Age estimates derived from the resolved stellar populations are in some agreement with those derived from measurements of $\mathrm{H\alpha}$ equivalent widths for nearby {\sc H ii} regions using single star stellar population synthesis models, but may potentially be in disagreement with binary population synthesis models.}
\item{The sites of stripped-envelope SNe, in particular Type Ib and Ic SNe, are associated with higher levels of extinction than previously assumed in analyses presented on progenitor constraints from pre-explosion images (although we note that, like other indirect measures, even this extinction may not be specifically applicable to the progenitor).}
\item{After correcting pre-explosion progenitor constraints for extinctions derived from the surrounding stellar populations, WR stars are still viable progenitors and not excluded by pre-explosion imaging.}
\item{The young ages and, hence, large initial masses inferred for the progenitors of Type Ibc SNe are in disagreement with the ejected masses inferred for these SNe from analysis of their light curves}.

\end{itemize}

%%%%%%%%%%%%%%%%%%%%%%%%%%%%%%%%
%ACKNOWLEDGMENTS
%%%%%%%%%%%%%%%%%%%%%%%%%%%%%%%%
\section*{Acknowledgments} 
The research of JRM is supported through a Royal Society University Research Fellowship.  JRM thanks J. Craig Wheeler for his kind hospitality at the University of Texas at Austin, during the preparation of this manuscript, and to both him and Paul Crowther for comments on an original draft.

Based on observations made with the NASA/ESA Hubble Space Telescope, obtained from the Data Archive at the Space Telescope Science Institute, which is operated by the Association of Universities for Research in Astronomy, Inc., under NASA contract NAS 5-26555. These observations are associated with programmes 8400, 9114, 9353, 8602, 10187, 10272, 10452, 10877, 11119, 11570, 11575, 11577, 12170, 12262, 12531, 12888, 13350, 13364, 13433, 13684 and 14668.

This research has made use of "Aladin sky atlas" developed at CDS, Strasbourg Observatory, France.Funding for SDSS-III has been provided by the Alfred P. Sloan Foundation, the Participating Institutions, the National Science Foundation, and the U.S. Department of Energy Office of Science. The SDSS-III web site is http://www.sdss3.org/.  

This paper has made use of observational data deposited in the Weizmann interactive supernova data repository - http://wiserep.weizmann.ac.il.

SDSS-III is managed by the Astrophysical Research Consortium for the Participating Institutions of the SDSS-III Collaboration including the University of Arizona, the Brazilian Participation Group, Brookhaven National Laboratory, Carnegie Mellon University, University of Florida, the French Participation Group, the German Participation Group, Harvard University, the Instituto de Astrofisica de Canarias, the Michigan State/Notre Dame/JINA Participation Group, Johns Hopkins University, Lawrence Berkeley National Laboratory, Max Planck Institute for Astrophysics, Max Planck Institute for Extraterrestrial Physics, New Mexico State University, New York University, Ohio State University, Pennsylvania State University, University of Portsmouth, Princeton University, the Spanish Participation Group, University of Tokyo, University of Utah, Vanderbilt University, University of Virginia, University of Washington, and Yale University. 
%%%%%%%%%%%%%%%%%%%%%%%%%%%%%%%%%%%%
%BIBLIOGRAPHY
%%%%%%%%%%%%%%%%%%%%%%%%%%%%%%%%%%%%
\bibliographystyle{mnras}
%\bibliography{/Users/justyn/Documents/Bibliography/main.bib}

\begin{thebibliography}{}
\makeatletter
\relax
\def\mn@urlcharsother{\let\do\@makeother \do\$\do\&\do\#\do\^\do\_\do\%\do\~}
\def\mn@doi{\begingroup\mn@urlcharsother \@ifnextchar [ {\mn@doi@}
  {\mn@doi@[]}}
\def\mn@doi@[#1]#2{\def\@tempa{#1}\ifx\@tempa\@empty \href
  {http://dx.doi.org/#2} {doi:#2}\else \href {http://dx.doi.org/#2} {#1}\fi
  \endgroup}
\def\mn@eprint#1#2{\mn@eprint@#1:#2::\@nil}
\def\mn@eprint@arXiv#1{\href {http://arxiv.org/abs/#1} {{\tt arXiv:#1}}}
\def\mn@eprint@dblp#1{\href {http://dblp.uni-trier.de/rec/bibtex/#1.xml}
  {dblp:#1}}
\def\mn@eprint@#1:#2:#3:#4\@nil{\def\@tempa {#1}\def\@tempb {#2}\def\@tempc
  {#3}\ifx \@tempc \@empty \let \@tempc \@tempb \let \@tempb \@tempa \fi \ifx
  \@tempb \@empty \def\@tempb {arXiv}\fi \@ifundefined
  {mn@eprint@\@tempb}{\@tempb:\@tempc}{\expandafter \expandafter \csname
  mn@eprint@\@tempb\endcsname \expandafter{\@tempc}}}

\bibitem[\protect\citeauthoryear{{Aldering}, {Humphreys}  \&
  {Richmond}}{{Aldering} et~al.}{1994}]{alder93j}
{Aldering} G.,  {Humphreys} R.~M.,   {Richmond} M.,  1994, \aj, \href
  {http://adsabs.harvard.edu/cgi-bin/nph-bib_query?bibcode=1994AJ....107..662A&amp;db_key=AST}
  {107, 662}

\bibitem[\protect\citeauthoryear{{Anderson} \& {Bedin}}{{Anderson} \&
  {Bedin}}{2010}]{2010PASP..122.1035A}
{Anderson} J.,  {Bedin} L.~R.,  2010, \mn@doi [\pasp] {10.1086/656399}, \href
  {http://adsabs.harvard.edu/abs/2010PASP..122.1035A} {122, 1035}

\bibitem[\protect\citeauthoryear{{Anderson} \& {James}}{{Anderson} \&
  {James}}{2008}]{2008MNRAS.390.1527A}
{Anderson} J.~P.,  {James} P.~A.,  2008, \mn@doi [\mnras]
  {10.1111/j.1365-2966.2008.13843.x}, \href
  {http://adsabs.harvard.edu/abs/2008MNRAS.390.1527A} {390, 1527}

\bibitem[\protect\citeauthoryear{{Anderson}, {Covarrubias}, {James}, {Hamuy}
  \& {Habergham}}{{Anderson} et~al.}{2010}]{2010MNRAS.407.2660A}
{Anderson} J.~P.,  {Covarrubias} R.~A.,  {James} P.~A.,  {Hamuy} M.,
  {Habergham} S.~M.,  2010, \mn@doi [\mnras]
  {10.1111/j.1365-2966.2010.17118.x}, \href
  {http://adsabs.harvard.edu/abs/2010MNRAS.407.2660A} {407, 2660}

\bibitem[\protect\citeauthoryear{{Anderson}, {Habergham}, {James}  \&
  {Hamuy}}{{Anderson} et~al.}{2012}]{2012MNRAS.424.1372A}
{Anderson} J.~P.,  {Habergham} S.~M.,  {James} P.~A.,   {Hamuy} M.,  2012,
  \mn@doi [\mnras] {10.1111/j.1365-2966.2012.21324.x}, \href
  {http://adsabs.harvard.edu/abs/2012MNRAS.424.1372A} {424, 1372}

\bibitem[\protect\citeauthoryear{{Anderson}, {James}, {Habergham}, {Galbany}
  \& {Kuncarayakti}}{{Anderson} et~al.}{2015}]{2015PASA...32...19A}
{Anderson} J.~P.,  {James} P.~A.,  {Habergham} S.~M.,  {Galbany} L.,
  {Kuncarayakti} H.,  2015, \mn@doi [\pasa] {10.1017/pasa.2015.19}, \href
  {http://adsabs.harvard.edu/abs/2015PASA...32...19A} {32, 19}

\bibitem[\protect\citeauthoryear{{Arcavi} et~al.,}{{Arcavi}
  et~al.}{2012}]{2012ATel.3881....1A}
{Arcavi} I.,  et~al., 2012, The Astronomer's Telegram, \href
  {http://adsabs.harvard.edu/abs/2012ATel.3881....1A} {3881}

\bibitem[\protect\citeauthoryear{{Arnett}}{{Arnett}}{1982}]{1982ApJ...253..785A}
{Arnett} W.~D.,  1982, \mn@doi [\apj] {10.1086/159681}, \href
  {http://adsabs.harvard.edu/abs/1982ApJ...253..785A} {253, 785}

\bibitem[\protect\citeauthoryear{{Asplund}, {Grevesse}, {Sauval}  \&
  {Scott}}{{Asplund} et~al.}{2009}]{2009ARA&A..47..481A}
{Asplund} M.,  {Grevesse} N.,  {Sauval} A.~J.,   {Scott} P.,  2009, \mn@doi
  [\araa] {10.1146/annurev.astro.46.060407.145222}, \href
  {http://adsabs.harvard.edu/abs/2009ARA%26A..47..481A} {47, 481}

\bibitem[\protect\citeauthoryear{{Badenes}, {Harris}, {Zaritsky}  \&
  {Prieto}}{{Badenes} et~al.}{2009}]{2009ApJ...700..727B}
{Badenes} C.,  {Harris} J.,  {Zaritsky} D.,   {Prieto} J.~L.,  2009, \mn@doi
  [\apj] {10.1088/0004-637X/700/1/727}, \href
  {http://adsabs.harvard.edu/abs/2009ApJ...700..727B} {700, 727}

\bibitem[\protect\citeauthoryear{{Barth}, {van Dyk}, {Filippenko}, {Leibundgut}
   \& {Richmond}}{{Barth} et~al.}{1996}]{1996AJ....111.2047B}
{Barth} A.~J.,  {van Dyk} S.~D.,  {Filippenko} A.~V.,  {Leibundgut} B.,
  {Richmond} M.~W.,  1996, \mn@doi [\aj] {10.1086/117940}, \href
  {http://adsabs.harvard.edu/cgi-bin/nph-bib_query?bibcode=1996AJ....111.2047B&db_key=AST}
  {111, 2047}

\bibitem[\protect\citeauthoryear{{Bastian}, {Gieles}, {Efremov}  \&
  {Lamers}}{{Bastian} et~al.}{2005}]{2005A&A...443...79B}
{Bastian} N.,  {Gieles} M.,  {Efremov} Y.~N.,   {Lamers} H.~J.~G.~L.~M.,  2005,
  \mn@doi [\aap] {10.1051/0004-6361:20053165}, \href
  {http://adsabs.harvard.edu/abs/2005A%26A...443...79B} {443, 79}

\bibitem[\protect\citeauthoryear{{Bersten} et~al.,}{{Bersten}
  et~al.}{2014}]{2014AJ....148...68B}
{Bersten} M.~C.,  et~al., 2014, \mn@doi [\aj] {10.1088/0004-6256/148/4/68},
  \href {http://adsabs.harvard.edu/abs/2014AJ....148...68B} {148, 68}

\bibitem[\protect\citeauthoryear{{Blondin}, {Filippenko}, {Foley}, {Li},
  {Dessart}  \& {Vaz}}{{Blondin} et~al.}{2008}]{2008CBET.1285....1B}
{Blondin} S.,  {Filippenko} A.~V.,  {Foley} R.~J.,  {Li} W.,  {Dessart} L.,
  {Vaz} A.,  2008, Central Bureau Electronic Telegrams, \href
  {http://adsabs.harvard.edu/abs/2008CBET.1285....1B} {1285}

\bibitem[\protect\citeauthoryear{{Boch} \& {Fernique}}{{Boch} \&
  {Fernique}}{2014}]{2014ASPC..485..277B}
{Boch} T.,  {Fernique} P.,  2014, in {Manset} N.,  {Forshay} P.,  eds,
  Astronomical Society of the Pacific Conference Series Vol. 485, Astronomical
  Data Analysis Software and Systems XXIII. p.~277

\bibitem[\protect\citeauthoryear{{Bonnarel} et~al.,}{{Bonnarel}
  et~al.}{2000}]{2000A&AS..143...33B}
{Bonnarel} F.,  et~al., 2000, \mn@doi [\aaps] {10.1051/aas:2000331}, \href
  {http://cdsads.u-strasbg.fr/abs/2000A%26AS..143...33B} {143, 33}

\bibitem[\protect\citeauthoryear{{Bressan}, {Marigo}, {Girardi}, {Salasnich},
  {Dal Cero}, {Rubele}  \& {Nanni}}{{Bressan}
  et~al.}{2012}]{2012MNRAS.427..127B}
{Bressan} A.,  {Marigo} P.,  {Girardi} L.,  {Salasnich} B.,  {Dal Cero} C.,
  {Rubele} S.,   {Nanni} A.,  2012, \mn@doi [\mnras]
  {10.1111/j.1365-2966.2012.21948.x}, \href
  {http://adsabs.harvard.edu/abs/2012MNRAS.427..127B} {427, 127}

\bibitem[\protect\citeauthoryear{{Cano}}{{Cano}}{2013}]{2013MNRAS.434.1098C}
{Cano} Z.,  2013, \mn@doi [\mnras] {10.1093/mnras/stt1048}, \href
  {http://adsabs.harvard.edu/abs/2013MNRAS.434.1098C} {434, 1098}

\bibitem[\protect\citeauthoryear{{Cao} et~al.,}{{Cao}
  et~al.}{2013a}]{2013ApJ...775L...7C}
{Cao} Y.,  et~al., 2013a, \mn@doi [\apjl] {10.1088/2041-8205/775/1/L7}, \href
  {http://adsabs.harvard.edu/abs/2013ApJ...775L...7C} {775, L7}

\bibitem[\protect\citeauthoryear{{Cao}, {Gorbikov}, {Arcavi}, {Ofek},
  {Gal-Yam}, {Nugent}  \& {Kasliwal}}{{Cao}
  et~al.}{2013b}]{2013ATel.5137....1C}
{Cao} Y.,  {Gorbikov} E.,  {Arcavi} I.,  {Ofek} E.,  {Gal-Yam} A.,  {Nugent}
  P.,   {Kasliwal} M.,  2013b, The Astronomer's Telegram, \href
  {http://adsabs.harvard.edu/abs/2013ATel.5137....1C} {5137}

\bibitem[\protect\citeauthoryear{{Cardelli}, {Clayton}  \& {Mathis}}{{Cardelli}
  et~al.}{1989}]{ccm89}
{Cardelli} J.~A.,  {Clayton} G.~C.,   {Mathis} J.~S.,  1989, \apj, \href
  {http://adsabs.harvard.edu/cgi-bin/nph-bib_query?bibcode=1989ApJ...345..245C&amp;db_key=AST}
  {345, 245}

\bibitem[\protect\citeauthoryear{{Carrasco} et~al.,}{{Carrasco}
  et~al.}{2013}]{2013CBET.3565....1C}
{Carrasco} F.,  et~al., 2013, Central Bureau Electronic Telegrams, \href
  {http://adsabs.harvard.edu/abs/2013CBET.3565....1C} {3565}

\bibitem[\protect\citeauthoryear{{Castelli} \& {Kurucz}}{{Castelli} \&
  {Kurucz}}{2004}]{2004astro.ph..5087C}
{Castelli} F.,  {Kurucz} R.~L.,  2004, ArXiv Astrophysics e-prints, \href
  {http://adsabs.harvard.edu/abs/2004astro.ph..5087C} {}

\bibitem[\protect\citeauthoryear{{Chen} et~al.,}{{Chen}
  et~al.}{2014}]{2014ApJ...790..120C}
{Chen} J.,  et~al., 2014, \mn@doi [\apj] {10.1088/0004-637X/790/2/120}, \href
  {http://adsabs.harvard.edu/abs/2014ApJ...790..120C} {790, 120}

\bibitem[\protect\citeauthoryear{{Chen} et~al.,}{{Chen}
  et~al.}{2017}]{2017arXiv170804618C}
{Chen} T.-W.,  et~al., 2017, \mn@doi [\apjl] {10.3847/2041-8213/aa8f40}, \href
  {http://adsabs.harvard.edu/abs/2017ApJ...849L...4C} {849, L4}

\bibitem[\protect\citeauthoryear{{Choi}, {Dotter}, {Conroy}, {Cantiello},
  {Paxton}  \& {Johnson}}{{Choi} et~al.}{2016}]{2016ApJ...823..102C}
{Choi} J.,  {Dotter} A.,  {Conroy} C.,  {Cantiello} M.,  {Paxton} B.,
  {Johnson} B.~D.,  2016, \mn@doi [\apj] {10.3847/0004-637X/823/2/102}, \href
  {http://adsabs.harvard.edu/abs/2016ApJ...823..102C} {823, 102}

\bibitem[\protect\citeauthoryear{{Chornock} \& {Filippenko}}{{Chornock} \&
  {Filippenko}}{2001}]{01biauc3}
{Chornock} R.,  {Filippenko} A.~V.,  2001, \iaucirc, \href
  {http://adsabs.harvard.edu/cgi-bin/nph-bib_query?bibcode=2001IAUC.7577....2C&db_key=AST}
  {7577, 2}

\bibitem[\protect\citeauthoryear{{Chornock}, {Filippenko}, {Li}, {Foley},
  {Stockton}, {Moran}, {Hodge}  \& {Merriman}}{{Chornock}
  et~al.}{2008}]{2008CBET.1298....1C}
{Chornock} R.,  {Filippenko} A.~V.,  {Li} W.,  {Foley} R.~J.,  {Stockton} A.,
  {Moran} E.~C.,  {Hodge} J.,   {Merriman} K.,  2008, Central Bureau Electronic
  Telegrams, \href {http://adsabs.harvard.edu/abs/2008CBET.1298....1C} {1298}

\bibitem[\protect\citeauthoryear{{Chugai}}{{Chugai}}{2000}]{2000AstL...26..797C}
{Chugai} N.~N.,  2000, \mn@doi [Astronomy Letters] {10.1134/1.1331160}, \href
  {http://adsabs.harvard.edu/abs/2000AstL...26..797C} {26, 797}

\bibitem[\protect\citeauthoryear{{Ciabattari} et~al.,}{{Ciabattari}
  et~al.}{2013}]{2013CBET.3557....1C}
{Ciabattari} F.,  et~al., 2013, Central Bureau Electronic Telegrams, \href
  {http://adsabs.harvard.edu/abs/2013CBET.3557....1C} {3557}

\bibitem[\protect\citeauthoryear{{Crockett} et~al.,}{{Crockett}
  et~al.}{2007}]{2007MNRAS.381..835C}
{Crockett} R.~M.,  et~al., 2007, \mn@doi [\mnras]
  {10.1111/j.1365-2966.2007.12283.x}, \href
  {http://adsabs.harvard.edu/abs/2007MNRAS.381..835C} {381, 835}

\bibitem[\protect\citeauthoryear{{Crockett} et~al.,}{{Crockett}
  et~al.}{2008a}]{2008MNRAS.391L...5C}
{Crockett} R.~M.,  et~al., 2008a, \mn@doi [\mnras]
  {10.1111/j.1745-3933.2008.00540.x}, \href
  {http://adsabs.harvard.edu/abs/2008MNRAS.391L...5C} {391, L5}

\bibitem[\protect\citeauthoryear{{Crockett} et~al.,}{{Crockett}
  et~al.}{2008b}]{2008ApJ...672L..99C}
{Crockett} R.~M.,  et~al., 2008b, \mn@doi [\apjl] {10.1086/527299}, \href
  {http://adsabs.harvard.edu/abs/2008ApJ...672L..99C} {672, L99}

\bibitem[\protect\citeauthoryear{{Crowther}}{{Crowther}}{2013}]{2013MNRAS.428.1927C}
{Crowther} P.~A.,  2013, \mn@doi [\mnras] {10.1093/mnras/sts145}, \href
  {http://adsabs.harvard.edu/abs/2013MNRAS.428.1927C} {428, 1927}

\bibitem[\protect\citeauthoryear{{Crowther} et~al.,}{{Crowther}
  et~al.}{2016}]{2016MNRAS.458..624C}
{Crowther} P.~A.,  et~al., 2016, \mn@doi [\mnras] {10.1093/mnras/stw273}, \href
  {http://adsabs.harvard.edu/abs/2016MNRAS.458..624C} {458, 624}

\bibitem[\protect\citeauthoryear{{Desroches}, {Wang}, {Ganeshalingam}  \&
  {Filippenko}}{{Desroches} et~al.}{2007}]{2007CBET.1001....2D}
{Desroches} L.-B.,  {Wang} X.,  {Ganeshalingam} M.,   {Filippenko} A.~V.,
  2007, Central Bureau Electronic Telegrams, \href
  {http://adsabs.harvard.edu/abs/2007CBET.1001....2D} {1001}

\bibitem[\protect\citeauthoryear{{Dessart}, {Hillier}, {Woosley}, {Livne},
  {Waldman}, {Yoon}  \& {Langer}}{{Dessart} et~al.}{2016}]{2016MNRAS.458.1618D}
{Dessart} L.,  {Hillier} D.~J.,  {Woosley} S.,  {Livne} E.,  {Waldman} R.,
  {Yoon} S.-C.,   {Langer} N.,  2016, \mn@doi [\mnras] {10.1093/mnras/stw418},
  \href {http://adsabs.harvard.edu/abs/2016MNRAS.458.1618D} {458, 1618}

\bibitem[\protect\citeauthoryear{{Dimai}, {Briganti}  \& {Brimacombe}}{{Dimai}
  et~al.}{2012}]{2012CBET.2993....1D}
{Dimai} A.,  {Briganti} F.,   {Brimacombe} J.,  2012, Central Bureau Electronic
  Telegrams, \href {http://adsabs.harvard.edu/abs/2012CBET.2993....1D} {2993,
  1}

\bibitem[\protect\citeauthoryear{{Dolphin}}{{Dolphin}}{2000}]{dolphhstphot}
{Dolphin} A.~E.,  2000, \pasp, \href
  {http://adsabs.harvard.edu/cgi-bin/nph-bib_query?bibcode=2000PASP..112.1383D&db_key=AST}
  {112, 1383}

\bibitem[\protect\citeauthoryear{{Dolphin}}{{Dolphin}}{2002}]{2002MNRAS.332...91D}
{Dolphin} A.~E.,  2002, \mn@doi [\mnras] {10.1046/j.1365-8711.2002.05271.x},
  \href {http://adsabs.harvard.edu/abs/2002MNRAS.332...91D} {332, 91}

\bibitem[\protect\citeauthoryear{{Drout} et~al.,}{{Drout}
  et~al.}{2011}]{2011ApJ...741...97D}
{Drout} M.~R.,  et~al., 2011, \mn@doi [\apj] {10.1088/0004-637X/741/2/97},
  \href {http://adsabs.harvard.edu/abs/2011ApJ...741...97D} {741, 97}

\bibitem[\protect\citeauthoryear{{Eisenstein} et~al.,}{{Eisenstein}
  et~al.}{2011}]{2011AJ....142...72E}
{Eisenstein} D.~J.,  et~al., 2011, \mn@doi [\aj] {10.1088/0004-6256/142/3/72},
  \href {http://adsabs.harvard.edu/abs/2011AJ....142...72E} {142, 72}

\bibitem[\protect\citeauthoryear{{Eldridge} \& {Maund}}{{Eldridge} \&
  {Maund}}{2016}]{2016MNRAS.461L.117E}
{Eldridge} J.~J.,  {Maund} J.~R.,  2016, \mn@doi [\mnras]
  {10.1093/mnrasl/slw099}, \href
  {http://adsabs.harvard.edu/abs/2016MNRAS.461L.117E} {461, L117}

\bibitem[\protect\citeauthoryear{{Eldridge} \& {Tout}}{{Eldridge} \&
  {Tout}}{2004}]{eld04}
{Eldridge} J.~J.,  {Tout} C.~A.,  2004, \mnras, \href
  {http://adsabs.harvard.edu/cgi-bin/nph-bib_query?bibcode=2004MNRAS.353...87E&db_key=AST}
  {353, 87}

\bibitem[\protect\citeauthoryear{{Eldridge}, {Fraser}, {Smartt}, {Maund}  \&
  {Crockett}}{{Eldridge} et~al.}{2013}]{2013MNRAS.436..774E}
{Eldridge} J.~J.,  {Fraser} M.,  {Smartt} S.~J.,  {Maund} J.~R.,   {Crockett}
  R.~M.,  2013, \mn@doi [\mnras] {10.1093/mnras/stt1612}, \href
  {http://adsabs.harvard.edu/abs/2013MNRAS.436..774E} {436, 774}

\bibitem[\protect\citeauthoryear{{Eldridge}, {Fraser}, {Maund}  \&
  {Smartt}}{{Eldridge} et~al.}{2015}]{2015MNRAS.446.2689E}
{Eldridge} J.~J.,  {Fraser} M.,  {Maund} J.~R.,   {Smartt} S.~J.,  2015,
  \mn@doi [\mnras] {10.1093/mnras/stu2197}, \href
  {http://adsabs.harvard.edu/abs/2015MNRAS.446.2689E} {446, 2689}

\bibitem[\protect\citeauthoryear{{Eldridge} et~al.,}{{Eldridge}
  et~al.}{2017}]{2017arXiv171002154E}
{Eldridge} J.~J.,  et~al., 2017, preprint, \href
  {http://adsabs.harvard.edu/abs/2017arXiv171002154E} {} (\mn@eprint {arXiv}
  {1710.02154})

\bibitem[\protect\citeauthoryear{{Elias-Rosa} et~al.,}{{Elias-Rosa}
  et~al.}{2013}]{2013MNRAS.436L.109E}
{Elias-Rosa} N.,  et~al., 2013, \mn@doi [\mnras] {10.1093/mnrasl/slt124}, \href
  {http://adsabs.harvard.edu/abs/2013MNRAS.436L.109E} {436, L109}

\bibitem[\protect\citeauthoryear{{Ensman} \& {Woosley}}{{Ensman} \&
  {Woosley}}{1988}]{1988ApJ...333..754E}
{Ensman} L.~M.,  {Woosley} S.~E.,  1988, \mn@doi [\apj] {10.1086/166785}, \href
  {http://adsabs.harvard.edu/abs/1988ApJ...333..754E} {333, 754}

\bibitem[\protect\citeauthoryear{{Feroz}, {Hobson}, {Cameron}  \&
  {Pettitt}}{{Feroz} et~al.}{2013}]{2013arXiv1306.2144F}
{Feroz} F.,  {Hobson} M.~P.,  {Cameron} E.,   {Pettitt} A.~N.,  2013, preprint,
  \href {http://adsabs.harvard.edu/abs/2013arXiv1306.2144F} {} (\mn@eprint
  {arXiv} {1306.2144})

\bibitem[\protect\citeauthoryear{{Filippenko} \& {Foley}}{{Filippenko} \&
  {Foley}}{2004}]{2004IAUC.8452....3F}
{Filippenko} A.~V.,  {Foley} R.~J.,  2004, \iaucirc, \href
  {http://adsabs.harvard.edu/abs/2004IAUC.8452....3F} {8452}

\bibitem[\protect\citeauthoryear{{Filippenko}, {Chornock}  \&
  {Modjaz}}{{Filippenko} et~al.}{2000}]{00ewiauc3}
{Filippenko} A.~V.,  {Chornock} R.,   {Modjaz} M.,  2000, \iaucirc, \href
  {http://adsabs.harvard.edu/cgi-bin/nph-bib_query?bibcode=2000IAUC.7547....2F&amp;db_key=AST}
  {7547, 2}

\bibitem[\protect\citeauthoryear{{Fiorentino}, {Musella}  \&
  {Marconi}}{{Fiorentino} et~al.}{2013}]{2013MNRAS.434.2866F}
{Fiorentino} G.,  {Musella} I.,   {Marconi} M.,  2013, \mn@doi [\mnras]
  {10.1093/mnras/stt1193}, \href
  {http://adsabs.harvard.edu/abs/2013MNRAS.434.2866F} {434, 2866}

\bibitem[\protect\citeauthoryear{{Folatelli} et~al.,}{{Folatelli}
  et~al.}{2014}]{2014ApJ...793L..22F}
{Folatelli} G.,  et~al., 2014, \mn@doi [\apjl] {10.1088/2041-8205/793/2/L22},
  \href {http://adsabs.harvard.edu/abs/2014ApJ...793L..22F} {793, L22}

\bibitem[\protect\citeauthoryear{{Folatelli}, {Bersten}, {Kuncarayakti},
  {Benvenuto}, {Maeda}  \& {Nomoto}}{{Folatelli}
  et~al.}{2015}]{2015ApJ...811..147F}
{Folatelli} G.,  {Bersten} M.~C.,  {Kuncarayakti} H.,  {Benvenuto} O.~G.,
  {Maeda} K.,   {Nomoto} K.,  2015, \mn@doi [\apj]
  {10.1088/0004-637X/811/2/147}, \href
  {http://adsabs.harvard.edu/abs/2015ApJ...811..147F} {811, 147}

\bibitem[\protect\citeauthoryear{{Folatelli} et~al.,}{{Folatelli}
  et~al.}{2016}]{2016ApJ...825L..22F}
{Folatelli} G.,  et~al., 2016, \mn@doi [\apjl] {10.3847/2041-8205/825/2/L22},
  \href {http://adsabs.harvard.edu/abs/2016ApJ...825L..22F} {825, L22}

\bibitem[\protect\citeauthoryear{{Fox} et~al.,}{{Fox}
  et~al.}{2014}]{2014ApJ...790...17F}
{Fox} O.~D.,  et~al., 2014, \mn@doi [\apj] {10.1088/0004-637X/790/1/17}, \href
  {http://adsabs.harvard.edu/abs/2014ApJ...790...17F} {790, 17}

\bibitem[\protect\citeauthoryear{{Freedman} et~al.,}{{Freedman}
  et~al.}{2001}]{2001ApJ...553...47F}
{Freedman} W.~L.,  et~al., 2001, \mn@doi [\apj] {10.1086/320638}, \href
  {http://adsabs.harvard.edu/abs/2001ApJ...553...47F} {553, 47}

\bibitem[\protect\citeauthoryear{{Fremling} et~al.,}{{Fremling}
  et~al.}{2016}]{2016A&A...593A..68F}
{Fremling} C.,  et~al., 2016, \mn@doi [\aap] {10.1051/0004-6361/201628275},
  \href {http://adsabs.harvard.edu/abs/2016A%26A...593A..68F} {593, A68}

\bibitem[\protect\citeauthoryear{{Gal-Yam}, {Shemmer}  \& {Dann}}{{Gal-Yam}
  et~al.}{2002}]{2002IAUC.7811....3G}
{Gal-Yam} A.,  {Shemmer} O.,   {Dann} J.,  2002, \iaucirc, \href
  {http://adsabs.harvard.edu/abs/2002IAUC.7811....3G} {7811}

\bibitem[\protect\citeauthoryear{{Gal-Yam} et~al.,}{{Gal-Yam}
  et~al.}{2005}]{2005ApJ...630L..29G}
{Gal-Yam} A.,  et~al., 2005, \mn@doi [\apjl] {10.1086/491622}, \href
  {http://adsabs.harvard.edu/abs/2005ApJ...630L..29G} {630, L29}

\bibitem[\protect\citeauthoryear{{Ganeshalingam}, {Swift}, {Serduke}  \&
  {Filippenko}}{{Ganeshalingam} et~al.}{2004}]{2004IAUC.8456....4G}
{Ganeshalingam} M.,  {Swift} B.~J.,  {Serduke} F.~J.~D.,   {Filippenko} A.~V.,
  2004, \iaucirc, \href
  {http://adsabs.harvard.edu/cgi-bin/nph-bib_query?bibcode=2004IAUC.8456....4G&db_key=AST}
  {8456, 4}

\bibitem[\protect\citeauthoryear{{Garcia}}{{Garcia}}{1993}]{1993IAUC.5731....1R}
{Garcia} F.,  1993, \iaucirc, \href
  {http://adsabs.harvard.edu/cgi-bin/nph-bib_query?bibcode=1993IAUC.5731....1R&db_key=AST}
  {5731, 1}

\bibitem[\protect\citeauthoryear{{Georgy}}{{Georgy}}{2012}]{2012A&A...538L...8G}
{Georgy} C.,  2012, \mn@doi [\aap] {10.1051/0004-6361/201118372}, \href
  {http://adsabs.harvard.edu/abs/2012A%26A...538L...8G} {538, L8}

\bibitem[\protect\citeauthoryear{{Gibson} \& {Stetson}}{{Gibson} \&
  {Stetson}}{2001}]{2001ApJ...547L.103G}
{Gibson} B.~K.,  {Stetson} P.~B.,  2001, \mn@doi [\apjl] {10.1086/318905},
  \href {http://adsabs.harvard.edu/abs/2001ApJ...547L.103G} {547, L103}

\bibitem[\protect\citeauthoryear{{Girardi}, {Bertelli}, {Bressan}, {Chiosi},
  {Groenewegen}, {Marigo}, {Salasnich}  \& {Weiss}}{{Girardi}
  et~al.}{2002}]{2002A&A...391..195G}
{Girardi} L.,  {Bertelli} G.,  {Bressan} A.,  {Chiosi} C.,  {Groenewegen}
  M.~A.~T.,  {Marigo} P.,  {Salasnich} B.,   {Weiss} A.,  2002, \mn@doi [\aap]
  {10.1051/0004-6361:20020612}, \href
  {http://adsabs.harvard.edu/abs/2002A%26A...391..195G} {391, 195}

\bibitem[\protect\citeauthoryear{{Gogarten}, {Dalcanton}, {Murphy}, {Williams},
  {Gilbert}  \& {Dolphin}}{{Gogarten} et~al.}{2009}]{2009ApJ...703..300G}
{Gogarten} S.~M.,  {Dalcanton} J.~J.,  {Murphy} J.~W.,  {Williams} B.~F.,
  {Gilbert} K.,   {Dolphin} A.,  2009, \mn@doi [\apj]
  {10.1088/0004-637X/703/1/300}, \href
  {http://adsabs.harvard.edu/abs/2009ApJ...703..300G} {703, 300}

\bibitem[\protect\citeauthoryear{{Granata} et~al.,}{{Granata}
  et~al.}{2016}]{2016ATel.8818....1G}
{Granata} V.,  et~al., 2016, The Astronomer's Telegram, \href
  {http://adsabs.harvard.edu/abs/2016ATel.8818....1G} {8818}

\bibitem[\protect\citeauthoryear{{Groh}, {Georgy}  \& {Ekstr{\"o}m}}{{Groh}
  et~al.}{2013a}]{2013A&A...558L...1G}
{Groh} J.~H.,  {Georgy} C.,   {Ekstr{\"o}m} S.,  2013a, \mn@doi [\aap]
  {10.1051/0004-6361/201322369}, \href
  {http://adsabs.harvard.edu/abs/2013A%26A...558L...1G} {558, L1}

\bibitem[\protect\citeauthoryear{{Groh}, {Meynet}, {Georgy}  \&
  {Ekstr{\"o}m}}{{Groh} et~al.}{2013b}]{2013A&A...558A.131G}
{Groh} J.~H.,  {Meynet} G.,  {Georgy} C.,   {Ekstr{\"o}m} S.,  2013b, \mn@doi
  [\aap] {10.1051/0004-6361/201321906}, \href
  {http://adsabs.harvard.edu/abs/2013A%26A...558A.131G} {558, A131}

\bibitem[\protect\citeauthoryear{{Habergham}, {Anderson}  \&
  {James}}{{Habergham} et~al.}{2010}]{2010ApJ...717..342H}
{Habergham} S.~M.,  {Anderson} J.~P.,   {James} P.~A.,  2010, \mn@doi [\apj]
  {10.1088/0004-637X/717/1/342}, \href
  {http://adsabs.harvard.edu/abs/2010ApJ...717..342H} {717, 342}

\bibitem[\protect\citeauthoryear{{Harutyunyan}, {Benetti}, {Pastorello},
  {Cappellaro}, {Tomasella}, {Ochner}  \& {Turatto}}{{Harutyunyan}
  et~al.}{2013}]{2013ATel.5165....1H}
{Harutyunyan} A.,  {Benetti} S.,  {Pastorello} A.,  {Cappellaro} E.,
  {Tomasella} L.,  {Ochner} P.,   {Turatto} M.,  2013, The Astronomer's
  Telegram, \href {http://adsabs.harvard.edu/abs/2013ATel.5165....1H} {5165}

\bibitem[\protect\citeauthoryear{{Hendry}, {Smartt}, {Maund}, {Pastorello},
  {Zampieri}, {Benetti}, {Turatto}  \& {et~al.}}{{Hendry}
  et~al.}{2005}]{2005MNRAS.359..906H}
{Hendry} M.~A.,  {Smartt} S.~J.,  {Maund} J.~R.,  {Pastorello} A.,  {Zampieri}
  L.,  {Benetti} S.,  {Turatto} M.,   {et~al.} 2005, \mnras, \href
  {http://adsabs.harvard.edu/cgi-bin/nph-bib_query?bibcode=2005MNRAS.359..906H&db_key=AST}
  {359, 906}

\bibitem[\protect\citeauthoryear{{Howerton} et~al.,}{{Howerton}
  et~al.}{2012}]{2012ATel.3967....1H}
{Howerton} S.,  et~al., 2012, The Astronomer's Telegram, \href
  {http://adsabs.harvard.edu/abs/2012ATel.3967....1H} {3967}

\bibitem[\protect\citeauthoryear{{Hunter} et~al.,}{{Hunter}
  et~al.}{2007}]{2007A&A...466..277H}
{Hunter} I.,  et~al., 2007, \mn@doi [\aap] {10.1051/0004-6361:20066148}, \href
  {http://adsabs.harvard.edu/abs/2007A%26A...466..277H} {466, 277}

\bibitem[\protect\citeauthoryear{{Izzard}, {Ramirez-Ruiz}  \& {Tout}}{{Izzard}
  et~al.}{2004}]{izzgrb}
{Izzard} R.~G.,  {Ramirez-Ruiz} E.,   {Tout} C.~A.,  2004, \mnras, \href
  {http://adsabs.harvard.edu/cgi-bin/nph-bib_query?bibcode=2004MNRAS.348.1215I&db_key=AST}
  {348, 1215}

\bibitem[\protect\citeauthoryear{{Jacobs}, {Rizzi}, {Tully}, {Shaya}, {Makarov}
   \& {Makarova}}{{Jacobs} et~al.}{2009}]{2009AJ....138..332J}
{Jacobs} B.~A.,  {Rizzi} L.,  {Tully} R.~B.,  {Shaya} E.~J.,  {Makarov} D.~I.,
   {Makarova} L.,  2009, \mn@doi [\aj] {10.1088/0004-6256/138/2/332}, \href
  {http://adsabs.harvard.edu/abs/2009AJ....138..332J} {138, 332}

\bibitem[\protect\citeauthoryear{Jeffreys}{Jeffreys}{1961}]{Jeffreys61}
Jeffreys H.,  1961, Theory of Probability, third edn.
Oxford, Oxford, England

\bibitem[\protect\citeauthoryear{{Jennings}, {Williams}, {Murphy}, {Dalcanton},
  {Gilbert}, {Dolphin}, {Fouesneau}  \& {Weisz}}{{Jennings}
  et~al.}{2012}]{2012ApJ...761...26J}
{Jennings} Z.~G.,  {Williams} B.~F.,  {Murphy} J.~W.,  {Dalcanton} J.~J.,
  {Gilbert} K.~M.,  {Dolphin} A.~E.,  {Fouesneau} M.,   {Weisz} D.~R.,  2012,
  \mn@doi [\apj] {10.1088/0004-637X/761/1/26}, \href
  {http://adsabs.harvard.edu/abs/2012ApJ...761...26J} {761, 26}

\bibitem[\protect\citeauthoryear{{Jennings}, {Williams}, {Murphy}, {Dalcanton},
  {Gilbert}, {Dolphin}, {Weisz}  \& {Fouesneau}}{{Jennings}
  et~al.}{2014}]{2014ApJ...795..170J}
{Jennings} Z.~G.,  {Williams} B.~F.,  {Murphy} J.~W.,  {Dalcanton} J.~J.,
  {Gilbert} K.~M.,  {Dolphin} A.~E.,  {Weisz} D.~R.,   {Fouesneau} M.,  2014,
  \mn@doi [\apj] {10.1088/0004-637X/795/2/170}, \href
  {http://adsabs.harvard.edu/abs/2014ApJ...795..170J} {795, 170}

\bibitem[\protect\citeauthoryear{{Jerkstrand}, {Ergon}, {Smartt}, {Fransson},
  {Sollerman}, {Taubenberger}, {Bersten}  \& {Spyromilio}}{{Jerkstrand}
  et~al.}{2015}]{2015A&A...573A..12J}
{Jerkstrand} A.,  {Ergon} M.,  {Smartt} S.~J.,  {Fransson} C.,  {Sollerman} J.,
   {Taubenberger} S.,  {Bersten} M.,   {Spyromilio} J.,  2015, \mn@doi [\aap]
  {10.1051/0004-6361/201423983}, \href
  {http://adsabs.harvard.edu/abs/2015A%26A...573A..12J} {573, A12}

\bibitem[\protect\citeauthoryear{{Johnson}, {Kochanek}  \& {Adams}}{{Johnson}
  et~al.}{2017}]{2017arXiv170703828}
{Johnson} S.~A.,  {Kochanek} C.~S.,   {Adams} S.~M.,  2017, \mn@doi [\mnras]
  {10.1093/mnras/stx2170}, \href
  {http://adsabs.harvard.edu/abs/2017MNRAS.472.3115J} {472, 3115}

\bibitem[\protect\citeauthoryear{{Kanbur}, {Ngeow}, {Nikolaev}, {Tanvir}  \&
  {Hendry}}{{Kanbur} et~al.}{2003}]{2003A&A...411..361K}
{Kanbur} S.~M.,  {Ngeow} C.,  {Nikolaev} S.,  {Tanvir} N.~R.,   {Hendry} M.~A.,
   2003, \mn@doi [\aap] {10.1051/0004-6361:20031373}, \href
  {http://adsabs.harvard.edu/abs/2003A%26A...411..361K} {411, 361}

\bibitem[\protect\citeauthoryear{{Kangas}, {Mattila}, {Kankare}, {Kotilainen},
  {V{\"a}is{\"a}nen}, {Greimel}  \& {Takalo}}{{Kangas}
  et~al.}{2013}]{2013MNRAS.436.3464K}
{Kangas} T.,  {Mattila} S.,  {Kankare} E.,  {Kotilainen} J.~K.,
  {V{\"a}is{\"a}nen} P.,  {Greimel} R.,   {Takalo} A.,  2013, \mn@doi [\mnras]
  {10.1093/mnras/stt1833}, \href
  {http://adsabs.harvard.edu/abs/2013MNRAS.436.3464K} {436, 3464}

\bibitem[\protect\citeauthoryear{{Kangas} et~al.,}{{Kangas}
  et~al.}{2017}]{2017A&A...597A..92K}
{Kangas} T.,  et~al., 2017, \mn@doi [\aap] {10.1051/0004-6361/201628705}, \href
  {http://adsabs.harvard.edu/abs/2017A%26A...597A..92K} {597, A92}

\bibitem[\protect\citeauthoryear{{Kankare} et~al.,}{{Kankare}
  et~al.}{2014}]{2014A&A...572A..75K}
{Kankare} E.,  et~al., 2014, \mn@doi [\aap] {10.1051/0004-6361/201424563},
  \href {http://adsabs.harvard.edu/abs/2014A%26A...572A..75K} {572, A75}

\bibitem[\protect\citeauthoryear{{Kasliwal}, {Howell}, {Fox}, {Quimby}  \&
  {Gal-Yam}}{{Kasliwal} et~al.}{2009}]{2009CBET.1955....1K}
{Kasliwal} M.~M.,  {Howell} J.~L.,  {Fox} D.~B.,  {Quimby} R.,   {Gal-Yam} A.,
  2009, Central Bureau Electronic Telegrams, \href
  {http://adsabs.harvard.edu/abs/2009CBET.1955....1K} {1955}

\bibitem[\protect\citeauthoryear{{Khandrika} \& {Li}}{{Khandrika} \&
  {Li}}{2007}]{2007CBET..997....1K}
{Khandrika} H.,  {Li} W.,  2007, Central Bureau Electronic Telegrams, \href
  {http://adsabs.harvard.edu/abs/2007CBET..997....1K} {997}

\bibitem[\protect\citeauthoryear{{Kinugasa}, {Kawakita}, {Ayani}, {Kawabata}
  \& {Yamaoka}}{{Kinugasa} et~al.}{2002}]{2002IAUC.7811....1K}
{Kinugasa} K.,  {Kawakita} H.,  {Ayani} K.,  {Kawabata} T.,   {Yamaoka} H.,
  2002, \iaucirc, \href {http://adsabs.harvard.edu/abs/2002IAUC.7811....1K}
  {7811}

\bibitem[\protect\citeauthoryear{{Kuncarayakti} et~al.,}{{Kuncarayakti}
  et~al.}{2013a}]{2013AJ....146...30K}
{Kuncarayakti} H.,  et~al., 2013a, \mn@doi [\aj] {10.1088/0004-6256/146/2/30},
  \href {http://adsabs.harvard.edu/abs/2013AJ....146...30K} {146, 30}

\bibitem[\protect\citeauthoryear{{Kuncarayakti} et~al.,}{{Kuncarayakti}
  et~al.}{2013b}]{2013AJ....146...31K}
{Kuncarayakti} H.,  et~al., 2013b, \mn@doi [\aj] {10.1088/0004-6256/146/2/31},
  \href {http://adsabs.harvard.edu/abs/2013AJ....146...31K} {146, 31}

\bibitem[\protect\citeauthoryear{{Langer}}{{Langer}}{2012}]{2012ARA&A..50..107L}
{Langer} N.,  2012, \mn@doi [\araa] {10.1146/annurev-astro-081811-125534},
  \href {http://adsabs.harvard.edu/abs/2012ARA%26A..50..107L} {50, 107}

\bibitem[\protect\citeauthoryear{{Leitherer} et~al.,}{{Leitherer}
  et~al.}{1999}]{1999ApJS..123....3L}
{Leitherer} C.,  et~al., 1999, \mn@doi [\apjs] {10.1086/313233}, \href
  {http://adsabs.harvard.edu/abs/1999ApJS..123....3L} {123, 3}

\bibitem[\protect\citeauthoryear{{Leloudas}, {Sollerman}, {Levan}, {Fynbo},
  {Malesani}  \& {Maund}}{{Leloudas} et~al.}{2010}]{2010A&A...518A..29L}
{Leloudas} G.,  {Sollerman} J.,  {Levan} A.~J.,  {Fynbo} J.~P.~U.,  {Malesani}
  D.,   {Maund} J.~R.,  2010, \mn@doi [\aap] {10.1051/0004-6361/200913753},
  \href {http://adsabs.harvard.edu/abs/2010A%26A...518A..29L} {518, A29+}

\bibitem[\protect\citeauthoryear{{Li}, {Yamaoka}  \& {Itagaki}}{{Li}
  et~al.}{2004}]{2004CBET..100....1L}
{Li} W.,  {Yamaoka} H.,   {Itagaki} K.,  2004, Central Bureau Electronic
  Telegrams, \href {http://adsabs.harvard.edu/abs/2004CBET..100....1L} {100}

\bibitem[\protect\citeauthoryear{{Li}, {Cenko}  \& {Filippenko}}{{Li}
  et~al.}{2009}]{2009CBET.1952....1L}
{Li} W.,  {Cenko} S.~B.,   {Filippenko} A.~V.,  2009, Central Bureau Electronic
  Telegrams, \href {http://adsabs.harvard.edu/abs/2009CBET.1952....1L} {1952}

\bibitem[\protect\citeauthoryear{{Lyman}, {Bersier}, {James}, {Mazzali},
  {Eldridge}, {Fraser}  \& {Pian}}{{Lyman} et~al.}{2016}]{2016MNRAS.457..328L}
{Lyman} J.~D.,  {Bersier} D.,  {James} P.~A.,  {Mazzali} P.~A.,  {Eldridge}
  J.~J.,  {Fraser} M.,   {Pian} E.,  2016, \mn@doi [\mnras]
  {10.1093/mnras/stv2983}, \href
  {http://adsabs.harvard.edu/abs/2016MNRAS.457..328L} {457, 328}

\bibitem[\protect\citeauthoryear{{Maeda}, {Mazzali}, {Deng}, {Nomoto},
  {Yoshii}, {Tomita}  \& {Kobayashi}}{{Maeda}
  et~al.}{2003}]{2003ApJ...593..931M}
{Maeda} K.,  {Mazzali} P.~A.,  {Deng} J.,  {Nomoto} K.,  {Yoshii} Y.,  {Tomita}
  H.,   {Kobayashi} Y.,  2003, \mn@doi [\apj] {10.1086/376591}, \href
  {http://adsabs.harvard.edu/abs/2003ApJ...593..931M} {593, 931}

\bibitem[\protect\citeauthoryear{{Makarov}, {Prugniel}, {Terekhova}, {Courtois}
   \& {Vauglin}}{{Makarov} et~al.}{2014}]{2014A&A...570A..13M}
{Makarov} D.,  {Prugniel} P.,  {Terekhova} N.,  {Courtois} H.,   {Vauglin} I.,
  2014, \mn@doi [\aap] {10.1051/0004-6361/201423496}, \href
  {http://adsabs.harvard.edu/abs/2014A%26A...570A..13M} {570, A13}

\bibitem[\protect\citeauthoryear{{Martin}, {Yamaoka}, {Monard}  \&
  {Africa}}{{Martin} et~al.}{2005}]{2005CBET..119....1M}
{Martin} R.,  {Yamaoka} H.,  {Monard} L.~A.~G.,   {Africa} S.,  2005, Central
  Bureau Electronic Telegrams, \href
  {http://adsabs.harvard.edu/abs/2005CBET..119....1M} {119}

\bibitem[\protect\citeauthoryear{{Massey}}{{Massey}}{2002}]{2002ApJS..141...81M}
{Massey} P.,  2002, \mn@doi [\apjs] {10.1086/338286}, \href
  {http://adsabs.harvard.edu/abs/2002ApJS..141...81M} {141, 81}

\bibitem[\protect\citeauthoryear{{Matheson}, {Jha}, {Challis}, {Kirshner}  \&
  {Berlind}}{{Matheson} et~al.}{2001}]{2001IAUC.7765....2M}
{Matheson} T.,  {Jha} S.,  {Challis} P.,  {Kirshner} R.,   {Berlind} P.,  2001,
  \iaucirc, \href {http://adsabs.harvard.edu/abs/2001IAUC.7765....2M} {7765}

\bibitem[\protect\citeauthoryear{{Maund}}{{Maund}}{2005}]{mythesis}
{Maund} J.~R.,  2005, PhD thesis, Institute of Astronomy, Cambridge, \url
  {http://www.dark-cosmology.dk/~justyn/thesis/}

\bibitem[\protect\citeauthoryear{{Maund}}{{Maund}}{2017}]{2017arXiv170401957M}
{Maund} J.~R.,  2017, preprint, \href
  {http://adsabs.harvard.edu/abs/2017arXiv170401957M} {} (\mn@eprint {arXiv}
  {1704.01957})

\bibitem[\protect\citeauthoryear{{Maund} \& {Ramirez-Ruiz}}{{Maund} \&
  {Ramirez-Ruiz}}{2016}]{2016MNRAS.456.3175M}
{Maund} J.~R.,  {Ramirez-Ruiz} E.,  2016, \mn@doi [\mnras]
  {10.1093/mnras/stv2760}, \href
  {http://adsabs.harvard.edu/abs/2016MNRAS.456.3175M} {456, 3175}

\bibitem[\protect\citeauthoryear{{Maund} \& {Smartt}}{{Maund} \&
  {Smartt}}{2005}]{2005astro.ph..1323M}
{Maund} J.~R.,  {Smartt} S.~J.,  2005, \mnras, \href
  {http://adsabs.harvard.edu/cgi-bin/nph-bib_query?bibcode=2005MNRAS.360..288M&db_key=AST}
  {360, 288}

\bibitem[\protect\citeauthoryear{{Maund}, {Smartt}, {Kudritzki},
  {Podsiadlowski}  \& {Gilmore}}{{Maund} et~al.}{2004}]{maund93j}
{Maund} J.~R.,  {Smartt} S.~J.,  {Kudritzki} R.~P.,  {Podsiadlowski} P.,
  {Gilmore} G.~F.,  2004, \nat, \href
  {http://adsabs.harvard.edu/cgi-bin/nph-bib_query?bibcode=2004Natur.427..129M&amp;db_key=AST}
  {427, 129}

\bibitem[\protect\citeauthoryear{{Maund}, {Smartt}  \& {Schweizer}}{{Maund}
  et~al.}{2005}]{2005ApJ...630L..33M}
{Maund} J.~R.,  {Smartt} S.~J.,   {Schweizer} F.,  2005, \mn@doi [\apjl]
  {10.1086/491620}, \href {http://adsabs.harvard.edu/abs/2005ApJ...630L..33M}
  {630, L33}

\bibitem[\protect\citeauthoryear{{Maund}, {Wheeler}, {Baade}, {Patat},
  {H{\"o}flich}, {Wang}  \& {Clocchiatti}}{{Maund}
  et~al.}{2009}]{2009ApJ...705.1139M}
{Maund} J.~R.,  {Wheeler} J.~C.,  {Baade} D.,  {Patat} F.,  {H{\"o}flich} P.,
  {Wang} L.,   {Clocchiatti} A.,  2009, \mn@doi [\apj]
  {10.1088/0004-637X/705/2/1139}, \href
  {http://adsabs.harvard.edu/abs/2009ApJ...705.1139M} {705, 1139}

\bibitem[\protect\citeauthoryear{{Maund} et~al.,}{{Maund}
  et~al.}{2011}]{2011ApJ...739L..37M}
{Maund} J.~R.,  et~al., 2011, \mn@doi [\apjl] {10.1088/2041-8205/739/2/L37},
  \href {http://adsabs.harvard.edu/abs/2011ApJ...739L..37M} {739, L37}

\bibitem[\protect\citeauthoryear{{Maund} et~al.,}{{Maund}
  et~al.}{2015}]{2015MNRAS.454.2580M}
{Maund} J.~R.,  et~al., 2015, \mn@doi [\mnras] {10.1093/mnras/stv2098}, \href
  {http://adsabs.harvard.edu/abs/2015MNRAS.454.2580M} {454, 2580}

\bibitem[\protect\citeauthoryear{{Maund}, {Pastorello}, {Mattila}, {Itagaki}
  \& {Boles}}{{Maund} et~al.}{2016}]{2016ApJ...833..128M}
{Maund} J.~R.,  {Pastorello} A.,  {Mattila} S.,  {Itagaki} K.,   {Boles} T.,
  2016, \mn@doi [\apj] {10.3847/1538-4357/833/2/128}, \href
  {http://adsabs.harvard.edu/abs/2016ApJ...833..128M} {833, 128}

\bibitem[\protect\citeauthoryear{{Mazzali}, {Maurer}, {Valenti}, {Kotak}  \&
  {Hunter}}{{Mazzali} et~al.}{2010}]{2010MNRAS.408...87M}
{Mazzali} P.~A.,  {Maurer} I.,  {Valenti} S.,  {Kotak} R.,   {Hunter} D.,
  2010, \mn@doi [\mnras] {10.1111/j.1365-2966.2010.17133.x}, \href
  {http://adsabs.harvard.edu/abs/2010MNRAS.408...87M} {408, 87}

\bibitem[\protect\citeauthoryear{{Meikle}, {Lucy}, {Smartt}, {Leibundgut},
  {Lundqvist}  \& {Ostensen}}{{Meikle} et~al.}{2002}]{2002IAUC.7811....2M}
{Meikle} P.,  {Lucy} L.,  {Smartt} S.,  {Leibundgut} B.,  {Lundqvist} P.,
  {Ostensen} R.,  2002, \iaucirc, \href
  {http://adsabs.harvard.edu/abs/2002IAUC.7811....2M} {7811}

\bibitem[\protect\citeauthoryear{{Milisavljevic} et~al.,}{{Milisavljevic}
  et~al.}{2013a}]{2013ApJ...770L..38M}
{Milisavljevic} D.,  et~al., 2013a, \mn@doi [\apjl]
  {10.1088/2041-8205/770/2/L38}, \href
  {http://adsabs.harvard.edu/abs/2013ApJ...770L..38M} {770, L38}

\bibitem[\protect\citeauthoryear{{Milisavljevic} et~al.,}{{Milisavljevic}
  et~al.}{2013b}]{2013ATel.5142....1M}
{Milisavljevic} D.,  et~al., 2013b, The Astronomer's Telegram, \href
  {http://adsabs.harvard.edu/abs/2013ATel.5142....1M} {5142}

\bibitem[\protect\citeauthoryear{{Modjaz}, {Kewley}, {Bloom}, {Filippenko},
  {Perley}  \& {Silverman}}{{Modjaz} et~al.}{2011}]{2011ApJ...731L...4M}
{Modjaz} M.,  {Kewley} L.,  {Bloom} J.~S.,  {Filippenko} A.~V.,  {Perley} D.,
  {Silverman} J.~M.,  2011, \mn@doi [\apjl] {10.1088/2041-8205/731/1/L4}, \href
  {http://adsabs.harvard.edu/abs/2011ApJ...731L...4M} {731, L4}

\bibitem[\protect\citeauthoryear{{Monard}, {Quimby}, {Gerardy}, {H\"{o}flich},
  {Wheeler}, {Chen}, {Smith}  \& {Bauer}}{{Monard}
  et~al.}{2004}]{2004IAUC.8454....1M}
{Monard} L.~A.~G.,  {Quimby} R.,  {Gerardy} C.,  {H\"{o}flich} P.,  {Wheeler}
  J.~C.,  {Chen} Y.-T.,  {Smith} H.~J.,   {Bauer} A.,  2004, \iaucirc, \href
  {http://adsabs.harvard.edu/cgi-bin/nph-bib_query?bibcode=2004IAUC.8454....1M&db_key=AST}
  {8454, 1}

\bibitem[\protect\citeauthoryear{{Monet}, {Levine}, {Canzian}  \& {et
  al.}}{{Monet} et~al.}{2003}]{monet03}
{Monet} D.~G.,  {Levine} S.~E.,  {Canzian} B.,   {et al.} 2003, \aj, \href
  {http://adsabs.harvard.edu/cgi-bin/nph-bib_query?bibcode=2003AJ....125..984M&amp;db_key=AST}
  {125, 984}

\bibitem[\protect\citeauthoryear{{Mostardi}, {Li}  \& {Filippenko}}{{Mostardi}
  et~al.}{2008}]{2008CBET.1280....1M}
{Mostardi} R.,  {Li} W.,   {Filippenko} A.~V.,  2008, Central Bureau Electronic
  Telegrams, \href {http://adsabs.harvard.edu/abs/2008CBET.1280....1M} {1280}

\bibitem[\protect\citeauthoryear{{Murphy}, {Jennings}, {Williams}, {Dalcanton}
  \& {Dolphin}}{{Murphy} et~al.}{2011}]{2011ApJ...742L...4M}
{Murphy} J.~W.,  {Jennings} Z.~G.,  {Williams} B.,  {Dalcanton} J.~J.,
  {Dolphin} A.~E.,  2011, \mn@doi [\apjl] {10.1088/2041-8205/742/1/L4}, \href
  {http://adsabs.harvard.edu/abs/2011ApJ...742L...4M} {742, L4}

\bibitem[\protect\citeauthoryear{{Nakano} et~al.,}{{Nakano}
  et~al.}{1996a}]{1996IAUC.6454....1N}
{Nakano} S.,  et~al., 1996a, \iaucirc, \href
  {http://adsabs.harvard.edu/abs/1996IAUC.6454....1N} {6454}

\bibitem[\protect\citeauthoryear{{Nakano}, {Aoki}, {Garnavich}, {Kirshner}  \&
  {Berlind}}{{Nakano} et~al.}{1996b}]{1996IAUC.6524....1N}
{Nakano} S.,  {Aoki} M.,  {Garnavich} P.,  {Kirshner} R.,   {Berlind} P.,
  1996b, \iaucirc, \href {http://adsabs.harvard.edu/abs/1996IAUC.6524....1N}
  {6524}

\bibitem[\protect\citeauthoryear{{Nakano}, {Itagaki}, {Kushida}, {Kushida}  \&
  {Dimai}}{{Nakano} et~al.}{2001}]{2001IAUC.7761....1N}
{Nakano} S.,  {Itagaki} K.,  {Kushida} Y.,  {Kushida} R.,   {Dimai} A.,  2001,
  \iaucirc, \href {http://adsabs.harvard.edu/abs/2001IAUC.7761....1N} {7761}

\bibitem[\protect\citeauthoryear{{Nakano}, {Hirose}, {Kushida}, {Kushida}  \&
  {Li}}{{Nakano} et~al.}{2002}]{2002IAUC.7810....1N}
{Nakano} S.,  {Hirose} Y.,  {Kushida} R.,  {Kushida} Y.,   {Li} W.,  2002,
  \iaucirc, \href {http://adsabs.harvard.edu/abs/2002IAUC.7810....1N} {7810}

\bibitem[\protect\citeauthoryear{{Nakano} et~al.,}{{Nakano}
  et~al.}{2012}]{2012CBET.3263....1N}
{Nakano} S.,  et~al., 2012, Central Bureau Electronic Telegrams, \href
  {http://adsabs.harvard.edu/abs/2012CBET.3263....1N} {3263}

\bibitem[\protect\citeauthoryear{{Nomoto}, {Suzuki}, {Shigeyama}, {Kumagai},
  {Yamaoka}  \& {Saio}}{{Nomoto} et~al.}{1993}]{1993Natur.364..507N}
{Nomoto} K.,  {Suzuki} T.,  {Shigeyama} T.,  {Kumagai} S.,  {Yamaoka} H.,
  {Saio} H.,  1993, \nat, \href
  {http://adsabs.harvard.edu/cgi-bin/nph-bib_query?bibcode=1993Natur.364..507N&db_key=AST}
  {364, 507}

\bibitem[\protect\citeauthoryear{{Nomoto}, {Iwamoto}, {Suzuki}, {Pols},
  {Yamaoka}, {Hashimoto}, {Hoflich}  \& {Van den Heuvel}}{{Nomoto}
  et~al.}{1996}]{nombin96}
{Nomoto} K.,  {Iwamoto} K.,  {Suzuki} T.,  {Pols} O.~R.,  {Yamaoka} H.,
  {Hashimoto} M.,  {Hoflich} P.,   {Van den Heuvel} E.~P.~J.,  1996, in IAU
  Symp. 165: Compact Stars in Binaries. p.~119

\bibitem[\protect\citeauthoryear{{Pastorello} et~al.,}{{Pastorello}
  et~al.}{2008}]{2008MNRAS.389..955P}
{Pastorello} A.,  et~al., 2008, \mn@doi [\mnras]
  {10.1111/j.1365-2966.2008.13618.x}, \href
  {http://adsabs.harvard.edu/abs/2008MNRAS.389..955P} {389, 955}

\bibitem[\protect\citeauthoryear{{Paxton}, {Bildsten}, {Dotter}, {Herwig},
  {Lesaffre}  \& {Timmes}}{{Paxton} et~al.}{2011}]{2011ApJS..192....3P}
{Paxton} B.,  {Bildsten} L.,  {Dotter} A.,  {Herwig} F.,  {Lesaffre} P.,
  {Timmes} F.,  2011, \mn@doi [\apjs] {10.1088/0067-0049/192/1/3}, \href
  {http://adsabs.harvard.edu/abs/2011ApJS..192....3P} {192, 3}

\bibitem[\protect\citeauthoryear{{Petrosian} et~al.,}{{Petrosian}
  et~al.}{2005}]{2005AJ....129.1369P}
{Petrosian} A.,  et~al., 2005, \mn@doi [\aj] {10.1086/427712}, \href
  {http://adsabs.harvard.edu/abs/2005AJ....129.1369P} {129, 1369}

\bibitem[\protect\citeauthoryear{{Phillips} et~al.,}{{Phillips}
  et~al.}{2013}]{2013ApJ...779...38P}
{Phillips} M.~M.,  et~al., 2013, \mn@doi [\apj] {10.1088/0004-637X/779/1/38},
  \href {http://adsabs.harvard.edu/abs/2013ApJ...779...38P} {779, 38}

\bibitem[\protect\citeauthoryear{{Pilyugin}, {V{\'{\i}}lchez}  \&
  {Contini}}{{Pilyugin} et~al.}{2004}]{metapil}
{Pilyugin} L.~S.,  {V{\'{\i}}lchez} J.~M.,   {Contini} T.,  2004, \aap, \href
  {http://adsabs.harvard.edu/cgi-bin/nph-bib_query?bibcode=2004A%26A...425..849P&amp;db_key=AST}
  {425, 849}

\bibitem[\protect\citeauthoryear{{Podsiadlowski}, {Hsu}, {Joss}  \&
  {Ross}}{{Podsiadlowski} et~al.}{1993}]{1993Natur.364..509P}
{Podsiadlowski} P.,  {Hsu} J.~J.~L.,  {Joss} P.~C.,   {Ross} R.~R.,  1993,
  \nat, \href
  {http://adsabs.harvard.edu/cgi-bin/nph-bib_query?bibcode=1993Natur.364..509P&db_key=AST}
  {364, 509}

\bibitem[\protect\citeauthoryear{{Poznanski}, {Prochaska}  \&
  {Bloom}}{{Poznanski} et~al.}{2012}]{2012MNRAS.426.1465P}
{Poznanski} D.,  {Prochaska} J.~X.,   {Bloom} J.~S.,  2012, \mn@doi [\mnras]
  {10.1111/j.1365-2966.2012.21796.x}, \href
  {http://adsabs.harvard.edu/abs/2012MNRAS.426.1465P} {426, 1465}

\bibitem[\protect\citeauthoryear{{Prentice} et~al.,}{{Prentice}
  et~al.}{2016}]{2016MNRAS.458.2973P}
{Prentice} S.~J.,  et~al., 2016, \mn@doi [\mnras] {10.1093/mnras/stw299}, \href
  {http://adsabs.harvard.edu/abs/2016MNRAS.458.2973P} {458, 2973}

\bibitem[\protect\citeauthoryear{{Prieto}}{{Prieto}}{2012}]{2012ATel.4502....1P}
{Prieto} J.~L.,  2012, The Astronomer's Telegram, \href
  {http://adsabs.harvard.edu/abs/2012ATel.4502....1P} {4502}

\bibitem[\protect\citeauthoryear{{Prieto}, {Stanek}  \& {Beacom}}{{Prieto}
  et~al.}{2008}]{2008ApJ...673..999P}
{Prieto} J.~L.,  {Stanek} K.~Z.,   {Beacom} J.~F.,  2008, \mn@doi [\apj]
  {10.1086/524654}, \href {http://adsabs.harvard.edu/abs/2008ApJ...673..999P}
  {673, 999}

\bibitem[\protect\citeauthoryear{{Puckett} et~al.,}{{Puckett}
  et~al.}{1994}]{1994IAUC.5961....1P}
{Puckett} T.,  et~al., 1994, \iaucirc, \href
  {http://adsabs.harvard.edu/abs/1994IAUC.5961....1P} {5961}

\bibitem[\protect\citeauthoryear{{Puckett}, {Langoussis}  \&
  {Garradd}}{{Puckett} et~al.}{2000}]{00ewiauc1}
{Puckett} T.,  {Langoussis} A.,   {Garradd} G.~J.,  2000, \iaucirc, \href
  {http://adsabs.harvard.edu/cgi-bin/nph-bib_query?bibcode=2000IAUC.7530....1P&amp;db_key=AST}
  {7530, 1}

\bibitem[\protect\citeauthoryear{{Pugh}, {Li}, {Manzini}  \& {Behrend}}{{Pugh}
  et~al.}{2004}]{2004IAUC.8452....2P}
{Pugh} H.,  {Li} W.,  {Manzini} F.,   {Behrend} R.,  2004, \iaucirc, \href
  {http://adsabs.harvard.edu/abs/2004IAUC.8452....2P} {8452}

\bibitem[\protect\citeauthoryear{{Qiu}, {Li}, {Qiao}  \& {Hu}}{{Qiu}
  et~al.}{1999}]{1999AJ....117..736Q}
{Qiu} Y.,  {Li} W.,  {Qiao} Q.,   {Hu} J.,  1999, \mn@doi [\aj]
  {10.1086/300731}, \href
  {http://adsabs.harvard.edu/cgi-bin/nph-bib_query?bibcode=1999AJ....117..736Q&db_key=AST}
  {117, 736}

\bibitem[\protect\citeauthoryear{{Reiland}, {Griga}, {Riou}, {Lamotte Bailey}
  \& {et~al.}}{{Reiland} et~al.}{2011}]{CBET2736}
{Reiland} T.,  {Griga} T.,  {Riou} A.,  {Lamotte Bailey} S.,   {et~al.} 2011,
  Central Bureau Electronic Telegrams, 2736, 1

\bibitem[\protect\citeauthoryear{{Reilly} et~al.,}{{Reilly}
  et~al.}{2016}]{2016MNRAS.457..288R}
{Reilly} E.,  et~al., 2016, \mn@doi [\mnras] {10.1093/mnras/stv3005}, \href
  {http://adsabs.harvard.edu/abs/2016MNRAS.457..288R} {457, 288}

\bibitem[\protect\citeauthoryear{{Riess} et~al.,}{{Riess}
  et~al.}{2011}]{2011ApJ...730..119R}
{Riess} A.~G.,  et~al., 2011, \mn@doi [\apj] {10.1088/0004-637X/730/2/119},
  \href {http://adsabs.harvard.edu/abs/2011ApJ...730..119R} {730, 119}

\bibitem[\protect\citeauthoryear{{Ross}, {Channa}, {Molloy}, {Zheng}  \&
  {Filippenko}}{{Ross} et~al.}{2016}]{2016ATel.8875....1R}
{Ross} T.~W.,  {Channa} S.,  {Molloy} J.~D.,  {Zheng} W.,   {Filippenko} A.~V.,
   2016, The Astronomer's Telegram, \href
  {http://adsabs.harvard.edu/abs/2016ATel.8875....1R} {8875}

\bibitem[\protect\citeauthoryear{{Rubin}, {Ford}  \& {D'Odorico}}{{Rubin}
  et~al.}{1970}]{1970ApJ...160..801R}
{Rubin} V.~C.,  {Ford} Jr. W.~K.,   {D'Odorico} S.,  1970, \mn@doi [\apj]
  {10.1086/150473}, \href {http://adsabs.harvard.edu/abs/1970ApJ...160..801R}
  {160, 801}

\bibitem[\protect\citeauthoryear{{Saha}, {Sandage}, {Thim}, {Labhardt},
  {Tammann}, {Christensen}, {Panagia}  \& {Macchetto}}{{Saha}
  et~al.}{2001}]{2001ApJ...551..973S}
{Saha} A.,  {Sandage} A.,  {Thim} F.,  {Labhardt} L.,  {Tammann} G.~A.,
  {Christensen} J.,  {Panagia} N.,   {Macchetto} F.~D.,  2001, \mn@doi [\apj]
  {10.1086/320223}, \href {http://adsabs.harvard.edu/abs/2001ApJ...551..973S}
  {551, 973}

\bibitem[\protect\citeauthoryear{{Saha}, {Thim}, {Tammann}, {Reindl}  \&
  {Sandage}}{{Saha} et~al.}{2006}]{2006ApJS..165..108S}
{Saha} A.,  {Thim} F.,  {Tammann} G.~A.,  {Reindl} B.,   {Sandage} A.,  2006,
  \mn@doi [\apjs] {10.1086/503800}, \href
  {http://adsabs.harvard.edu/abs/2006ApJS..165..108S} {165, 108}

\bibitem[\protect\citeauthoryear{{Sahu}, {Anupama}  \& {Gurugubelli}}{{Sahu}
  et~al.}{2009}]{2009CBET.1955....2S}
{Sahu} D.~K.,  {Anupama} G.~C.,   {Gurugubelli} U.~K.,  2009, Central Bureau
  Electronic Telegrams, \href
  {http://adsabs.harvard.edu/abs/2009CBET.1955....2S} {1955}

\bibitem[\protect\citeauthoryear{{Sahu}, {Gurugubelli}, {Anupama}  \&
  {Nomoto}}{{Sahu} et~al.}{2011}]{2011MNRAS.413.2583S}
{Sahu} D.~K.,  {Gurugubelli} U.~K.,  {Anupama} G.~C.,   {Nomoto} K.,  2011,
  \mn@doi [\mnras] {10.1111/j.1365-2966.2011.18326.x}, \href
  {http://adsabs.harvard.edu/abs/2011MNRAS.413.2583S} {413, 2583}

\bibitem[\protect\citeauthoryear{{Sana} et~al.,}{{Sana}
  et~al.}{2012}]{2012Sci...337..444S}
{Sana} H.,  et~al., 2012, \mn@doi [Science] {10.1126/science.1223344}, \href
  {http://adsabs.harvard.edu/abs/2012Sci...337..444S} {337, 444}

\bibitem[\protect\citeauthoryear{{Schlafly} \& {Finkbeiner}}{{Schlafly} \&
  {Finkbeiner}}{2011}]{2011ApJ...737..103S}
{Schlafly} E.~F.,  {Finkbeiner} D.~P.,  2011, \mn@doi [\apj]
  {10.1088/0004-637X/737/2/103}, \href
  {http://adsabs.harvard.edu/abs/2011ApJ...737..103S} {737, 103}

\bibitem[\protect\citeauthoryear{{Schmidt} \& {Salvo}}{{Schmidt} \&
  {Salvo}}{2005}]{2005CBET..122....1S}
{Schmidt} B.,  {Salvo} M.,  2005, Central Bureau Electronic Telegrams, \href
  {http://adsabs.harvard.edu/abs/2005CBET..122....1S} {122}

\bibitem[\protect\citeauthoryear{{Schweizer} et~al.,}{{Schweizer}
  et~al.}{2008}]{2008AJ....136.1482S}
{Schweizer} F.,  et~al., 2008, \mn@doi [\aj] {10.1088/0004-6256/136/4/1482},
  \href {http://adsabs.harvard.edu/abs/2008AJ....136.1482S} {136, 1482}

\bibitem[\protect\citeauthoryear{{Silverman}, {Cenko}  \&
  {Filippenko}}{{Silverman} et~al.}{2011}]{CBET2736a}
{Silverman} J.~M.,  {Cenko} S.~B.,   {Filippenko} A.~V.,  2011, Central Bureau
  Electronic Telegrams, 2736, 4

\bibitem[\protect\citeauthoryear{{Skilling}}{{Skilling}}{2004}]{2004AIPC..735..395S}
{Skilling} J.,  2004, in {R.~Fischer, R.~Preuss, \& U.~V.~Toussaint} ed.,
  American Institute of Physics Conference Series Vol. 735, American Institute
  of Physics Conference Series. pp 395--405, \mn@doi{10.1063/1.1835238}

\bibitem[\protect\citeauthoryear{{Smartt}}{{Smartt}}{2015}]{2015PASA...32...16S}
{Smartt} S.~J.,  2015, \mn@doi [\pasa] {10.1017/pasa.2015.17}, \href
  {http://adsabs.harvard.edu/abs/2015PASA...32...16S} {32, 16}

\bibitem[\protect\citeauthoryear{{Smartt}, {Eldridge}, {Crockett}  \&
  {Maund}}{{Smartt} et~al.}{2009}]{2008arXiv0809.0403S}
{Smartt} S.~J.,  {Eldridge} J.~J.,  {Crockett} R.~M.,   {Maund} J.~R.,  2009,
  \mn@doi [\mnras] {10.1111/j.1365-2966.2009.14506.x}, \href
  {http://adsabs.harvard.edu/abs/2009MNRAS.395.1409S} {395, 1409}

\bibitem[\protect\citeauthoryear{{Smith}}{{Smith}}{2014}]{2014ARA&A..52..487S}
{Smith} N.,  2014, \mn@doi [\araa] {10.1146/annurev-astro-081913-040025}, \href
  {http://adsabs.harvard.edu/abs/2014ARA%26A..52..487S} {52, 487}

\bibitem[\protect\citeauthoryear{{Smith}, {Li}, {Filippenko}  \&
  {Chornock}}{{Smith} et~al.}{2011}]{2011MNRAS.412.1522S}
{Smith} N.,  {Li} W.,  {Filippenko} A.~V.,   {Chornock} R.,  2011, \mn@doi
  [\mnras] {10.1111/j.1365-2966.2011.17229.x}, \href
  {http://adsabs.harvard.edu/abs/2011MNRAS.412.1522S} {412, 1522}

\bibitem[\protect\citeauthoryear{{Soderberg} et~al.,}{{Soderberg}
  et~al.}{2012}]{2012ATel.3968....1S}
{Soderberg} A.,  et~al., 2012, The Astronomer's Telegram, \href
  {http://adsabs.harvard.edu/abs/2012ATel.3968....1S} {3968}

\bibitem[\protect\citeauthoryear{{Sorce}, {Tully}, {Courtois}, {Jarrett},
  {Neill}  \& {Shaya}}{{Sorce} et~al.}{2014}]{2014MNRAS.444..527S}
{Sorce} J.~G.,  {Tully} R.~B.,  {Courtois} H.~M.,  {Jarrett} T.~H.,  {Neill}
  J.~D.,   {Shaya} E.~J.,  2014, \mn@doi [\mnras] {10.1093/mnras/stu1450},
  \href {http://adsabs.harvard.edu/abs/2014MNRAS.444..527S} {444, 527}

\bibitem[\protect\citeauthoryear{{Taddia} et~al.,}{{Taddia}
  et~al.}{2017}]{2017arXiv170707614T}
{Taddia} F.,  et~al., 2017, preprint, \href
  {http://adsabs.harvard.edu/abs/2017arXiv170707614T} {} (\mn@eprint {arXiv}
  {1707.07614})

\bibitem[\protect\citeauthoryear{{Takaki} et~al.,}{{Takaki}
  et~al.}{2013}]{2013ApJ...772L..17T}
{Takaki} K.,  et~al., 2013, \mn@doi [\apjl] {10.1088/2041-8205/772/2/L17},
  \href {http://adsabs.harvard.edu/abs/2013ApJ...772L..17T} {772, L17}

\bibitem[\protect\citeauthoryear{{Tak{\'a}ts} \& {Vink{\'o}}}{{Tak{\'a}ts} \&
  {Vink{\'o}}}{2006}]{2006MNRAS.372.1735T}
{Tak{\'a}ts} K.,  {Vink{\'o}} J.,  2006, \mn@doi [\mnras]
  {10.1111/j.1365-2966.2006.10974.x}, \href
  {http://adsabs.harvard.edu/abs/2006MNRAS.372.1735T} {372, 1735}

\bibitem[\protect\citeauthoryear{{Tang}, {Bressan}, {Rosenfield}, {Slemer},
  {Marigo}, {Girardi}  \& {Bianchi}}{{Tang} et~al.}{2014}]{2014MNRAS.445.4287T}
{Tang} J.,  {Bressan} A.,  {Rosenfield} P.,  {Slemer} A.,  {Marigo} P.,
  {Girardi} L.,   {Bianchi} L.,  2014, \mn@doi [\mnras]
  {10.1093/mnras/stu2029}, \href
  {http://adsabs.harvard.edu/abs/2014MNRAS.445.4287T} {445, 4287}

\bibitem[\protect\citeauthoryear{{Taubenberger} et~al.,}{{Taubenberger}
  et~al.}{2009}]{2009MNRAS.397..677T}
{Taubenberger} S.,  et~al., 2009, \mn@doi [\mnras]
  {10.1111/j.1365-2966.2009.15003.x}, \href
  {http://adsabs.harvard.edu/abs/2009MNRAS.397..677T} {397, 677}

\bibitem[\protect\citeauthoryear{{Theureau}, {Hanski}, {Coudreau}, {Hallet}  \&
  {Martin}}{{Theureau} et~al.}{2007}]{2007A&A...465...71T}
{Theureau} G.,  {Hanski} M.~O.,  {Coudreau} N.,  {Hallet} N.,   {Martin} J.-M.,
   2007, \mn@doi [\aap] {10.1051/0004-6361:20066187}, \href
  {http://adsabs.harvard.edu/abs/2007A%26A...465...71T} {465, 71}

\bibitem[\protect\citeauthoryear{{Tomasella } et~al.,}{{Tomasella }
  et~al.}{2014}]{2014AN....335..841T}
{Tomasella } L.,  et~al., 2014, \mn@doi [Astronomische Nachrichten]
  {10.1002/asna.201412068}, \href
  {http://adsabs.harvard.edu/abs/2014AN....335..841T} {335, 841}

\bibitem[\protect\citeauthoryear{{Trotta}}{{Trotta}}{2008}]{2008ConPh..49...71T}
{Trotta} R.,  2008, \mn@doi [Contemporary Physics] {10.1080/00107510802066753},
  \href {http://adsabs.harvard.edu/abs/2008ConPh..49...71T} {49, 71}

\bibitem[\protect\citeauthoryear{{Tully} \& {Fisher}}{{Tully} \&
  {Fisher}}{1988}]{1988ang..book.....T}
{Tully} R.~B.,  {Fisher} J.~R.,  1988, {Catalog of Nearby Galaxies}.
Cambridge University Press

\bibitem[\protect\citeauthoryear{{Tully}, {Rizzi}, {Shaya}, {Courtois},
  {Makarov}  \& {Jacobs}}{{Tully} et~al.}{2009}]{2009AJ....138..323T}
{Tully} R.~B.,  {Rizzi} L.,  {Shaya} E.~J.,  {Courtois} H.~M.,  {Makarov}
  D.~I.,   {Jacobs} B.~A.,  2009, \mn@doi [\aj] {10.1088/0004-6256/138/2/323},
  \href {http://adsabs.harvard.edu/abs/2009AJ....138..323T} {138, 323}

\bibitem[\protect\citeauthoryear{{Tully} et~al.,}{{Tully}
  et~al.}{2013}]{2013AJ....146...86T}
{Tully} R.~B.,  et~al., 2013, \mn@doi [\aj] {10.1088/0004-6256/146/4/86}, \href
  {http://adsabs.harvard.edu/abs/2013AJ....146...86T} {146, 86}

\bibitem[\protect\citeauthoryear{{Valenti} et~al.,}{{Valenti}
  et~al.}{2011}]{2011MNRAS.416.3138V}
{Valenti} S.,  et~al., 2011, \mn@doi [\mnras]
  {10.1111/j.1365-2966.2011.19262.x}, \href
  {http://adsabs.harvard.edu/abs/2011MNRAS.416.3138V} {416, 3138}

\bibitem[\protect\citeauthoryear{{Van Dyk} et~al.,}{{Van Dyk}
  et~al.}{1999a}]{1999PASP..111..313V}
{Van Dyk} S.~D.,  et~al., 1999a, \mn@doi [\pasp] {10.1086/316331}, \href
  {http://adsabs.harvard.edu/abs/1999PASP..111..313V} {111, 313}

\bibitem[\protect\citeauthoryear{{Van Dyk}, {Peng}, {Barth}  \&
  {Filippenko}}{{Van Dyk} et~al.}{1999b}]{1999AJ....118.2331V}
{Van Dyk} S.~D.,  {Peng} C.~Y.,  {Barth} A.~J.,   {Filippenko} A.~V.,  1999b,
  \aj, \href
  {http://adsabs.harvard.edu/cgi-bin/nph-bib_query?bibcode=1999AJ....118.2331V&db_key=AST}
  {118, 2331}

\bibitem[\protect\citeauthoryear{{Van Dyk}, {Garnavich}, {Filippenko},
  {H{\"o}flich}, {Kirshner}, {Kurucz}  \& {Challis}}{{Van Dyk}
  et~al.}{2002}]{2002PASP..114.1322V}
{Van Dyk} S.~D.,  {Garnavich} P.~M.,  {Filippenko} A.~V.,  {H{\"o}flich} P.,
  {Kirshner} R.~P.,  {Kurucz} R.~L.,   {Challis} P.,  2002, \pasp, \href
  {http://adsabs.harvard.edu/cgi-bin/nph-bib_query?bibcode=2002PASP..114.1322V&db_key=AST}
  {114, 1322}

\bibitem[\protect\citeauthoryear{{Van Dyk} et~al.,}{{Van Dyk}
  et~al.}{2011}]{2011ApJ...741L..28V}
{Van Dyk} S.~D.,  et~al., 2011, \mn@doi [\apjl] {10.1088/2041-8205/741/2/L28},
  \href {http://adsabs.harvard.edu/abs/2011ApJ...741L..28V} {741, L28}

\bibitem[\protect\citeauthoryear{{Van Dyk}, {Cenko}, {Silverman}, {Miller},
  {Filippenko}, {Bloom}  \& {Nugent}}{{Van Dyk}
  et~al.}{2012}]{2012ATel.3971....1V}
{Van Dyk} S.~D.,  {Cenko} S.~B.,  {Silverman} J.~M.,  {Miller} A.~A.,
  {Filippenko} A.~V.,  {Bloom} J.~S.,   {Nugent} P.~E.,  2012, The Astronomer's
  Telegram, \href {http://adsabs.harvard.edu/abs/2012ATel.3971....1V} {3971}

\bibitem[\protect\citeauthoryear{{Van Dyk} et~al.,}{{Van Dyk}
  et~al.}{2013}]{2013ApJ...772L..32V}
{Van Dyk} S.~D.,  et~al., 2013, \mn@doi [\apjl] {10.1088/2041-8205/772/2/L32},
  \href {http://adsabs.harvard.edu/abs/2013ApJ...772L..32V} {772, L32}

\bibitem[\protect\citeauthoryear{{Van Dyk} et~al.,}{{Van Dyk}
  et~al.}{2014}]{2014AJ....147...37V}
{Van Dyk} S.~D.,  et~al., 2014, \mn@doi [\aj] {10.1088/0004-6256/147/2/37},
  \href {http://adsabs.harvard.edu/abs/2014AJ....147...37V} {147, 37}

\bibitem[\protect\citeauthoryear{{Van Dyk}, {de Mink}  \& {Zapartas}}{{Van Dyk}
  et~al.}{2016}]{2016ApJ...818...75V}
{Van Dyk} S.~D.,  {de Mink} S.~E.,   {Zapartas} E.,  2016, \mn@doi [\apj]
  {10.3847/0004-637X/818/1/75}, \href
  {http://adsabs.harvard.edu/abs/2016ApJ...818...75V} {818, 75}

\bibitem[\protect\citeauthoryear{{Wheeler}, {Harkness}, {Clocchiatti},
  {Benetti}, {Brotherton}, {Depoy}  \& {Elias}}{{Wheeler}
  et~al.}{1994}]{1994ApJ...436L.135W}
{Wheeler} J.~C.,  {Harkness} R.~P.,  {Clocchiatti} A.,  {Benetti} S.,
  {Brotherton} M.~S.,  {Depoy} D.~L.,   {Elias} J.,  1994, \mn@doi [\apjl]
  {10.1086/187651}, \href {http://adsabs.harvard.edu/abs/1994ApJ...436L.135W}
  {436, L.135}

\bibitem[\protect\citeauthoryear{{Wheeler}, {Johnson}  \&
  {Clocchiatti}}{{Wheeler} et~al.}{2015}]{2015MNRAS.450.1295W}
{Wheeler} J.~C.,  {Johnson} V.,   {Clocchiatti} A.,  2015, \mn@doi [\mnras]
  {10.1093/mnras/stv650}, \href
  {http://adsabs.harvard.edu/abs/2015MNRAS.450.1295W} {450, 1295}

\bibitem[\protect\citeauthoryear{{Whitmore}, {Zhang}, {Leitherer}, {Fall},
  {Schweizer}  \& {Miller}}{{Whitmore} et~al.}{1999}]{1999AJ....118.1551W}
{Whitmore} B.~C.,  {Zhang} Q.,  {Leitherer} C.,  {Fall} S.~M.,  {Schweizer} F.,
    {Miller} B.~W.,  1999, \aj, \href
  {http://adsabs.harvard.edu/cgi-bin/nph-bib_query?bibcode=1999AJ....118.1551W&db_key=AST}
  {118, 1551}

\bibitem[\protect\citeauthoryear{{Williams} et~al.,}{{Williams}
  et~al.}{2009}]{2009AJ....137..419W}
{Williams} B.~F.,  et~al., 2009, \mn@doi [\aj] {10.1088/0004-6256/137/1/419},
  \href {http://adsabs.harvard.edu/abs/2009AJ....137..419W} {137, 419}

\bibitem[\protect\citeauthoryear{{Williams} et~al.,}{{Williams}
  et~al.}{2014a}]{2014ApJS..215....9W}
{Williams} B.~F.,  et~al., 2014a, \mn@doi [\apjs] {10.1088/0067-0049/215/1/9},
  \href {http://cdsads.u-strasbg.fr/abs/2014ApJS..215....9W} {215, 9}

\bibitem[\protect\citeauthoryear{{Williams}, {Peterson}, {Murphy}, {Gilbert},
  {Dalcanton}, {Dolphin}  \& {Jennings}}{{Williams}
  et~al.}{2014b}]{2014ApJ...791..105W}
{Williams} B.~F.,  {Peterson} S.,  {Murphy} J.,  {Gilbert} K.,  {Dalcanton}
  J.~J.,  {Dolphin} A.~E.,   {Jennings} Z.~G.,  2014b, \mn@doi [\apj]
  {10.1088/0004-637X/791/2/105}, \href
  {http://adsabs.harvard.edu/abs/2014ApJ...791..105W} {791, 105}

\bibitem[\protect\citeauthoryear{{Xiao}, {Eldridge}, {Stanway}  \&
  {Galbany}}{{Xiao} et~al.}{2017}]{2017arXiv170503606}
{Xiao} L.,  {Eldridge} J.~J.,  {Stanway} E.,   {Galbany} L.,  2017, preprint,
  \href {http://adsabs.harvard.edu/abs/2017arXiv170503606X} {} (\mn@eprint
  {arXiv} {1705.03606})

\bibitem[\protect\citeauthoryear{{Xu} \& {Qiu}}{{Xu} \& {Qiu}}{2001}]{01biauc1}
{Xu} D.~W.,  {Qiu} Y.~L.,  2001, \iaucirc, \href
  {http://adsabs.harvard.edu/cgi-bin/nph-bib_query?bibcode=2001IAUC.7555....2X&db_key=AST}
  {7555, 2}

\bibitem[\protect\citeauthoryear{{Yamanaka}, {Itoh}, {Ui}, {Arai}, {Nagashima}
  \& {Kajiawa}}{{Yamanaka} et~al.}{2011}]{CBET2736b}
{Yamanaka} M.,  {Itoh} R.,  {Ui} T.,  {Arai} A.,  {Nagashima} M.,   {Kajiawa}
  K.,  2011, Central Bureau Electronic Telegrams, 2736, 6

\bibitem[\protect\citeauthoryear{{Yoon}, {Gr{\"a}fener}, {Vink}, {Kozyreva}  \&
  {Izzard}}{{Yoon} et~al.}{2012}]{2012A&A...544L..11Y}
{Yoon} S.-C.,  {Gr{\"a}fener} G.,  {Vink} J.~S.,  {Kozyreva} A.,   {Izzard}
  R.~G.,  2012, \mn@doi [\aap] {10.1051/0004-6361/201219790}, \href
  {http://adsabs.harvard.edu/abs/2012A%26A...544L..11Y} {544, L11}

\bibitem[\protect\citeauthoryear{{Zapartas} et~al.,}{{Zapartas}
  et~al.}{2017}]{2017ApJ...842..125Z}
{Zapartas} E.,  et~al., 2017, \mn@doi [\apj] {10.3847/1538-4357/aa7467}, \href
  {http://adsabs.harvard.edu/abs/2017ApJ...842..125Z} {842, 125}

\bibitem[\protect\citeauthoryear{{van den Bergh}, {Li}  \& {Filippenko}}{{van
  den Bergh} et~al.}{2005}]{2005PASP..117..773V}
{van den Bergh} S.,  {Li} W.,   {Filippenko} A.~V.,  2005, \mn@doi [\pasp]
  {10.1086/431435}, \href {http://adsabs.harvard.edu/abs/2005PASP..117..773V}
  {117, 773}

\makeatother
\end{thebibliography}

%%%%%%%%%%%%%%%%%%%%%%%%%%%%%%%%%%%%
%APPENDICES
%%%%%%%%%%%%%%%%%%%%%%%%%%%%%%%%%%%%
%%%%%%%%%%%%%%%%%%%%%%%%%%%%%%%%
%APPENDICES
%%%%%%%%%%%%%%%%%%%%%%%%%%%%%%%%
\appendix
\newpage
\section{Colour-magnitude diagrams}
\label{sec:app:cmd}
\begin{figure}
\begin{center}
\includegraphics[width=6.0cm, angle=270]{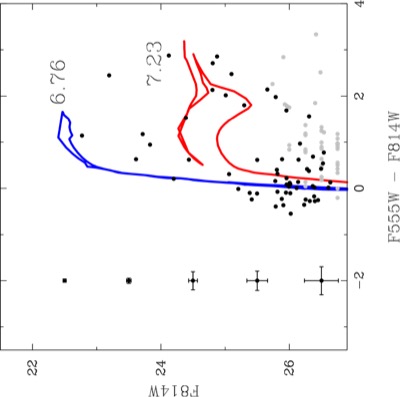}
\end{center}
\caption{Colour-magnitude diagram for the Type Ic SN 1996aq in NGC 5584.   Black points indicate stars detected in both filters, while grey points indicate those stars detected in only one of the filters.  The column of points with error bars on the left-hand side shows the average uncertainties for stars at that brightness.}
\label{fig:obs:96aq:cmd}
\end{figure}

\begin{figure}
\begin{center}
\includegraphics[width=6.0cm, angle=270]{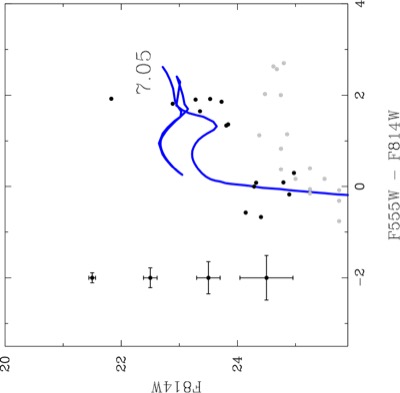}
\end{center}
\caption{The same as Fig. \ref{fig:obs:96aq:cmd} but for the Type IIb SN 1996cb in NGC~3510.   }
\label{fig:obs:96cb:cmd}
\end{figure}

\begin{figure}
\begin{center}
\includegraphics[width=6.0cm, angle=270]{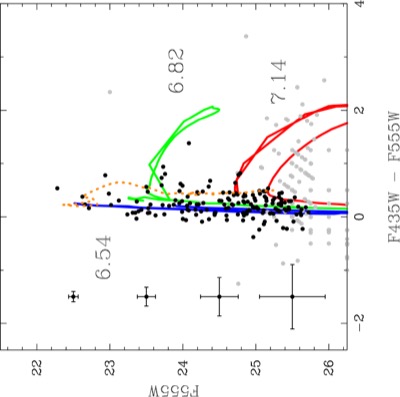}
\includegraphics[width=6.0cm, angle=270]{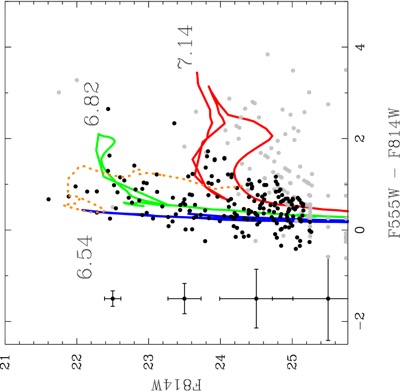}
\end{center}
\caption{The same as Fig. \ref{fig:obs:96aq:cmd} but for ACS observations of the Type Ic SN 2000ew in NGC~3810.  The orange dashed line shows the locus of STARBURST99 models with ages in the range ${6} \leq \tau  \leq {8}$ for an initial cluster mass of $10^{4}M_{\odot}$.}
\label{fig:obs:00ew_ACS:cmd}
\end{figure}

\begin{figure}
\begin{center}
\includegraphics[width=6.0cm, angle=270]{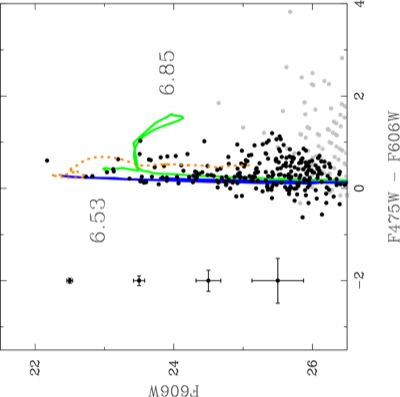}
\caption{The same as Fig. \ref{fig:obs:96aq:cmd} but for WFC3 observations of the Type Ic SN 2000ew in NGC~3810.  The orange dashed line shows the locus of STARBURST99 models with ages in the range $6 \leq \tau \leq 8$ for an initial cluster mass of $10^{4}M_{\odot}$.}
\end{center}
\label{fig:obs:00ew_WFC3:cmd}
\end{figure}

\begin{figure}
\begin{center}
\includegraphics[width=6.0cm, angle=270]{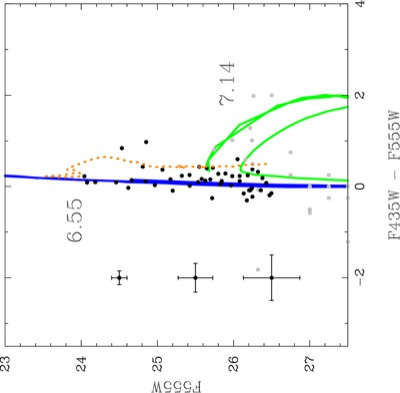}
\includegraphics[width=6.0cm, angle=270]{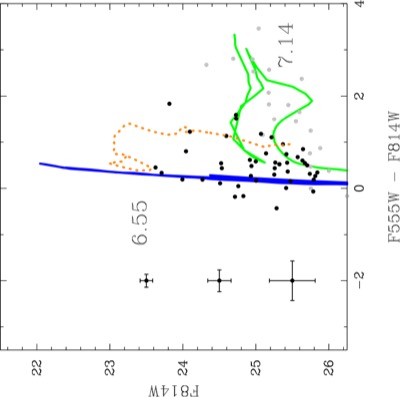}
\end{center}
\caption{The same as Fig. \ref{fig:obs:96aq:cmd} but for the Type Ic SN 2001B in IC~391.  The orange dashed line shows the locus of STARBURST99 models with ages in the range $6 \leq \tau \leq 8$ for an initial cluster mass of $10^{4}M_{\odot}$.}
\label{fig:obs:01B:cmd}
\end{figure}

\begin{figure}
\begin{center}
\includegraphics[width=6.0cm, angle=270]{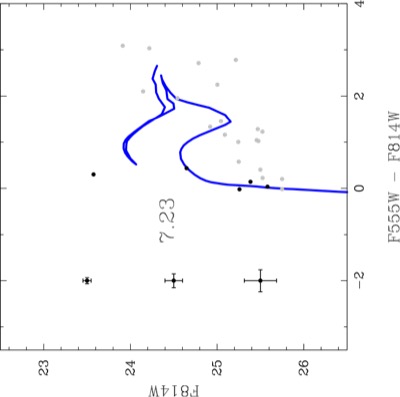}
\end{center}
\caption{The same as Fig. \ref{fig:obs:96aq:cmd} but for the Type IIb SN 2001gd in NGC~5033. }
\label{fig:obs:01gd:cmd}
\end{figure}

\begin{figure}
\begin{center}
\includegraphics[width=6.0cm, angle=270]{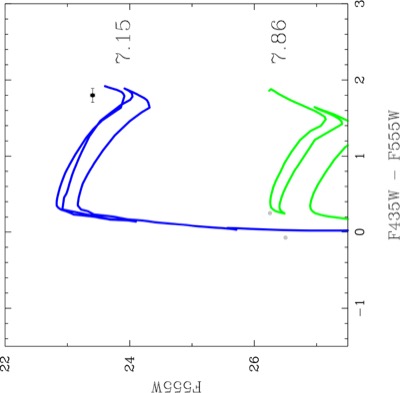}
\includegraphics[width=6.0cm, angle=270]{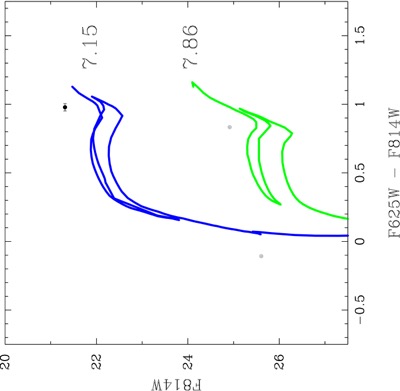}
\end{center}
\caption{The same as Fig. \ref{fig:obs:96aq:cmd} but for the Type Ic SN 2002ap in M74. }
\label{fig:obs:02ap:cmd}
\end{figure}

\begin{figure}
\begin{center}
\includegraphics[width=6.0cm, angle=270]{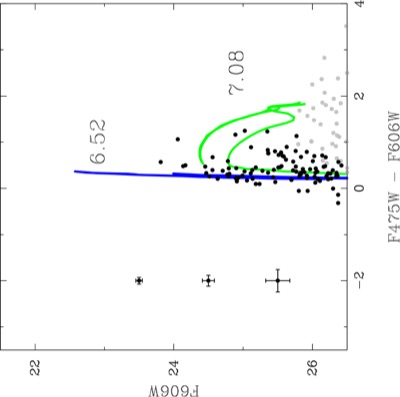}
\caption{The same as Fig. \ref{fig:obs:96aq:cmd} but for the Type Ib SN 2004gn in NGC 4527.}
\end{center}
\label{fig:obs:04gn:cmd}
\end{figure}

\begin{figure}
\begin{center}
\includegraphics[width=6.0cm, angle=270]{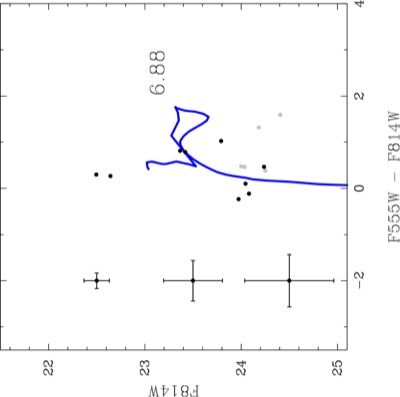}
\end{center}
\caption{The same as Fig. \ref{fig:obs:96aq:cmd} but for the Type IIb SN 2004gq in NGC~1832.}
\label{fig:obs:04gq:cmd}
\end{figure}

\begin{figure}
\begin{center}
\includegraphics[width=6.0cm, angle=270]{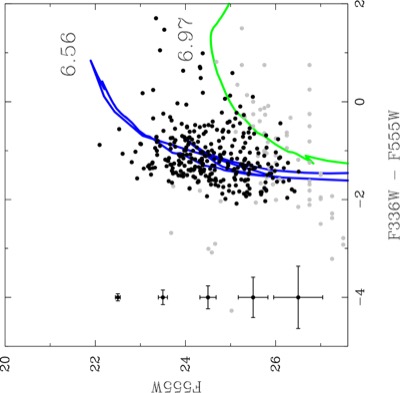}
\includegraphics[width=6.0cm, angle=270]{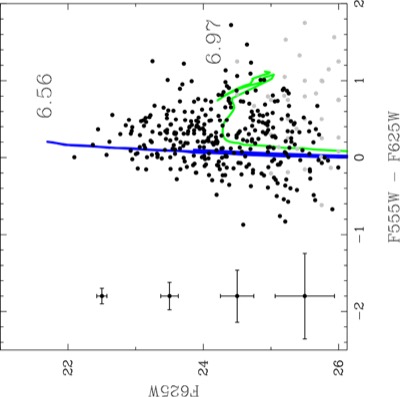}
\includegraphics[width=6.0cm, angle=270]{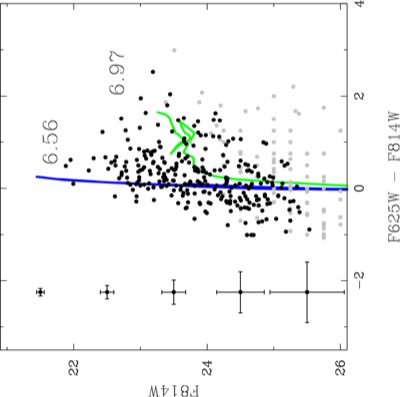}
\end{center}
\caption{The same as Fig. \ref{fig:obs:96aq:cmd} but for the Type Ic SN 2004gt in NGC 4038.}
\label{fig:obs:04gt:cmd}
\end{figure}

\begin{figure}
\begin{center}
\includegraphics[width=6.0cm, angle=270]{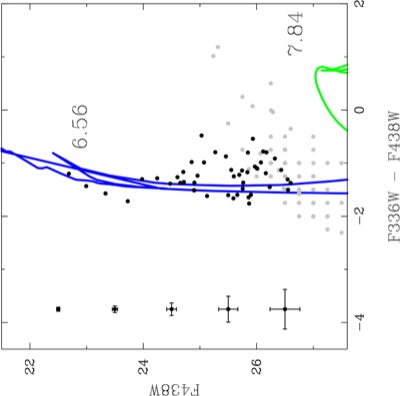}
\includegraphics[width=6.0cm, angle=270]{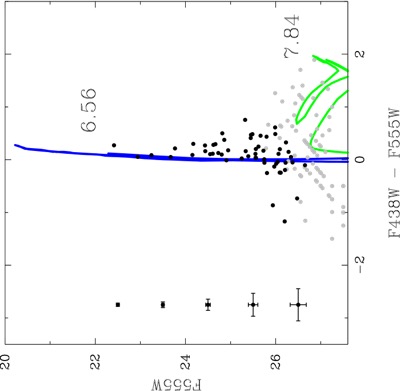}
\includegraphics[width=6.0cm, angle=270]{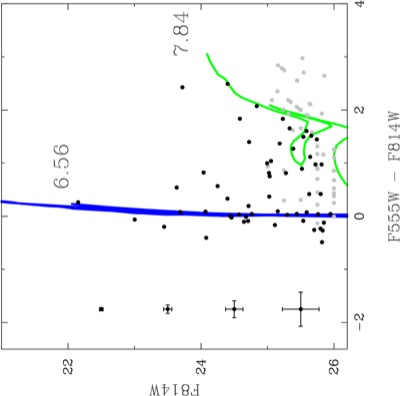}
\end{center}
\caption{The same as Fig. \ref{fig:obs:96aq:cmd} but for the Type Ic SN 2005at in NGC 6744.}
\label{fig:obs:05at:cmd}
\end{figure}

\begin{figure}
\begin{center}
\includegraphics[width=6.0cm, angle=270]{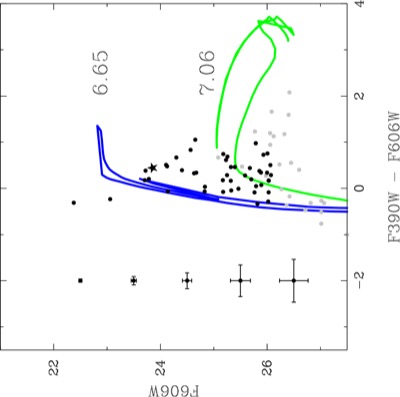}
\includegraphics[width=6.0cm, angle=270]{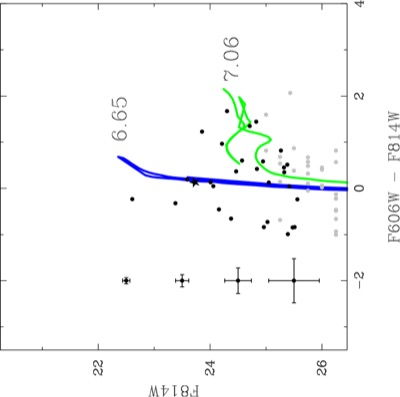}
\end{center}
\caption{The same as Fig. \ref{fig:obs:96aq:cmd} but for the Type Ib SN 2007fo in NGC 7714.}
\label{fig:obs:07fo:cmd}
\end{figure}

\begin{figure*}
\begin{center}
\includegraphics[width=6.0cm, angle=270]{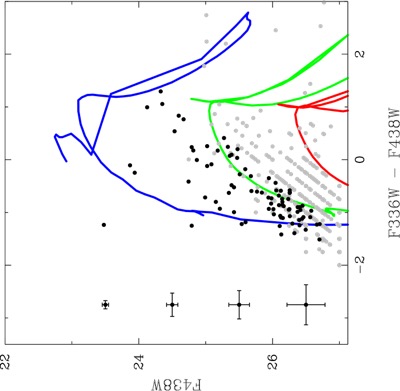}
\includegraphics[width=6.0cm, angle=270]{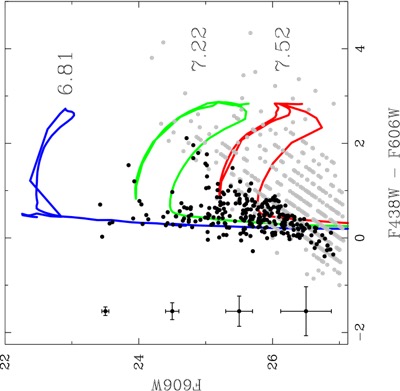}
\includegraphics[width=6.0cm, angle=270]{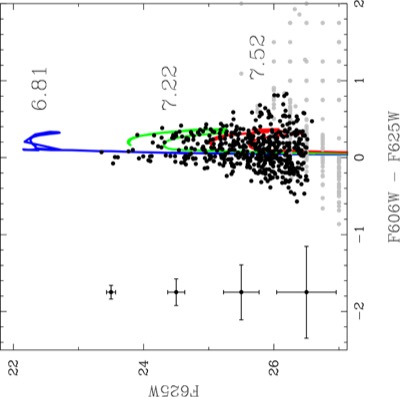}
\includegraphics[width=6.0cm, angle=270]{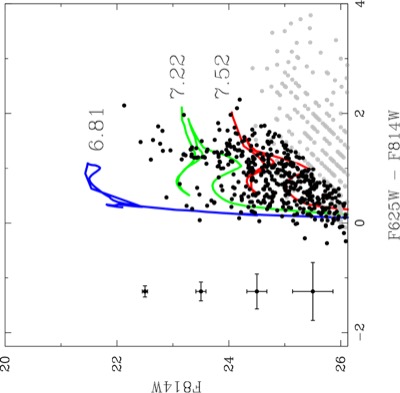}
\end{center}
\caption{The same as Fig. \ref{fig:obs:96aq:cmd} but for the Type IIb 2008ax in NGC 4490.}
\label{fig:obs:08ax:cmd}
\end{figure*}

\begin{figure}
\begin{center}
\includegraphics[width=6.0cm, angle=270]{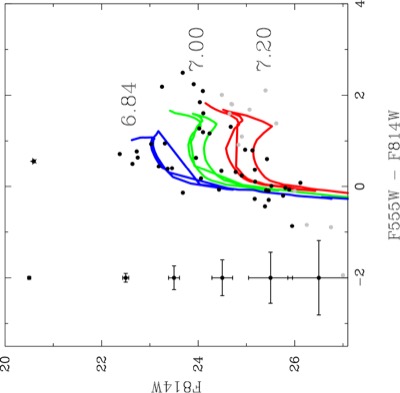}
\end{center}
\caption{The same as Fig. \ref{fig:obs:96aq:cmd} but for the Type Ib 2009jf in NGC 7479. The starred point ($\bigstar$) indicates a likely cluster excluded from the analysis.}
\label{fig:obs:09jf:cmd}
\end{figure}

\begin{figure}
\begin{center}
\includegraphics[width=6.0cm, angle=270]{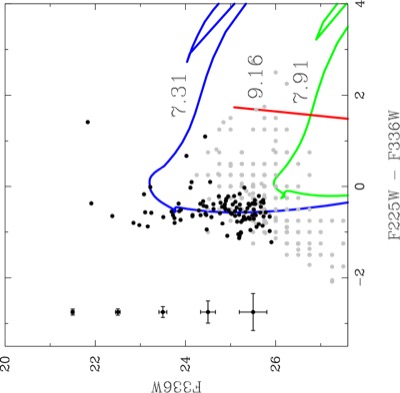}
\includegraphics[width=6.0cm, angle=270]{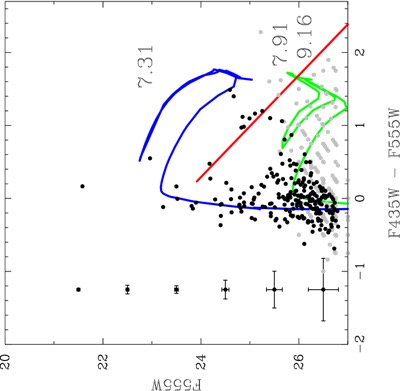}
\includegraphics[width=6.0cm, angle=270]{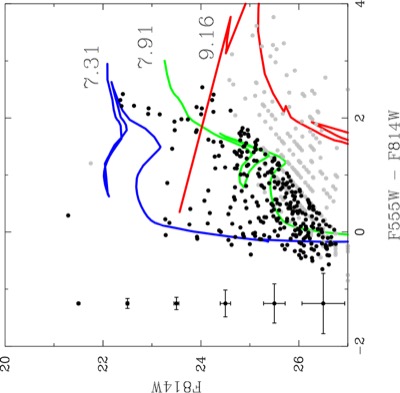}
\end{center}
\caption{The same as Fig. \ref{fig:obs:96aq:cmd} but for the Type IIb 2011dh in M51.}
\label{fig:obs:11dh:cmd}
\end{figure}

Colour magnitude diagrams for all observations for SNe 1996aq, 1996cb, 2000ew, 2001B, 2001gd, 2004gn, 2004gq, 2004gt, 2005at, 2007fo, 2008ax, 2009jf, 2011dh, 2012au, 2012fh, iPTF13bvn, 2013df, 2013dk and 2016bau are presented in Figs. \ref{fig:obs:96aq:cmd} - \ref{fig:obs:16bau:cmd}. 

\begin{figure}
\begin{center}
\includegraphics[width=6.0cm, angle=270]{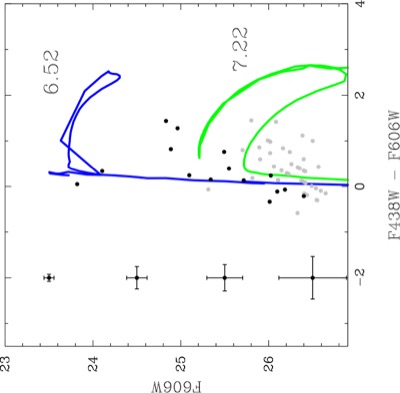}
\end{center}
\caption{The same as Fig. \ref{fig:obs:96aq:cmd} but for theType Ib 2012au in NGC4790.}
\label{fig:obs:12fh:cmd}
\end{figure}

\begin{figure}
\begin{center}
\includegraphics[width=6.0cm, angle=270]{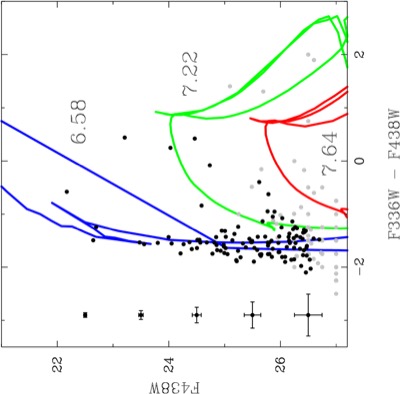}
\includegraphics[width=6.0cm, angle=270]{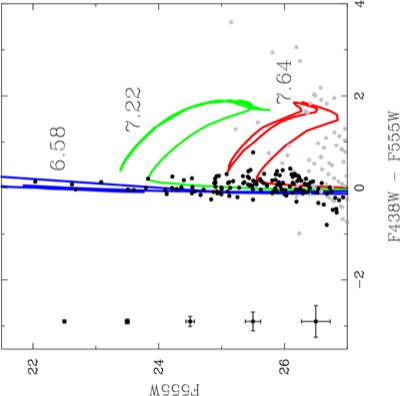}
\includegraphics[width=6.0cm, angle=270]{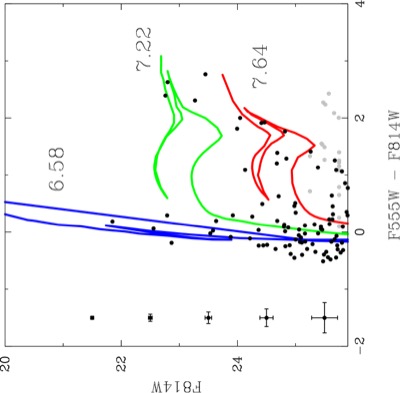}
\end{center}
\caption{The same as Fig. \ref{fig:obs:96aq:cmd} but for theType Ib 2012fh in NGC3344.}
\label{fig:obs:12fh:cmd}
\end{figure}

\begin{figure}
\begin{center}
\includegraphics[width=6.0cm, angle=270]{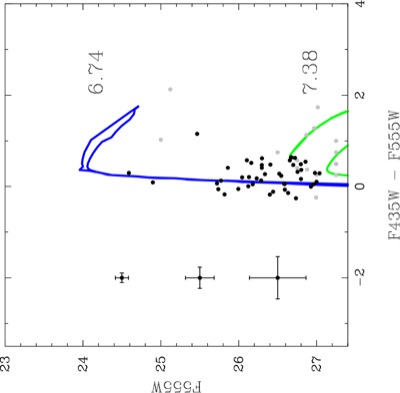}
\includegraphics[width=6.0cm, angle=270]{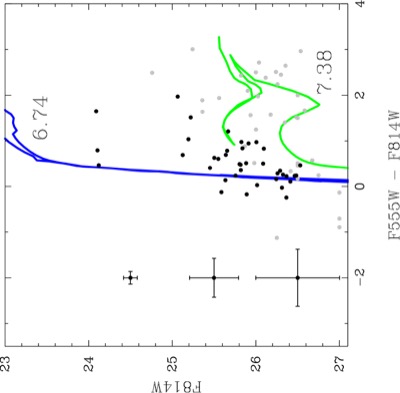}
\end{center}
\caption{The same as Fig. \ref{fig:obs:96aq:cmd} but for the Type Ib SN iPTF13bvn in NGC5806.}
\label{fig:obs:13bvn:cmd}
\end{figure}

\begin{figure}
\begin{center}
\includegraphics[width=6.0cm, angle=270]{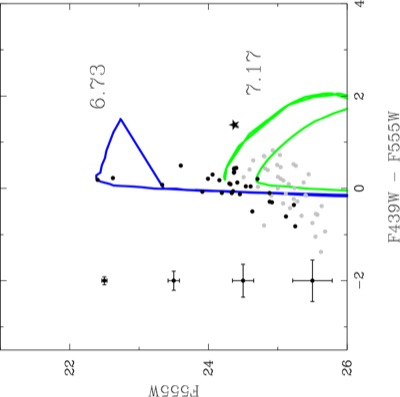}
\includegraphics[width=6.0cm, angle=270]{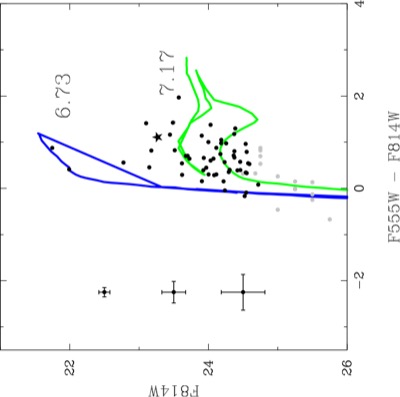}
\end{center}
\caption{The same as Fig. \ref{fig:obs:96aq:cmd} but for the Type IIb 2013df in NGC 4414.  The progenitor source is indicated by the black star ($\star$).}
\label{fig:obs:13df:cmd}
\end{figure}

\begin{figure}
\begin{center}
\includegraphics[width=6.0cm, angle=270]{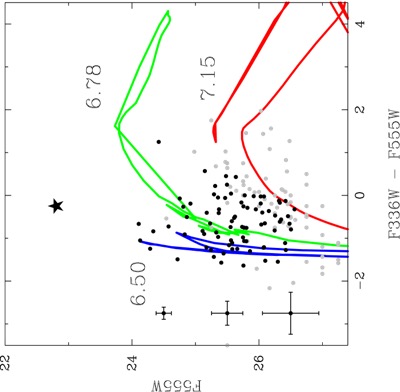}
\includegraphics[width=6.0cm, angle=270]{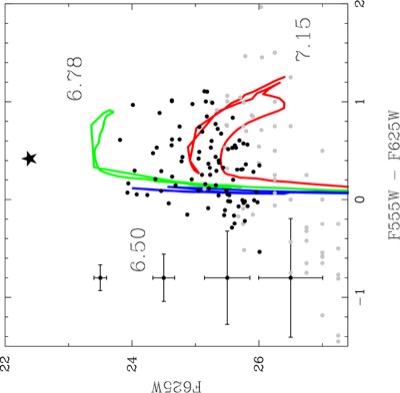}
\includegraphics[width=6.0cm, angle=270]{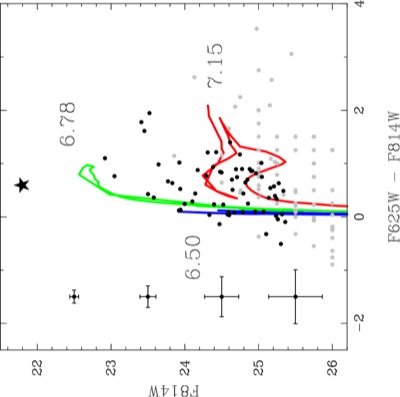}
\end{center}
\caption{The same as Fig. \ref{fig:obs:96aq:cmd} but for the Type Ic 2013dk in NGC 4038.  The bright cluster, in close proximity to the position of SN~2013dk, as identified by \citet{2013MNRAS.436L.109E}, is indicated by the black star ($\star$). }
\label{fig:obs:13dk:cmd}
\end{figure}

\begin{figure}
\begin{center}
\includegraphics[width=6.0cm, angle=270]{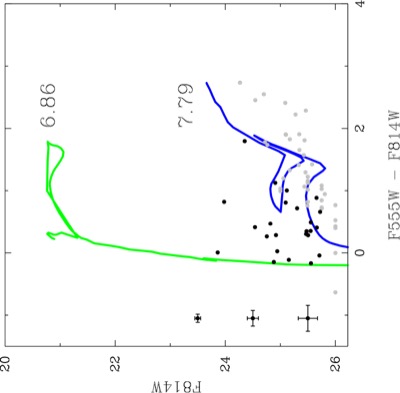}
\end{center}
\caption{The same as Fig. \ref{fig:obs:96aq:cmd} but for the Type Ib 2016bau in NGC 3631.}
\label{fig:obs:16bau:cmd}
\end{figure}

%%%%%%%%%%%%%%%%%%%%%%%%%%%%%
%APPENDIX B - SPATIAL DISTRIBUTION
%APPENDIX B - SPATIAL DISTRIBUTION
%APPENDIX B - SPATIAL DISTRIBUTION
%%%%%%%%%%%%%%%%%%%%%%%%%%%%%
\newpage
\section{Models of  the spatial distribution of stars}
\label{sec:app:spatial}
In the most general case of a SN occuring in a uniformly distributed field population of stars, one would expect the ratios of the number counts of stars within 100, 150 and $300\,\mathrm{pc}$ to have the fixed values of $N_{100}/N_{150} = 2.25$ and $N_{150}/N_{300} = 4$, simply due to the increase in surface area contained within apertures of increasing size.  To evaluate the presence of possible associations at the SN locations, above a background field star density, we consider such an association to be radially symmetric and described by a two-dimensional Gaussian distribution, where the number density of stars is given by: 
\begin{equation}
f_{assoc}(x,y,\Delta y, \sigma) \propto \frac{1}{2\pi\sigma^{2}}\exp\left[{-\frac{1}{2}\left( \frac{x^{2} + (y+ \Delta y)^{2}}{\sigma^2}\right)}\right]
\end{equation}
where $\Delta y$ is the offset of the SN location from the centre of the association and $\sigma$ parameterises the spatial extent of the association.   Due to the need to consider SN positions offset from the centre of any nearby association, we eschew a radial gaussian profile in order to make the process of integrating the profile easier.

\begin{figure}
\includegraphics[width=8cm,angle=270]{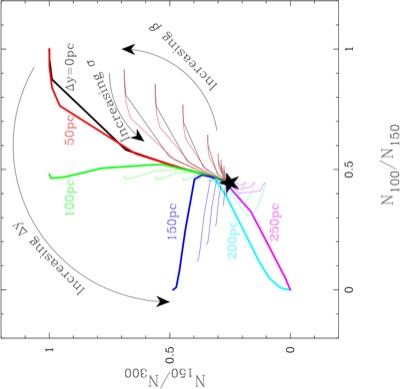}
\caption{The behaviour of the ratio of the number of sources measured in apertures of size $100$, $150$ and $300\,\mathrm{pc}$ for different association properties, parameterised by the offset of the SN location from the centre of the association ($\Delta y$), the spatial width of the association ($\sigma$) and the relative number of association members to the field population ($\beta$).  The heavy lines indicate the expected number count ratios with no background field star population (i.e. $\beta \rightarrow \infty$).   Tracks are shown for offsets $\Delta y = 0$, $50$, $100$, $150$, $200$, and $250\,\mathrm{pc}$, $\beta  = 0.2$, $0.5$, $0.75$, $1$, $2$, $5$, $10$ and $20$ and  $10 \leq \sigma \leq 500\,\mathrm{pc}$}
\label{fig:res:leaf}
\end{figure}

We may also consider a background field contribution to the number counts in the respective apertures, where the corresponding number density of stars is:
\begin{equation}
f_{field}(x, y) = \mathrm{const.}
\end{equation}.
The number of stars $N$ that are therefore expected to arise in an aperture $D$ with radius $R$, offset from the centre of the association by $\Delta y$, is given by the integral over the area of the circular aperture:

\begin{equation}
{N} \propto \int\int_{D(r = R)}\left(\alpha\beta f_{assoc}(x,y,\Delta y, \sigma) + f_{field}\right)\mathrm{d}x\mathrm{d}y					
\label{eqn:res:spatial}
\end{equation}
where $\alpha$ normalises the number of stars arising from the association with the number of field stars inside an aperture within a set radius, chosen as $50\,\mathrm{pc}$ and $\beta$ is the ratio of stars arising from the association and those in the field.  As $\beta$ tends to $0$, for less populous associations, the values for the two ratios will tend towards the field values.  In Figure \ref{fig:res:clus_sep} we show the observed ratios of number counts for the sample of SNe considered here.  Using the form of Equation \ref{eqn:res:spatial} we can identify different regimes for the spatial extent and proximity to the SN position for a possible massive star association.  The consequences of increasing the spatial offset ($\Delta y$), the spatial breadth of the stellar association ($\sigma$) and the number of stars in the association relative to the background field population ($\beta$) are shown on Fig. \ref{fig:res:leaf}.

\end{document}